\documentclass[11pt]{article}
\usepackage{verbatim,amsmath,amssymb}
\usepackage{epsfig,color}
\usepackage[font=small,labelfont=bf]{caption}
\usepackage{geometry}
\usepackage{setspace}
\usepackage{natbib}
\usepackage{amsmath,amssymb,amsthm,mathrsfs,amsfonts,dsfont}   
\usepackage{hyperref}
\usepackage{float}
\usepackage{graphicx}
\usepackage{scalerel}
\usepackage{todonotes}
\usepackage{longtable}

\makeatletter
\newcommand\footnoteref[1]{\protected@xdef\@thefnmark{\ref{#1}}\@footnotemark}
\makeatother

\definecolor{darkblue}{rgb}{0,0,1}
\hypersetup{pdftex=true, colorlinks=true, breaklinks=true, linkcolor=darkblue, menucolor=darkblue, pagecolor=darkblue, citecolor=darkblue, urlcolor=darkblue}

\newtheoremstyle{rem}
{6pt}
{6pt}
{\small}
{}
{\bf}
{:}
{.5em}
{}

\theoremstyle{rem}
\newtheorem{remark}{Remark}[section]

\input{neco_updated.tex}

\renewcommand{\thefootnote}{\fnsymbol{footnote}}

\begin{document}

\begin{center}
\Large{\bf{A general theory for anisotropic Kirchhoff-Love shells with in-plane bending of embedded fibers}}\\

\end{center}

\begin{center}
\large{Thang X. Duong$^{a,b}$, Vu N. Khi{\^e}m$^b$, Mikhail Itskov$^{b,}$\footnote[1]{\label{note0} {corresponding authors, email: itskov@km.rwth-aachen.de; sauer@aices.rwth-aachen.de}}, and Roger A. Sauer$^{a,c,d,}$\footnoteref{note0}}\\
\vspace{4mm}

\small{$^a$\textit{Aachen Institute for Advanced Study in Computational Engineering Science (AICES), RWTH Aachen
University, Templergraben 55, 52056 Aachen, Germany}}

\small{$^b$\textit{Department of Continuum Mechanics, RWTH Aachen
University, Templergraben 55, 52056 Aachen, Germany}}

\small{$^c$\textit{Faculty of Civil and Environmental Engineering, Gda{\'n}sk University of Technology,\\ ul. Narutowicza 11/12, 80-233 Gda{\'n}sk, Poland}}

\small{$^d$\textit{Department of Mechanical Engineering, Indian Institute of Technology Kanpur, UP 208016, India}}

\end{center}

\renewcommand*{\thefootnote}{\arabic{footnote}}
\setcounter{footnote}{0}
\begin{center}
Published\footnote{This pdf is the personal version of an article whose journal version is available at \href{https://journals.sagepub.com/doi/full/10.1177/10812865221104427}{https://journals.sagepub.com}} in Mathematics and Mechanics of Solids, \href{https://doi.org/10.1177/10812865221104427}{DOI: 10.1177/10812865221104427}\\
Submitted on 12 October 2021; Last revised on 13 May 2022; Accepted on 14 May 2022
\end{center}


%
%
%


\rule{\linewidth}{.15mm}
{\bf Abstract:} This work presents a generalized Kirchhoff-Love shell theory that can  explicitly capture  fiber-induced anisotropy not only in stretching and out-of-plane bending, but also in in-plane bending. This setup is particularly suitable for heterogeneous  and fibrous materials such as textiles, biomaterials, composites and pantographic structures. The presented theory is a direct extension of classical Kirchhoff-Love shell theory to incorporate the in-plane bending resistance of fibers. It also extends  existing second-gradient Kirchhoff-Love shell theory for initially straight fibers to initially curved fibers.  To describe the additional kinematics of multiple fiber families, a so-called in-plane curvature tensor -- which is symmetric and of second order -- is proposed. The effective stress tensor and the in-plane and out-of-plane moment tensors are then identified from the mechanical power balance. These tensors are all second order and symmetric in general. Constitutive equations for hyperelastic materials are derived from different expressions of the mechanical power balance. The  weak form is also presented as it is required for computational shell formulations based on rotation-free finite element discretizations. 
 

\vspace{-4mm}
\rule{\linewidth}{.15mm}
%


%
%
%

%
\vspace{-4mm}
\section*{List of important symbols}
%
\begin{longtable}[l]{ l l }
$ \bone $ & identity tensor in $\bbR^3$ \\ 
$\ba_\alpha$, $\ba^\alpha$  & co- and contravariant tangent vectors at surface point $\bx\in\sS$; $\alpha = 1, 2$ \\ 
$\bA_\alpha$, $\bA^\alpha$ & co- and contravariant tangent vectors at surface point $\bX\in\sS_0$; $\alpha = 1, 2$ \\  
$ \ba_{\alpha,\beta}$ & parametric derivative of $\ba_\alpha$ w.r.t.~$\xi^\beta$ \\ 
$ \ba_{\alpha;\beta}$ & covariant derivative of $\ba_\alpha$ w.r.t.~$\xi^\beta$ \\  
$ a_{\alpha\beta} $, $ a^{\alpha\beta}$  & co- and contravariant surface metric at surface point $\bx\in\sS$ \\ 
$ A_{\alpha\beta}$, $ A^{\alpha\beta}$ & co- and contravariant surface metric at surface point $\bX\in\sS_0$ \\ 
$ \bb $ & out-of-plane curvature tensor of surface $\sS$ at surface point $\bx\in\sS$ \\ 
$ \bbbar $ & in-plane curvature tensor at fiber point $\bx$ of fiber $\sC$ embedded in $\sS$ \\ 
$ b_{\alpha\beta}$ & covariant components of tensor $\bb$ at surface point $\bx\in\sS$ \\ 
$ \bar{b}_{\alpha\beta}$ & covariant components of tensor $\bbbar$ at fiber point $\bx\in\sC\subset\sS$\\ 
$ \bb_\mathrm{0} $ & out-of-plane curvature tensor of surface $\sS_0$ at surface point $\bX\in\sS_0$ \\ 
$ \bbbar_\mathrm{0} $ & in-plane curvature tensor of fiber $\sC_0$ at fiber point $\bX\in\sC_0\subset\sS_0$ \\ 
$ B_{\alpha\beta}$ & covariant components of tensor $\bb_\mathrm{0}$ at surface point $\bX\in\sS_0$ \\ 
$ \bar{B}_{\alpha\beta}$ & covariant components of tensor $\bbbar_\mathrm{0}$ at fiber point $\bX\in\sC_0\subset\sS_0$ \\ 
$ \beta_\bullet $ & material parameters for fiber bending and torsion \\
$\bc$ & in-plane fiber director vector of fiber $\sC$ at fiber point $\bx\in\sC\subset\sS$\\
$\bc_\mathrm{0}$ & in-plane fiber director vector of fiber $\sC_0$ at fiber point $\bX\in\sC_0\subset\sS_0$ \\
$c_\alpha$, $c^\alpha$ & co- and contravariant components of vector $\bc$ at fiber point $\bx\in\sC\subset\sS$ \\
$c_{\alpha;\beta}$, $c^{\alpha}_{;\beta}$  & covariant derivatives of $c_\alpha$ and $c^\alpha$\\
$ \bC $ & right Cauchy-Green tensor of the shell mid-surface\\
$ \sC $ & a curve representing a fiber embedded in shell surface $\sS$\\
$\sC_0$ & initial configuration of fiber curve $\sC$ embedded in shell surface $\sS_0$\\
$\bD$ & rate of surface deformation tensor \\
$ \delta ... $ & variation of $...$ \\
$ \delta_\alpha^\beta  $ & surface Kronecker delta  \\ 
$ \epsilon_\bullet $ & material parameters for fiber stretching and shearing \\
$ \bE $ & Green-Lagrange strain tensor of the shell mid-surface \\
$ \bff $ & prescribed surface loads \\
$ f^\alpha $ & in-plane components of $ \bff$ \\
$ \bF $ & deformation gradient of the shell mid-surface\\
$\gamma_{ij}$ & nominal angle between current fiber configurations $\sC_i$ and $\sC_j$ in $\sS$; $i \neq j$ \\
$\gamma_{ij}^0$ & nominal angle between reference fiber configurations $\sC_{0i}$ and $\sC_{0j}$ in $\sS_0$; $i \neq j$ \\
$\hat\gamma_{ij}$ & absolute angle between fiber $\sC_i$ and $\sC_j$ ; $i \neq j$ \\
$ G_\mathrm{in} $ & inertial virtual work \\
$ G_\mathrm{ext} $ & external virtual work \\
$ G_\mathrm{int} $ & internal virtual work \\
$ \Gamma^\gamma_{\alpha\beta} $ & surface Christoffel symbols of the second kind on $\sS$\\
$\bar{ \Gamma}^\gamma_{\alpha\beta} $ & surface Christoffel symbols of the second kind on $\sS_0$ \\
$ \Gamma^\mrc_{\alpha\beta} $ & projection of vectors $\ba_{\alpha,\beta}$ in direction $\bc$ \\
$ \Gamma^{\uell}_{\alpha\beta} $ & projection of vectors $\ba_{\alpha,\beta}$ in  direction $\bell$\\
$\bH$ & collective symbol for various structural tensors \\
$ H $ & mean curvature of surface $\sS$ at surface point $\bx\in\sS$ \\ 
$h^{\alpha\beta}$ & collective symbol for components of various structural tensors \\
$ i $ &  fiber index; added to all fiber-related quantities in case of multiple fibers\\
$ \bi $ & surface identity tensor on $\sS$\\
$ \bI $ & surface identity tensor on $\sS_0$ \\
$ I_1 $ & first invariant of $ \bC $ \\
$\sI$ & parametrized interface obtained by cutting through $\sS$ \\ 
$ j $ &  fiber index; added to all fiber-related quantities in case of multiple fibers\\
$ J $ & surface area change \\
$\mathrm{k}_\mrn$ & absolute change in normal curvature $\kappa_\mrn$ at point $\bx$ of fiber $\sC\subset\sS$\\
$\mathrm{k}_\mrg$ & absolute change in geodesic curvature $\kappa_\mrg$ at point $\bx$ of fiber $\sC\subset\sS$\\
$k_\mrn$ & stretch-excluded change in normal curvature $\kappa_\mrn$ at point $\bx$ of fiber $\sC\subset\sS$\\
$k_\mrg$ & stretch-excluded change in geodesic curvature $\kappa_\mrg$ at point $\bx$ of fiber $\sC\subset\sS$\\
$ \kappa $ & Gaussian curvature of surface $\sS$ at surface point $\bx\in\sS$ \\ 
$ \kappa_\mrn $  & normal curvature of current fiber configuration $\sC$ at $\bx\in\sC\subset\sS$ \\ 
$ \kappa_\mrn^0 $  & normal curvature of reference fiber configuration $\sC_0$ at $\bX\in\sC_0\subset\sS_0$\\ 
$ \kappa_\mrg $& geodesic curvature of current fiber configuration $\sC$ at $\bx\in\sC\subset\sS$ \\ 
$ \kappa_\mrg^0 $ & geodesic curvature of reference fiber configuration $\sC_0$ at $\bX\in\sC_0\subset\sS_0$ \\[0.55mm] 
$ \kappa_\mrg^\Gamma $ & contribution of the Christoffel symbol to geodesic curvature $\kappa_\mrg$ of fiber $\sC$   \\ [0.55mm] 
$ \kappa_\mrg^{\mathrm{L}} $ & contribution of the gradient of $\bL$ to geodesic curvature $\kappa_\mrg$ of fiber $\sC$   \\ 
$ \kappa_\mrp $ & principal curvature of fiber $\sC$ at fiber point $\bx\in\sC\subset\sS$ \\ 
%
$K$ & kinetic energy of surface $\sS$ \\ 
$K_\mrn$ & nominal change in normal curvature $\kappa_\mrn$ at point $\bx$ of fiber $\sC\subset\sS$\\
$K_\mrg$ & nominal change in geodesic curvature $\kappa_\mrg$ at point $\bx$ of fiber $\sC\subset\sS$\\
$ \bK $ & relative change of the out-of-plane curvature tensor  \\ 
$ \bar\bK $ & relative change of the in-plane curvature tensor  \\ 
$ \bell $ & normalized tangent vector of fiber $\sC$ at fiber point $\bx\in\sC\subset\sS$\\ 
$\ell_\alpha$, $\ell^\alpha$ & co- and contravariant components of $\bell$ in $\sS$; $\alpha = 1, 2$ \\
$\ellab$ & contravariant components of structural tensor $\bell\otimes\bell$ in $\sS$\\
$\ell_{\alpha;\beta}$, $\ell^{\alpha}_{;\beta}$ & covariant derivatives of $\ell_\alpha$ and $\ell^\alpha$\\
%
$\lambda$ & stretch of fiber $\sC$ at fiber point $\bx\in\sC\subset\sS$ \\
$\bL$ & normalized tangent vector of fiber $\sC_0$ at fiber point $\bX\in\sC_0\subset\sS_0$\\
$L_\alpha$, $L^\alpha$ & co- and contravariant components of $\bL$; $\alpha = 1, 2$ \\
$L^{\alpha\beta}$ & contravariant components of structural tensor $\bL\otimes\bL$\\
$\hat{L}^\alpha_{,\beta}$, $\hat{L}_{\alpha,\beta}$  & parametric derivatives $L^\alpha_{,\beta}$ and ${L}_{\alpha,\beta}$ scaled by inverted fiber stretch $\lambda^{-1}$\\ 
$\Lambda$ & $= \lambda^2$; square of stretch of fiber $\sC\subset\sS$ \\
%
$\bmhat$ &  moment vector acting on a cut $\sI$ normal to $\bnu$\\ 
$\bmhat^\alpha$ & moment vector acting on a cut $\sI$ normal to $\ba^\alpha$ \\
$\bm$  & component of $\bmhat$ causing out-of-plane bending and twisting \\
$\bmbar$ & component of $\bmhat$ causing in-plane bending \\ 
$m^{\alpha\beta}$ &  components of moment tensor $\bmuhat$ with basis $\ba_\alpha\otimes\ba_\beta$ \\ 
$\bar{m}^{\alpha}$ &  components of moment tensor $\bmuhat$ with basis $\ba_\alpha\otimes\bn$ \\ 
$m_\tau$, $m_\nu$, $\bar{m}$ & components of moment vector $\bmhat$ in directions $\btau$, $\bnu$, and $\bn$\\ 
$ \mu $ & surface shear modulus \\
$ \bar{\mu} $ & component of tensor $\bmuhat_{\!\mathrm{fib}}$ with basis $\bell\otimes\bn$, causing in-plane bending of fiber $\sC$ \\
$ \bar{\mu}_0 $ & moment component $\bar{\mu}$ scaled by $J$ \\
$ \bmu $  & stress couple tensor associated with out-of-plane bending and twisting \\
$ \bmu_0 $ & stress couple tensor obtained by the pull-back of tensor $J\,\bmu$.  \\
$ \bmubar $ & stress couple tensor associated with in-plane bending of fiber $\sC$\\
$ \bmubar_0 $ & stress couple tensor obtained by the pull-back of tensor $J\,\bmubar$.   \\
$ \bmuhat $ & (total) internal moment tensor at $\bx\in\sS$ \\
$ \bmuhat_{\!\mathrm{fib}} $ & (total) internal moment tensor within fiber $\sC$ \\
%
$\bM$ & stress couple vector associated with out-of-plane bending at cut $\sI\perp \bnu$ \\ 
$\bM^\alpha$ & stress couple vectors associated with out-of-plane bending at cut $\sI \perp\ba^\alpha$ \\ 
$\bMbar$ & stress couple vector associated with in-plane bending of fiber $\sC$ at cut $\sI \perp\bnu$ \\ 
$\bMbar^\alpha$ & stress couple vectors associated with in-plane bending at cut $\sI \perp\ba^\alpha$ \\ 
$ M^{\alpha\beta} $ & contravariant components of stress couple tensor $-\bmu$  \\ 
$ M_0^{\alpha\beta} $ & stress couple components $ M^{\alpha\beta}$  scaled by $J$ \\ 
$ \bar{M}^{\alpha\beta} $ & contravariant components of stress couple tensor $-\bmubar$  \\ 
$ \bar{M}_{0}^{\alpha\beta} $ &stress couple components $ \bar{M}^{\alpha\beta}$  scaled by $J$ \\ 
$\bar{M}^{\alpha\beta}_{0\gamma}$ & stress couple components for in-plane bending in second-gradient shell theory \\ 
$ \bn $ & unit surface normal vector of $\sS$ at surface point $\bx\in\sS$ \\
$ \bn_\mrp $ & principal normal vector of fiber $\sC$ at fiber point $\bx\in\sC\subset\sS$ \\
$n_\mrf$ & number of fiber families \\
$ \bN $ & unit surface normal vector of $\sS_0$ at surface point $\bX\in\sS_0$ \\ 
$ N^{\alpha\beta} $ & in-plane contravariant components of Cauchy stress tensor $\bsig$ \\
$\nabla_{\!\mrs}\, \bullet$ & $=\bullet_{,\alpha}\otimes\ba^\alpha$; surface gradient operator  \\
$\bar\nabla_{\!\mrs}\, \bullet$ & $=\bi\,\nabla_{\!\mrs}\, \bullet$; projected surface gradient operator   \\
$ \boldsymbol{\bnu} $ & unit normal on cut $\sI$  \\ 
$ \nu_\alpha $, $\nu^\alpha$ & co- and contravariant components of $\bnu$ \\
$ \xi^\alpha $ & curvilinear coordinates; $\alpha = 1, 2$ \\ 
%
$\dot{w}_{\mathrm{int}}$ & internal stress power per current area \\ 
$\bP$ & first Piola-Kirchhoff stress tensor of the shell surface \\ 
$P_{\mathrm{int}}$, $P_{\mathrm{ext}}$ & surface internal and external power \\
$ \sP $ & parametric domain spanned by $\xi^1$ and $\xi^2$ \\
$ \sR $, $ \sR_0 $  & arbitrary simply-connected sub-region of surface $\sS$ or $\sS_0$\\
$ q_i $ & Lagrange multiplier for the inextensibility constraint for fiber $\sC_i$ \\
$ \rho $ & areal mass density of surface $\sS$ \\
$s$ & parameter coordinate of fiber $\sC$ \\
$\sS$ & current configuration of the shell surface \\ 
$\sS_0$ & reference configuration of the shell surface \\ 
$\partial\sS$ & boundary of $\sS$ \\
$ \bS $ & second Piola-Kirchhoff stress tensor of the shell surface \\
$ S^\alpha $ & contravariant, out-of-plane shear stress components \\ %
$ S^\gamma_{\alpha\beta} $ & change in Christoffel symbols from $\sS_0$ to $\sS$ \\
$ \bsig $ &  Cauchy stress tensor within the shell \\
$ \bsig_{\!\mathrm{fib}} $ &  Cauchy stress tensor within fiber $\sC$ \\ 
%
$\tilde\sigma^{\alpha\beta} $ & effective membrane stress associated with in-plane curvature measure $c^\beta\ell_{\beta;\alpha}$ \\
$\sigma^{\alpha\beta} $ & effective membrane stress associated with in-plane curvature measure $\kappa_\mrg$ or $\bar{b}_{\alpha\beta}$ \\
$ t $ & time variable \\ 
%
$\mathrm{t}_\mrg$ & absolute change in geodesic torsion $\tau_\mrg$ at point $\bx$ of fiber $\sC\subset\sS$\\
$t_\mrg$ & stretch-excluded change in geodesic torsion $\tau_\mrg$ at point $\bx$ of fiber $\sC\subset\sS$\\
$T_\mrg$ & nominal change in geodesic torsion $\tau_\mrg$ at point $\bx$ of fiber $\sC\subset\sS$\\
$ \bT $ & traction vector acting on cut $\sI$ normal to $\bnu$ \\
$\bT^\alpha$ & traction vectors acting on cut $\sI$ normal to $ \ba^\alpha $ \\
$T^\alpha$ & in-plane contravarint components of traction vector $\bT$ \\ 
$T^3$ &  component of traction vector $\bT$ in direction $\bn$ \\ 
$ \btau $ & unit vector along cut $\sI$\\
$ \btau^\alpha $ & effective stress vector work-conjugate to $\dot\ba_\alpha$ in second-gradient shell theory\\
$ \hat\btau $ & Kirchhoff stress tensor of the shell surface \\
$\tau_\mrg$& geodesic torsion of current fiber configuration $\sC$ at $\bx\in\sC\subset\sS$  \\
$\tau_\mrg^0$ & geodesic torsion of reference fiber configuration $\sC_0$ at $\bX\in\sC_0\subset\sS_0$  \\
$ \tau_\alpha $, $\tau^\alpha$ & co- and contravariant components of vector $\btau$ \\
$ \tau^{\alpha\beta} $ & contravariant components of Kirchhoff stress tensor $ \hat\btau $ of the shell \\
$ \bv $ & velocity, i.e. the material time derivative of $\bx$ \\ 
$ \sV $ & space of admissible variations $\delta\bx$ \\
$ W $ & strain energy density function per reference area \\
$ \bx $ & current position of a surface point on the current shell surface $\sS$ \\ 
$ {\bx_\mrc} $ & {function describing curve  $\sC$} \\ 
$ \bX $ & initial position of $\bx$ on the reference shell surface $\sS_0$ \\
%
$\dot\bullet$ & material time derivative\\
\end{longtable}
\setcounter{table}{0}

\section{Introduction}\label{s:intro}

Fiber reinforced composites have become an important material in the sports, automotive, marine, and aerospace industry owing to their high specific stiffness-to-weight ratio, which allows for lightweight designs. To produce such composites, fabric sheets are formed by warp and weft yarns (i.e.~in a bundle of fibers) loosely linked together by different technologies, resulting for example in woven fabrics or non-crimp fabrics. The fabrics are then  draped (molded) into desired shapes before injecting liquid resin (adhesives).  After the resin  solidifies, the fibers are strongly bonded together in the final product.
  
In this work, we are interested in continuum models for fabric sheets both with and without matrix material. Such models are required for the description and simulation of fiber-reinforced shell structures and draping processes.  Geometrically, fabric structures can be modeled by a surface with embedded curves representing yarns. From the microscopic point of view, the resistance (in-plane and out-of-plane) of fabrics results from the deformation of yarns and their interaction, i.e.~their linkage and contact. In particular, axial stretching of fibers is associated with (anisotropic) membrane resistance, and the linkage between yarns offers  shear resistance. Twisting of a yarn
can be assumed to be fully associated with the second fundamental form of the yarn-embedding surface \citep{Steigmann2015}. Bending of a yarn can have both in-plane and out-of-plane components and is characterized by the corresponding curvatures. The out-of-plane curvature is associated with the second fundamental form, while the in-plane curvature is associated with the gradient of the surface metric.

Fabric sheets can be modeled as thin shells from the macroscopic point of view, and Kirchhoff-Love kinematics together with plane stress conditions are usually adopted. Membrane deformation is characterized by stretching and shearing, while out-of-plane deformation is characterized by bending and twisting. In Kirchhoff-Love shell models, two kinematical quantities -- the surface metric and  the second fundamental form -- are used for the two deformation types.  In the literature, the general case of arbitrarily large deformations and nonlinear material behavior of shells has been treated extensively. See e.g.~the texts of \cite{naghdi82,pietraszkiewicz89,libai} and references therein. Here, we refer to  
this nonlinear case as classical Kirchhoff-Love shell theory. Note that although Kirchhoff-Love shell theory is mostly discussed for solids, its application can also be extended to liquid shells \citep{steigmann99b}. The incorporation of material anisotropy  in classical Kirchhoff-Love shells due to embedded fibers -- for both stretching and  out-of-plane bending -- is straightforward, see e.g.~\cite{TEPOLE2015}, \cite{WU201823} and references therein. As shown by \citet{Roohbakhshan2017}, classical Kirchhoff-Love shell theory also admits  complex anisotropic bending models, e.g.~due to fibers not located at the mid-surface.

While classical Kirchhoff-Love shell theory can also be regarded as a special case of Cosserat theory \citep{steigmann99}\footnote{From another point of view, classical  Kirchhoff-Love shell theory  also falls in the category of  high gradient theories due to the high gradient term in the bending energy.}, it has its own development history and has the advantage of simplicity and intuitiveness when following its argument structure. Therefore, it facilitates building corresponding computational as well as physically-based constitutive models.  This motivated \citet{shelltheo} to provide a unified formulation for both liquid and solid shells (within the framework of classical Kirchhoff-Love shell theory). Their work aimed at providing a concise, yet general  theoretical framework for  corresponding computational rotation-free shell formulations \citep{solidshell,liquidshell}. 
%
%

%

It should be noted that most existing shell models, including classical Kirchhoff-Love shell theory, focus on out-of-plane bending, while the in-plane response is still based on the classical Cauchy continuum. That is, one assumes that there is no moment (or stress couple) causing in-plane bending at a material point.
This assumption is usually sufficient when only the overall material behavior is of interest, as e.g.~in the draping simulations of \citet{KhiemNCF2018}.
However, 
it fails to capture deformations governed by in-plane fiber bending, which are important for obtaining accurate and convergent numerical results. An example of these are  the localized shear bands  \citep{Ferretti2014,Boisse17} in the bias-extension test. 
In this case, simulations with finite element shell models based on the in-plane Cauchy  continuum will fail to converge to a finite width of the shear band under mesh refinement. Another example is the asymmetric deformation of  woven fabrics with different  in-plane fiber bending stiffness for different  fiber families \citep{Madeo2016,Barbagallo17}.


Similar effects of the in-plane bending stiffness can also be found in pantographic structures \citep{dellisola_large_2016,placidi_second_2016,dellisola_advances_2019} and fibrous composites, such as biological tissues \citep{Gasser2006}.  Thus, a more general model that considers the in-plane bending response is required for fabrics, fibrous composites and pantographic structures.

The first theoretical work considering in-plane bending was presented by \cite{Wang86, Wang87} to model cloth and cable networks. Their theory is  a special form of finite-deformation fibrous plate theory with inextensible fibers that contain bending couples that are proportional to their curvature.
Since in-plane bending is related to the in-plane components of the second displacement gradient, it can be captured by more general continuum theories, such as Cosserat theories (see e.g.~\cite{Mindlin1962,Koiter63b,toupin_theories_1964}), and gradient theories (see e.g.~\cite{Green64,Mindlin64,Mindlin65,Germain73}). For 3D fiber-reinforced solids, \cite{spencer_finite_2007}  introduced explicitly the bending resistance of embedded fibers  in the context of nonlinear second-gradient theory. 
A computational model based on \cite{spencer_finite_2007} has  been developed by \cite{Asmanoglo17}. Starting from Cosserat theory, \cite{steigmann_theory_2012} derives  a fiber-reinforced solid model that includes fiber bending, twisting, and stretching.

Concerned with in-plane bending for thin structures, 
%
%
\citet{Steigmann2015} presented a continuum model for woven fabric sheets  modeled as orthotropic plates, which treats fibers as Kirchhoff-Love rods that are distributed continuously across the sheet. Although the concept of stress couples from Cosserat theory is used, the model of \citet{Steigmann2015} can be categorized as second-gradient theory, since their material model depends on the first and second displacement gradients.  
Following this, \cite{Steigmann2018} further developed a second-gradient shell model that explicitly includes general fiber bending, twisting, and stretching. There, a concise set of equilibrium equations, boundary conditions, and material symmetries are discussed. 
However, to the best of our knowledge, the theory has not yet been fully formulated for the general case of more than two fiber families with initially curved fibers. \\[2.5mm]
%
%
Focusing on Kirchhoff-Love shells, \cite{Balobanov19} presented a new shell model derived from the second displacement gradient theory of \cite{Mindlin64}.  A corresponding computational formulation was also discussed in \cite{Balobanov19}. However, explicit in-plane fiber bending is not considered and the weak form requires at least $C^2$-continuity of the geometry.  Recently, \cite{Schulte2020} applied the second-gradient theory of \cite{Steigmann2018} directly to Kirchhoff-Love shells, and presented the first rotation-free computational shell formulation accounting for in-plane bending using  $C^1$-continuous discretization.

In this contribution, we propose a general Kirchhoff-Love shell theory that explicitly incorporates fiber bending (both in-plane and out-of-plane), geodesic fiber twisting, and stretching. Unlike the existing approaches that derive the Kirchhoff-Love shell from a more general theory, here our theory is constructed directly from Kirchhoff-Love thin shell assumptions without introducing extra degrees-of-freedom, such as independent directors or micro-displacements. It is shown that the in-plane fiber bending contribution can be formulated analogously to its out-of-plane counterpart. All the definitions of stresses and moments, and the equilibrium equations thus follow in the same manner as in classical Kirchhoff-Love shell theory. Unlike the second-gradient theory of \cite{Steigmann2018}, the stress couple (or double force) in our theory is fully equivalent to the bending moment under Kirchhoff-Love assumptions. 

Our approach provides several advantages over existing second-gradient shell theories, such as those of~\cite{Steigmann2015} and \cite{Steigmann2018}. In particular, it allows us to identify work-conjugated pairs of symmetric stress and symmetric strain measures for all terms, including a new in-plane stress couple and corresponding in-plane curvature tensor. Instead of using third order tensors for  in-plane bending, as is done in current second-gradient shell theories, the stress and strain tensors in our theory are all of second order. Their invariants thus can be easily identified and geometrically interpreted, which is advantageous for constructing constitutive models. Furthermore, our theory admits a wide range of constitutive models for straight as well as initially curved fibers without limitation on the number of fiber families and the angles between them.

Besides, the presented work also aims at providing a general formulation that is suitable for a straightforward isogeometric finite element implementation \citep{shelltextileIGA}. Since the in-plane bending term in our theory is analogous to the out-of-plane bending term, existing finite element formulations for out-of-plane bending can be easily extended to in-plane bending. As for existing formulations, the corresponding weak form requires at least $C^1$-continuous surface discretizations in the framework of rotation-free finite element formulations. For the purpose of verifying finite element implementations, we further provide the analytical solution for several nonlinear benchmark examples.
Here, we are restricting ourselves to hyperelastic material models for the fabrics. Inelastic behavior, e.g.~due to inter- and intra-ply fiber sliding, will be considered in future work.

In summary, our approach contains the following novelties and merits compared to earlier works:


$\bullet$ A concise shell theory with in-plane bending that is a direct extension of classical  Kirchhoff-Love thin shell theory, 

$\bullet$ a general theory that can admit a wide range of 
constitutive models for straight or initially curved fibers with no limitation on the number of fiber families and the angles between them,

$\bullet$ the introduction of a new symmetric in-plane curvature tensor,

$\bullet$ the identification of work-conjugated pairs of symmetric stress and symmetric strain measures,

$\bullet$ the weak form as it is required for rotation-free finite element formulations,

$\bullet$ the analytical solution for several nonlinear benchmark examples that include different modes of fiber deformation and are useful for verifying computational formulations. 

The following presentation is structured as follows:  Sec.~\ref{s:kinematics} summarizes the kinematics of thin shells with embedded curves and introduces the in-plane curvature tensor to capture the  in-plane curvature of fibers. With this, the balance laws are presented in Sec.~\ref{s:balance}. Different choices of work-conjugated pairs are then discussed in Sec.~\ref{s:workconjugation}. Sec.~\ref{s:examples} gives some examples of constitutive models for the proposed theory. Sec.~\ref{s:weakform} then presents the weak form. Several analytical benchmark examples supporting the proposed theory are presented in Sec.~\ref{s:analyticalexamples}.  The paper is concluded by Sec.~\ref{s:conclusion}.

\section{Kinematics for thin shells with embedded curves} {\label{s:kinematics}}
In this section, nonlinear Kirchhoff-Love shell kinematics \citep{naghdi82}  is extended to deforming surfaces with embedded curves. The extended kinematics allows  to capture not only stretching and out-of-plane bending, but also in-plane bending of the embedded curves. Fiber-induced anisotropy is allowed in all these modes of deformation. The description is  presented fully in the general framework of curvilinear coordinates. The variation of different kinematical quantities can be found in Appendix~\ref{s:variation}.



\subsection{Geometric description}

The mid-surface of a thin shell at time $t$ is modeled  in three dimensional space as a 2D manifold, denoted by $\sS$.  In curvilinear coordinates, $\sS$ is described by the one-to-one mapping of a point $(\xi^1\,,\xi^2)$ in parameter space {$\sP$} to the point $\bx\in\sS$ as
\eqb{l}
\bx=  \bx(\xi^\alpha,t)~,\quad $with$ \quad \alpha=1,2~.
\label{e:x_shell}
\eqe
The (covariant) tangent vectors along the convective coordinate curves $\xi^\alpha$ at any point $\bx\in\sS$ can  be defined by
\eqb{l} 
 \ba_\alpha:=\ds\pa{\bx}{\xi^\alpha} = \bx_{,\alpha}~,
 \label{e:tangentsl}
\eqe
where the comma denotes the parametric derivative. The unit normal vector can then be defined by
\eqb{l}
\bn := \ds\frac{\ba_1\times\ba_2}{\norm{\ba_1\times\ba_2}}~.
\label{e:normalmsl}
\eqe

From the tangent vectors $\ba_\alpha$ in Eq.~\eqref{e:tangentsl},  the so-called dual tangent vectors, denoted by $\ba^\alpha$, are defined by 
\eqb{lll}
 \ba^\alpha\cdot\ba_\beta = \delta^\alpha_\beta~,
 \label{e:def_dualvector}
 \eqe
with $\delta^\alpha_\beta$ being the Kronecker delta.   The covariant and dual tangent vectors are related to each other by\footnote{In this paper, the summation convention is applied for repeated Greek indices taking values from 1 to 2.}  $\ba_{\alpha}=a_{\alpha\beta}\,\ba^\beta$ and $\ba^{\alpha}=a^{\alpha\beta}\,\ba_\beta~$, where
 \eqb{l} 
a_{\alpha\beta}:=\ba_\alpha\cdot\ba_\beta~,\quad\quad a^{\alpha\beta}:=\ba^\alpha\cdot\ba^\beta
\label{e:a_ab}
\eqe
denote  the co- and contravariant surface metric, respectively.
\begin{figure}[htp]
\begin{center} \unitlength1cm
\begin{picture}(0,6.8)
\put(-6.5,0.0){\includegraphics[width=0.8\textwidth]{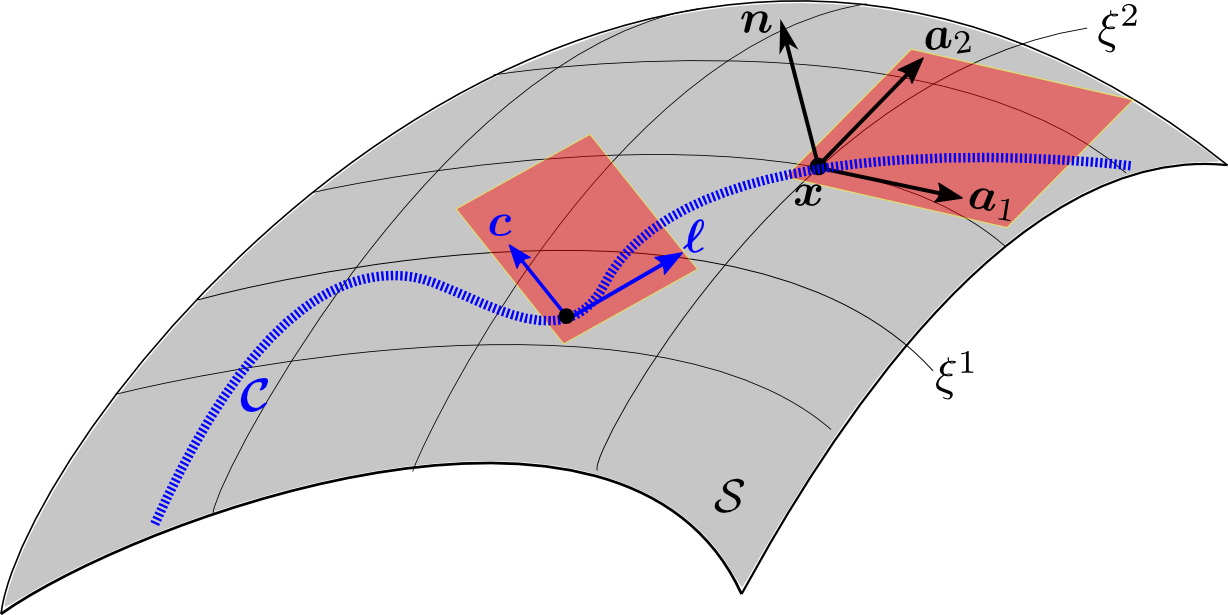}}
\end{picture}
\caption{A fiber bundle represented by curve $\sC$ is embedded in shell surface $\sS$. The red planes illustrate tangent planes}
\label{f:kin}
\end{center}
\vspace{-0.7cm}
\end{figure} 

Further, in order to model fibrous thin shells, a fiber (or a bundle of fibers) is geometrically represented by a curve $\sC$ defined by the points $\bx = {\bx_\mrc}(s,t)$, with $\dif s = \norm{\dif \bx}$, embedded in surface $\sS$ (see Fig.~\ref{f:kin}).  The distribution of fibers is considered continuous such that a homogenized shell theory is obtained. The normalized tangent vector of $\sC$, defined by
\eqb{l} 
{\bell}:= \ds \frac{\partial {{\bx_\mrc}}}{\partial s} = \ell_\alpha\,\ba^\alpha = \ell^\alpha\,\ba_\alpha~,
\label{e:belldef}
\eqe
represents the fiber direction at location $s$. Here, $\ell_\alpha:= \bell\cdot\ba_\alpha $ and $ \ell^\alpha := \bell\cdot\ba^\alpha$ denote the covariant and contravariant components of vector $\bell$ in the convective coordinate system, respectively.
 {Assuming} that the deformation of fibers satisfies Euler-Bernoulli kinematics, the in-plane director for the fiber $\sC$ can be defined by
\eqb{l} 
\bc := \bn \times{\bell} = c_\alpha\,\ba^\alpha =  c^\alpha\,\ba_\alpha ~,
\label{e:vectorcdef}
\eqe
where, $c_\alpha:= \bc\cdot\ba_\alpha $ and $ c^\alpha := \bc\cdot\ba^\alpha$ are the covariant and contravariant components of vector $\bc$, respectively.
With Eqs.~\eqref{e:belldef} and \eqref{e:vectorcdef}, bases $\{\ba_1,\ba_2,\bn\}$ and  $\{\ba^1,\ba^2,\bn\}$ can be represented by the local Cartesian basis $\{\bell,\bc,\bn\}$ as
\eqb{l} 
\ba_\alpha = \ell_\alpha\,\bell + c_\alpha\,\bc~, \quad $and$ \quad \ba^\alpha = \ell^\alpha\,\bell + c^\alpha\,\bc~.
\label{e:transformbase}
\eqe
The surface identity tensor $\bi$ and the full identity $\bone$ in $\mathbb{R}^3$ can then be written as
\eqb{l} 
\bi := \ba_\alpha\otimes\ba^\alpha = \ba^\alpha\otimes\ba_\alpha = \bc\otimes\bc + \bell\otimes\bell~,
\eqe
and
\eqb{l} 
\bone= \bi + \bn\otimes\bn~.
\eqe


\subsection{Surface curvature}
The curvature of surface $\sS$ can be described by the symmetric second order  tensor
\eqb{l} 
\bb :=  b_{\alpha\beta}\,\ba^\alpha\otimes\ba^\beta = b_{\alpha}^\beta\,\ba^\alpha\otimes\ba_\beta =  b^{\alpha\beta}\,\ba_\alpha\otimes\ba_\beta~.
\label{e:tensorb}
\eqe
The components  $b_{\alpha\beta}$ can be computed from the derivative of surface normal $\bn$ as
\eqb{l} 
{b}_{\alpha\beta}:= -\bn_{,\alpha}\cdot\ba_\beta ~,
\label{e:bab}
\eqe
which leads to Weingarten's formula
\eqb{l} 
\bn_{,\alpha} =  - {b}_{\alpha\beta}\,\ba^\beta~.
\label{e:bncoma_alpha}
\eqe
Alternatively, components  $b_{\alpha\beta}$ can be extracted from the derivatives of tangent vectors $\ba_{\alpha}$ as
\eqb{l} 
{b}_{\alpha\beta}:=\bn\cdot\ba_{\alpha,\beta} = \bn\cdot\ba_{\alpha;\beta}~,
\label{e:bab2}
\eqe
where
\eqb{lll} 
 \ba_{\alpha,\beta} := \ds\pa{\ba_\alpha}{\xi^\beta} = \bx_{,\alpha\beta}~, \quad$and$\quad
 \ba_{\alpha;\beta} := ( \bn\otimes\bn)\, \ba_{\alpha,\beta} = b_{\alpha\beta}\,\bn
 \label{e:curvector}
\eqe
are the parametric and covariant derivative of $\ba_\alpha$,  respectively.\footnote{They only differ for quantities with free indices. Index-free quantities, such as $\bn$, satisfy $\bn_{,\alpha} =\bn_{;\alpha}$. }

Unlike $\bn_{,\alpha}$, which always lies in the tangent plane, $\ba_{\alpha,\beta}$ can have both tangential and normal components. With respect to basis $\{\ba_1,\ba_2,\bn\}$ the vectors $\ba_{\alpha,\beta}$ can be expressed as 
\eqb{l} 
 \ba_{\alpha,\beta}= \Gamma^\gamma_{\alpha\beta}\,\ba_\gamma + b_{\alpha\beta}\,\bn~,
 \label{e:gaussform}
\eqe
where the tangential components
\eqb{l} 
\Gamma^\gamma_{\alpha\beta}:=\ba_{\alpha,\beta}\cdot\ba^\gamma  
\label{e:Gamaab}
\eqe
are known as the surface Christoffel symbols. They are symmetric in indices $\alpha$ and $\beta$. Using transformation (\ref{e:transformbase}.2), they can be expressed as
\eqb{l} 
\Gamma^\gamma_{\alpha\beta}= c^\gamma\, \Gamma^{\mrc}_{\alpha\beta} + \ell^\gamma\, \Gamma^{\uell}_{\alpha\beta}~,  
\eqe
where we have defined
\eqb{lll} 
\Gamma_{\!\alpha\beta}^{\mrc} \dis \bc\cdot\ba_{\alpha,\beta} = c_\gamma\,\Gamma^\gamma_{\alpha\beta} ~, \\[2mm]
  \Gamma_{\!\alpha\beta}^{\uell} \dis \bell\cdot\ba_{\alpha,\beta}=\ell_\gamma\,\Gamma^\gamma_{\alpha\beta}~.
\label{e:Gama_c_and_ell}
\eqe

The curvature tensor $\bb$ defined in Eq.~\eqref {e:tensorb} has the two invariants
\eqb{l} 
H := \ds\frac{1}{2}\,\tr_{\!\mrs}\bb =   \ds\frac{1}{2}\,b^\alpha_\alpha~,\quad $and$ \quad \kappa= \det_\mrs \bb := \det[b_\alpha^\beta]~.
\eqe
They correspond to the mean and Gaussian curvatures, respectively.
%
The curvature tensor $\bb$ fully describes the out-of-plane curvature of surface $\sS$. Thus, we can extract from it the curvature of $\sS$ along any direction. For example,
\eqb{l} 
\kappa_\mrn:=\bb:\bell\otimes\bell = b_{\alpha\beta}\,\ellab ~, \quad~$with$~\quad \ellab:=\ell^\alpha\,\ell^\beta~, 
\label{e:Kn_bab}
\eqe
represents the curvature of $\sS$ in direction $\bell$. Hence, $\kappa_\mrn$ also expresses the so-called normal curvature of the curve $\sC$ embedded in $\sS$. Further,
\eqb{l} 
\tau_\mrg:=\bb:\bell\otimes\bc =\bb:\bc\otimes\bell  = b_{\alpha\beta}\,\ell^\alpha\,c^\beta~, 
\label{e:taug_bab}
\eqe
denotes the so-called geodesic torsion of $\sC$. \\[1.5mm] 
The mentioned invariants $H$, $\kappa$, $\kappa_\mrn$, and $\tau_\mrg$ are included in Tab.~\ref{t:invariantKbar}. They are  useful in constructing  material models for (both isotropic and anisotropic) out-of-plane bending.
%
\begin{table}[!ht]
 \small
\begin{center}
\def\arraystretch{1.6}\tabcolsep=14pt
\begin{tabular}{|r|l|l|l| }
\hline 
 \parbox[m][][t]{1.2cm}{Invariant~~~} & \parbox[t]{3.9cm}{tensor notation}&\parbox[t]{2.5cm}{index notation} &\parbox[t]{4cm}{geometrical meaning} \\ \hline\hline
 \multicolumn{4}{ |c| }{out-of-plane and in-plane curvature tensors: $\bb=b_{\alpha\beta}\,\ba^\alpha\otimes\ba^\beta$ and $\bbbar=\bar{b}_{\alpha\beta}\,\ba^\alpha\otimes\ba^\beta$}  \\\hline
$H:= $ & $\frac{1}{2}\,\mathrm{tr}_\mrs\,\bb$ &  $\frac{1}{2}\,{b}^\alpha_\alpha$ & mean curvature of $\sS$ \\ \hline
$\kappa:= $ & $\det_\mrs\bb$ & $\det[{b}_\alpha^\beta]$  & {Gaussian curvature of $\sS$} \\ \hline
$\kappa_\mrn:=$ & $\bb: \bell\otimes \bell = \bell_{,s}\cdot\bn$ &  ${b}_{\alpha\beta}\,\ellab$ & {normal curvature of $\sC\in\sS$}\\\hline
$\tau_\mrg:=$ & $\bb: \bell\otimes \bc = \bb: \bc\otimes \bell  $ &  $ {b}_{\alpha\beta}\,\ell^\alpha\,c^\beta$ &geodesic torsion of $\sC\in\sS$\\\hline
$\kappa_\mrg:=$ & 
$\bbbar:\bell\otimes\bell$
 &  $\bar{b}_{\alpha\beta}\,\ellab  $ &geodesic curvature of $\sC\in\sS$\\\hline
$\kappa_\mrp:=$ & 
$ \norm{\bell_{,s}}$ & $\sqrt{\kappa_\mrg^2 + \kappa_\mrn^2}$ & principal curvature of $\sC\in\sS$\\\hline
\multicolumn{4}{ |c| }{measures for the change in normal curvature $\kappa_\mrn$ of $\sC\in\sS$}  \\\hline
${\mrk_{\mrn}:=}$ & 
$ {\bb}\!:\!\bell\otimes\bell - {\bb_0}\!:\!\bL\otimes\bL $ & $  {\kappa_\mrn - {\kappa}^0_\mrn}$ & absolute change\\\hline
$k_{\mrn}:=$ & 
$  {\bb}\!:\!\lambda\bell\otimes\bell - {\bb_0}\!:\!\bL\otimes\bL $ & $  {\kappa_\mrn\, \lambda - {\kappa}^0_\mrn}$ & {stretch-excluded change}\\\hline
$K_{\mrn}:=$ & 
$ {\bK}:\bL\otimes\bL $ & $  \kappa_\mrn\, {\lambda^2} - {\kappa}_\mrn^0$ & {nominal change}\\\hline
%
\multicolumn{4}{ |c| }{{measures for the change in geodesic torsion $\tau_\mrg$ of $\sC\in\sS$}}  \\\hline
${\mrt_{\mrg}:=}$ & 
$ {\bb}\!:\!\bell\otimes\bc - {\bb_0}\!:\!\bL\otimes\bc_0 $ & $  {\tau_\mrg - {\tau}^0_\mrg}$ & {absolute change}\\\hline
${t_{\mrg}:=}$ & 
$ {\bb}\!:\! \lambda\bell\otimes\bc - {\bb_0}\!:\!\bL\otimes\bc_0 $ & $  {\tau_\mrg\, \lambda - {\tau}^0_\mrg}$ & {stretch-excluded change}\\\hline
$T_{\mrg}:=$ & 
$ {\bK}:\bL\otimes\bc_0 $ & $  { b_{\alpha\beta}\,c_0^\alpha\,L^\beta} - {\tau}_\mrg^0$ & {nominal change}\\\hline
%
\multicolumn{4}{ |c| }{{measures for the change in geodesic curvature $\kappa_\mrg$ of $\sC\in\sS$}}  \\\hline
${\mrk_{\mrg}:=}$ & 
$ {\bbbar}\!:\!\bell\otimes\bell - {\bbbar_0}\!:\!\bL\otimes\bL $ & $  {\kappa_\mrg - {\kappa}^0_\mrg}$ & {absolute change}\\\hline
${k_{\mrg}:=}$ & 
$  {\bbbar}\!:\!\lambda\bell\otimes\bell - {\bbbar_0}\!:\!\bL\otimes\bL $ & $  {\kappa_\mrg\, \lambda - {\kappa}^0_\mrg}$ & {stretch-excluded change }\\\hline
$K_{\mrg}:=$ & 
$ \bar{\bK}:\bL\otimes\bL $ & $  \kappa_\mrg\, {\lambda^2} - {\kappa}^0_\mrg$ & {nominal change}\\\hline
\end{tabular}
\end{center}
\caption{Various curvature measures of the surface $\sS$ and of a fiber family $\sC$ embedded in $\sS$. Note, that the magnitude of these measures is invariant, but their sign (except for $\kappa$ and $\kappa_\mrp$)  depends on the direction of directors $\bn$ and/or $\bc$.  All these measures can be shown to be frame invariant under superimposed rigid body motions of $\sS$, see Appendix~\ref{s:frameinvariance}.
}
\label{t:invariantKbar}
\end{table}

\subsection{Curvature of embedded curves} 
We aim at capturing any geodesic torsion, normal and in-plane curvature of an embedded curve $\sC\in\sS$. 
As seen in Eqs.~\eqref{e:Kn_bab} and \eqref{e:taug_bab}, the normal curvature  and geodesic torsion  of $\sC$  can already be described via the out-of-plane curvature tensor \eqref{e:tensorb}. The in-plane curvature, on the other hand, does not follow from tensor $\bb$. Instead, it  can be extracted from the so-called curvature vector of $\sC$ that is defined by 
{the directional derivative of $\bell$ in direction $\bell$, i.e.}
\eqb{l}
\bell_{,s}:= \ds \frac{\partial \bell}{\partial s} =  (\nabla_{\!\mrs}\,\bell)\,\bell = \ell^\alpha\,\bell_{,\alpha}~. 
\label{e:ells2}
\eqe
Here and henceforth,  $\nabla_{\!\mrs}\,\bullet:=  \bullet_{,\alpha}\otimes\ba^\alpha$ denotes the surface gradient operator.
\footnote{\label{note3}Note, that $\nabla_{\!\mrs}\bullet$ can have both in-plane and out-of-plane components. 
} Following Eq.~\eqref{e:belldef} and considering Eqs.~\eqref{e:def_dualvector} and \eqref{e:gaussform}, derivative $\bell_{,\alpha}$ can be expressed as
\eqb{lll}
\bell_{,\alpha} = \ell^\beta_{;\alpha}\,\ba_\beta + \ell^\beta\,b_{\beta\alpha}\,\bn~=  \ell_{\beta;\alpha}\,\ba^\beta + \ell^\beta\,b_{\beta\alpha}\,\bn~,
\label{e:ell_alp0}
\eqe
where the semicolon denotes the covariant derivatives 
%
\eqb{lll}
\ell^\beta_{;\alpha} \dis \bell_{,\alpha}\cdot\ba^\beta  = \ell^\beta_{,\alpha} + \ell^\gamma\,\Gamma_{\gamma\alpha}^\beta~, \\[2mm]
 \ell_{\beta;\alpha} \dis  \bell_{,\alpha}\cdot\ba_\beta = \ell_{\beta,\alpha} - \ell_\gamma\,\Gamma_{\beta\alpha}^\gamma~.
\label{e:covardervEll}
\eqe
%
%
The magnitude $\norm{\bell_{,s}}=:\kappa_\mrp$ is called the principal curvature of $\sC$,\footnote{since it is the sole principal invariant of the first order tensor $\bell_{,s}$~.} and the direction $\bell_{,s}/\kappa_\mrp=:\bn_\mrp$ is referred to as the principal normal to $\sC$. Note that $\bn_{\mrp}$ is normal to the curve but not necessarily normal to the surface $\sS$.

In principle, vector $\bell_{,s}$ can be expressed in any basis, which then induces different curvatures from the corresponding components. Here, we express $\bell_{,s}$ in
 %
%
the basis  $\{\bell,~\bc,~\bn\}$, i.e.
\eqb{l}
\bell_{,s} = \kappa_\mrp\,\bn_\mrp = \kappa_\mrg\,\bc + \kappa_\mrn\,\bn~,
\label{e:ells4}
\eqe
where
\eqb{l}
\kappa_\mrn:=\bn\cdot\bell_{,s} ~,
\eqe
denotes the normal curvature of $\sC$. By inserting \eqref{e:ells2} and \eqref{e:ell_alp0}, $\kappa_\mrn$ can be shown to be identical to Eq.~\eqref{e:Kn_bab}.  On the other hand, 
\eqb{l}
\kappa_\mrg:=\bc\cdot\bell_{,s}~
\label{e:kgdefinitionnew}
\eqe
is the in-plane (i.e.~geodesic) curvature of $\sC$. It can be computed  by inserting  \eqref{e:ells2} and \eqref{e:ell_alp0} into Eq.~\eqref{e:kgdefinitionnew}, giving
\eqb{l}
\kappa_\mrg= \ell_{\alpha;\beta}\, c^\alpha\,\ell^\beta = \ell^\alpha_{;\beta}\, c_\alpha\,\ell^\beta = (\bar\nabla_{\!\mrs} \bell):\bc\otimes\bell~,
\label{e:kg0}
\eqe
where
\eqb{l}
\bar\nabla_{\!\mrs}\bell:= \bi \,\, \nabla_{\!\mrs}\bell=  \ell_{\alpha;\beta} \,\ba^\alpha\otimes\ba^\beta =  \ell^\alpha_{;\beta} \,\ba_\alpha\otimes\ba^\beta
\label{e:def_lambdap}
\eqe
denotes the projected surface gradient of $\bell$.\footnote{Unlike $\nabla_{\!\mrs}\bullet$, $\bar\nabla_{\!\mrs}\bullet$ has only in-plane components. }

\begin{remark}
From Eq.~\eqref{e:ells4} follows
\eqb{lll}
\kappa_\mrp^2 = \kappa_\mrg^2 + \kappa_\mrn^2~.
\label{e:kpvskg}
\eqe
\end{remark}

\begin{remark}
The three scalars $\kappa_\mrn$,  $\kappa_\mrg$, and $\tau_\mrg$ are associated with the three bending modes of $\sC\in\sS$: normal (i.e.~out-of-plane) bending, geodesic (i.e.~in-plane) bending, and geodesic torsion, respectively.
\end{remark}

\subsection{Definition of the in-plane curvature tensor} {\label{s:geodesiccurvature}}

In principle, we can use the second order tensor $\bar\nabla_{\!\mrs}\bell$ to characterize the in-plane curvature as the counterpart to tensor $\bb$ from Eq.~\eqref{e:tensorb}, which characterizes  the out-of-plane curvature. Tensor $\bar\nabla_{\!\mrs}\bell$ is, however, unsymmetric. We thus construct an alternative tensor by rewriting  Eq.~\eqref{e:kg0} and using  identity  $\bc_{,s}\cdot\bell = -\bc\cdot\bell_{,s}$ that follows from $\bc\cdot\bell = 0$, so that
\eqb{l}
\kappa_\mrg :=  -\ell_\alpha^\beta \,c^\alpha_{;\beta} = -\ellab\,c_{\alpha;\beta} = \bbbar:\bell\otimes\bell~,
\label{e:kg1}
\eqe
where 
\eqb{lll}
c^\beta_{;\alpha} := c^\beta_{,\alpha} + c^\gamma\,\Gamma_{\gamma\alpha}^\beta~,\quad $and$\quad
c_{\beta;\alpha} := c_{\beta,\alpha} - c_\gamma\,\Gamma_{\beta\alpha}^\gamma~,
\eqe
similar to Eq.~\eqref{e:covardervEll}. Here, we have defined the so-called in-plane curvature tensor of $\sC$ as
 \eqb{l}
 \bbbar:=  -\frac{1}{2} \left[\bar\nabla_{\!\mrs} \bc  + (\bar\nabla_{\!\mrs} \bc)^\mrT \right] = 
 -\frac{1}{2}\,(c_{\alpha;\beta} + c_{\beta;\alpha})\,\ba^\alpha\otimes\ba^\beta = \bar{b}_{\alpha\beta}\,\ba^\alpha\otimes\ba^\beta~.
\label{e:binplane} 
 \eqe
Components $\bar{b}_{\alpha\beta}$ thus can be computed from
\eqb{l}
\bar{b}_{\alpha\beta} =  \bbbar:\ba_\alpha\otimes\ba_\beta  = -\frac{1}{2} ( c_{\alpha;\beta} + c_{\beta;\alpha}) =  -\frac{1}{2} ( \bc_{,\alpha}\cdot\ba_\beta + \bc_{,\beta}\cdot\ba_\alpha) ~,
\label{e:binplane2} 
\eqe
where 
\eqb{rlll}
\bc_{,\alpha} = \bc_{;\alpha} \is c^\beta\,\ba_{\beta;\alpha} +  c^\beta_{;\alpha}\,\ba_\beta~\\[3mm]
\is c_\beta\,\ba^\beta_{;\alpha} +  c_{\beta;\alpha}\,\ba^\beta~\\[3mm]
\is  b_{\alpha\beta}\,c^\beta\,\bn - c^\beta\,\ell_{\beta;\alpha}\,\bell~.
\label{e:coderivbc}
\eqe
The last equation is obtained from identities $\bc\cdot\bn = 0$, $\bell\cdot\bn=0$, Eqs.~\eqref{e:ell_alp0} and \eqref{e:bncoma_alpha}. The quantities $\bar{b}_{\alpha\beta}$, $\kappa_\mrg$, $c_{\beta;\alpha}$, and $\ell_{\beta;\alpha}$ can be shown to be frame invariant under superimposed rigid body motions of $\sS$, see Appendix \ref{s:frameinvariance}.

 

\begin{remark}
Since vector $\bell$ is normalized, we have the identity
\eqb{lll}
\bell\cdot\bell_{,\alpha}= \ell_\beta\,\ell^\beta_{;\alpha} =   \ell^\beta\,\ell_{\beta;\alpha} = 0,
\eqe
due to Eq.~\eqref{e:ell_alp0}. Further, equating Eqs.~\eqref{e:kg0} and \eqref{e:kg1} gives the relation
 \eqb{lll}
\ell^\beta\,c_{\beta;\alpha} = - c^\beta\,\ell_{\beta;\alpha},
\label{e:c_ab_vs_l_ab}
\eqe
and from Eq.~\eqref{e:coderivbc}, we find
\eqb{lll}
c_{\beta;\alpha} \is \ba_\beta\cdot \bc_{,\alpha} =  -  c^\gamma\, \ell_\beta\,\ell_{\gamma;\alpha}~\\[2mm]
 c^{\beta}_{;\alpha}~ \is  \ba^\beta\cdot \bc_{,\alpha} = -  c^\gamma\, \ell^\beta\,\ell_{\gamma;\alpha}~.
\label{e:c_a_comma_b}
\eqe
\end{remark}

\subsection{Shell deformation}

To characterize the shell deformation, a reference configuration $\sS_0$ is chosen. The tangent vectors $\bA_\alpha$, the normal vector $\bN$, the surface metric $A_{\alpha\beta}$, the out-of-plane curvature tensor $\bb_0:=B_{\alpha\beta}\bA^\alpha\otimes\bA^\beta$ are defined on $\sS_0$ similarly to Eqs~\eqref{e:tangentsl}, \eqref{e:normalmsl}, \eqref{e:a_ab}, and \eqref{e:tensorb}, respectively.The fiber embedded within the shell surface is denoted by $\sC_0$ in the reference configuration. Also,
 the normalized  fiber direction $\bL=L^\alpha\bA_\alpha=L_\alpha\bA^\alpha$, the in-plane fiber director $\bc_0 = c^0_\alpha\,\bA^\alpha$, and the in-plane curvature tensor $\bbbar_0:= \bar{B}_{\alpha\beta}\,\bA^\alpha\otimes\bA^\beta = -\frac{1}{2}\,(c^0_{\alpha;\beta} + c^0_{\beta;\alpha})\,\bA^\alpha\otimes\bA^\beta$ are defined similarly to Eqs.~\eqref{e:belldef}, \eqref{e:vectorcdef}, and \eqref{e:binplane}, respectively.


Having $a_{\alpha\beta}$, $b_{\alpha\beta}$, and $\bar{b}_{\alpha\beta}$  the deformation of a shell can now be characterized by the  following three tensors: 

1. The surface deformation gradient:
\eqb{l}
\bF:=\ba_\alpha\otimes\bA^\alpha~.
\eqe
This tensor can be used to map the reference fiber direction $\bL$ to $\bell$ as
\eqb{l}
\lambda\,\bell = \bF\,\bL = L^\alpha\,\ba_\alpha~,
\label{e:bellhatF}
\eqe
where $\lambda$ is the fiber stretch. Comparing \eqref{e:belldef} and \eqref{e:bellhatF} gives
\eqb{l}
 \ell^\alpha = L^\alpha\,\lambda^{-1}~.
\label{e:bellF}
\eqe
 From \eqref{e:covardervEll} and \eqref{e:bellF} follows that
\eqb{lll}
\ell_{\alpha;\beta} = \hat{L}_{\alpha,\beta} - \ell_{\alpha\gamma}\,\big( \hat{L}^\gamma_{,\beta} + \ell^\delta\,\Gamma_{\delta\beta}^\gamma\big) + \ell^\gamma\,  \ba_\alpha  \cdot\ba_{\gamma,\beta}~,
\label{e:dbellaF}
\eqe
where
%
\eqb{l}
\hat{L}^\alpha_{,\beta}:=\lambda^{-1}\,L^\alpha_{,\beta}~,\quad $and$  \quad  \hat{L}_{\alpha,\beta} := a_{\alpha\gamma}\, \hat{L}^\gamma_{,\beta}~.
\label{e:Lhatdefine}
\eqe

Inserting \eqref{e:dbellaF} into Eq.~\eqref{e:c_a_comma_b} gives
%
\eqb{lll}
c_{\beta;\alpha} \is   - \ell_{\beta}\,   \big(c^\gamma\,  \hat{L}_{\gamma,\alpha} + \ell^\gamma\,\Gamma^\mrc_{\gamma\alpha}\big)~,\\[2mm]
c^{\beta}_{;\alpha}~ \is   -  \ell^{\beta}\, \big(c^\gamma\,  \hat{L}_{\gamma,\alpha} + \ell^\gamma\,\Gamma^\mrc_{\gamma\alpha}\big)~.
\label{e:coderivbcDef}
\eqe
With the surface deformation gradient, the right Cauchy-Green surface tensor is defined by $\bC:=\bF^\mrT\,\bF = a_{\alpha\beta}\bA^\alpha\otimes\bA^\beta$. Tab.~\ref{t:invariantC} lists some invariants induced by $\bC$ that can be useful for constructing material models. The  Green-Lagrange surface strain tensor, which represents the change of the surface metric, is then defined by
\eqb{l}
\bE  :=\ds\frac{1}{2}(\bC - \bI) =\ds \frac{1}{2}\,(a_{\alpha\beta} - A_{\alpha\beta})\, \bA^\alpha\otimes\bA^\beta= E_{\alpha\beta}\,\bA^\alpha\otimes\bA^\beta~.
\label{e:bEtensor}
\eqe

2. The change of the out-of-plane curvature tensor:
\eqb{l}
\bK :=  \bF^T\,\bb\,\bF - \bb_0 = 
  (b_{\alpha\beta} - B_{\alpha\beta})\,\bA^\alpha\otimes\bA^\beta= K_{\alpha\beta}\,\bA^\alpha\otimes\bA^\beta~.
\label{e:Kten}
\eqe

3. The change of the in-plane curvature tensor:
\eqb{l}
\bar\bK :=  \bF^T\,\bbbar\,\bF - \bbbar_0 =  ( \bar{b}_{\alpha\beta} - \bar{B}_{\alpha\beta})\,\bA^\alpha\otimes\bA^\beta = \bar{K}_{\alpha\beta}\,\bA^\alpha\otimes\bA^\beta~.
\label{e:Ktenb}
\eqe
Here, $\bar{b}_{\alpha\beta}$ 
can be computed from Eq.~\eqref{e:binplane2}   taking into account Eq.~\eqref{e:coderivbcDef}. This gives
\eqb{l}
\bar{b}_{\alpha\beta} = \frac{1}{2}\, \ell^\gamma ( \ell_\alpha\,\Gamma^{\mrc}_{\gamma\beta} + \ell_\beta\,\Gamma^{\mrc}_{\gamma\alpha})  + \frac{1}{2}\, c^\gamma\,\big(\ell_\alpha\, \hat{L}_{\gamma,\beta} + \ell_\beta\, \hat{L}_{\gamma,\alpha}\big) ~.  
\label{e:bbarab_ok}
\eqe

Accordingly, the geodesic curvature follows from Eq.~\eqref{e:kg1} as
\eqb{l}
\kappa_\mrg := \bar{b}_{\alpha\beta}\,\ellab =  \ellab\,\Gamma_{\alpha\beta}^\gamma\,c_\gamma + \lambda^{-1}\, c_\alpha\,\ell^\beta\,{L}^{\alpha}_{,\beta}~.
\label{e:kgok0}
\eqe
Further, by using relation ${L}^{\alpha}_{,\beta} =  \bL_{,\beta}\cdot\bA^\alpha - L^\gamma\bar{\Gamma}_{\gamma\beta}^\alpha$, similar to  Eq.~(\ref{e:covardervEll}.1) --  where $ \bar{\Gamma}_{\alpha\beta}^\gamma:=\bA^\gamma\cdot\bA_{\alpha,\beta}$ denote the Christoffel symbols of the initial configuration -- Eq.~\eqref{e:kgok0} can be rewritten as
\eqb{l}
\kappa_\mrg =  \underbrace{\ellab\,c_\gamma\,S_{\alpha\beta}^\gamma}_{=:\,\kappa_\mrg^\mathrm{\Gamma}} \,+ \, \underbrace{\lambda^{-1}\, c_\alpha\,\ell^\beta\,L^\alpha_{;\beta}}_{=:\,\kappa_\mrg^{\mathrm{L}}}~,
\label{e:kgok}
\eqe
where  $S_{\alpha\beta}^\gamma:=\Gamma_{\alpha\beta}^\gamma - \bar{\Gamma}_{\alpha\beta}^\gamma $ and $L^\alpha_{;\beta}:= \bA^\alpha\cdot\bL_{,\beta}$~.


\begin{remark}
As seen in Eq.~\eqref{e:kgok},  the geodesic curvature involves not only the change in the Christoffel symbols (term $\kappa_\mrg^{\Gamma}$), but also the  gradient of $\bL$ (term  $\kappa_\mrg^\mathrm{L}$). For initially straight fiber, the geodesic curvature becomes
\eqb{lll}
\kappa_\mrg:= \kappa_\mrg^\mathrm{\Gamma} = \ellab\,c_\gamma\,S_{\alpha\beta}^\gamma~
\label{e:kg2}
\eqe
as  $\kappa_\mrg^\mathrm{L}$ vanishes. But for initially curved fibers, where $\kappa_\mrg^\mathrm{L}\neq 0$, expression \eqref{e:kgok} should be used. This point will be demonstrated in Sec.~\ref{s:criticalexample}.


\end{remark}

%
\begin{remark}
To measure the change in the curvatures, one can use the invariants
\eqb{llll}
\mrk_\mrn \dis {\bb}\!:\!\bell\otimes\bell - {\bb_0}\!:\!\bL\otimes\bL = \kappa_\mrn - \kappa_\mrn^0~,\\[3mm]
\mrk_\mrg\dis {\bbbar}\!:\!\bell\otimes\bell - {\bbbar_0}\!:\!\bL\otimes\bL= \kappa_\mrg - \kappa^0_\mrg~,\\[3mm]
\mrt_\mrg \dis {\bb}\!:\!\bell\otimes\bc - {\bb_0}\!:\!\bL\otimes\bc_0 = \tau_\mrg - \tau_\mrg^0~,
\label{e:abscurvature}
\eqe
called the \textit{absolute change} in the normal curvature, geodesic curvature, and geodesic torsion, respectively. Here, $\bullet^0$ denotes the corresponding quantities in the initial configuration.
\end{remark}

\begin{remark}
However, the  curvature changes \eqref{e:abscurvature} are not ideal for constitutive models, as they can lead to  unphysical stress couples that respond to fiber stretching even when there is no bending.
An example is pure dilatation, e.g.~due to  thermal expansion or hydrostatic stress states as is discussed in example~\ref{s:example_diskexpan}. To exclude fiber stretching, we define the invariants
\eqb{llll}
k_\mrn \dis {\bb}\!:\!\lambda\bell\otimes\bell - {\bb_0}\!:\!\bL\otimes\bL = \kappa_\mrn\,\lambda - \kappa_\mrn^0~,\\[3mm]
k_\mrg\dis {\bbbar}\!:\!\lambda\bell\otimes\bell - {\bbbar_0}\!:\!\bL\otimes\bL= \kappa_\mrg\lambda - \kappa^0_\mrg~,\\[3mm]
t_\mrg \dis {\bb}\!:\!\lambda\bell\otimes\bc - {\bb_0}\!:\!\bL\otimes\bc_0 = \tau_\mrg\,\lambda - \tau_\mrg^0~,
\label{e:dilcurvature}
\eqe
called the \textit{stretch-excluded change} in the normal curvature, geodesic curvature, and geodesic torsion, respectively.
%
\end{remark}

\begin{remark}
Both curvature changes \eqref{e:abscurvature} and  \eqref{e:dilcurvature}  
 can cause fiber tension apart from fiber bending.
 %
Therefore, one can use the so-called \textit{nominal change} in the normal curvature, geodesic curvature, and geodesic torsion, defined by
\eqb{llll}
K_\mrn \dis \bK:\bL\otimes\bL = \kappa_\mrn\, \lambda^2 - \kappa^0_\mrn~,\\[3mm]
K_\mrg\dis \bar\bK:\bL\otimes\bL = \kappa_\mrg\, {\lambda^2} - \kappa^0_\mrg~,\\[3mm]
T_\mrg \dis \bK:\bL\otimes\bc_0 = b_{\alpha\beta}\,c_0^\alpha\,L^\beta - \tau^0_\mrg~,
\label{e:normcurvature}
\eqe
where $\bK$ and $\bKbar$ are defined by \eqref{e:Kten} and \eqref{e:Ktenb}, respectively. These invariants can also be found e.g.~in \cite{Steigmann2015} and \cite{Schulte2020}. Since the measures \eqref{e:normcurvature} do not cause an axial tension in the fibers, their material tangents  simplify significantly. Note however that, like  \eqref{e:abscurvature}, they can cause unphysical stress couples responding to fiber stretching.
\end{remark}

\begin{remark}
The mentioned curvature measures \eqref{e:abscurvature}, \eqref{e:dilcurvature}, and \eqref{e:normcurvature} are also listed in Tab.~\ref{t:invariantKbar}. It should be noted that all  these measures are equivalent for inextensible fibers, i.e.~$\lambda=1$, which is usually assumed for textile composites. 
\end{remark}

\begin{table}[!ht]
 \small
\begin{center}

\def\arraystretch{1.6}\tabcolsep=2.7pt
\begin{tabular}{|r|l|l|l| }
\hline
 \parbox[t][][t]{1.55cm}{Invariant~~~} & \parbox[t]{4.5cm}{tensor notation}&\parbox[t]{3cm}{index notation} &\parbox[t]{3.6cm}{geometrical relevance} \\ \hline\hline
\multicolumn{4}{ |c| }{ (in-plane) right Cauchy-Green surface strain tensor $\bC =  a_{\alpha\beta}\,\bA^\alpha\otimes\bA^\beta$}  \\\hline
$I_1:=$ & $\mathrm{tr}_\mrs\, \bC$ &  $ A^{\alpha\beta}\,a_{\alpha\beta}$ & {surface shearing of $\sS$} \\ \hline
$I_2=J^2 := $ & $\mathrm{det}_\mrs\bC = (\mathrm{det}_\mrs\bF )^2 $ &  $\det[a_{\alpha\beta}]/  \det[{A}_{\alpha\beta}]$ & {surface area change of $\sS$} \\\hline
${\Lambda}_i= \lambda_i^2:=$ & $\bC: \bL_i\otimes \bL_i$~~ (no sum in $i$) &  $a_{\alpha\beta}\,L_i^{\alpha\beta}$ & {
 stretching of $\sC_i$} \\\hline
${\gamma}_{ij}:=$ & $\bC: \bL_i\otimes \bL_j $ &  $a_{\alpha\beta}\,L_i^{\alpha}\,L_j^{\beta}$ & {nominal angle between $\sC_i$ \& $\sC_j$, $i\neq j$}\\\hline
${\hat{\gamma}_{ij}:=}$ & $\bell_1\cdot\bell_2 $ &  $a_{\alpha\beta}\,\ell_i^{\alpha}\,\ell_j^{\beta}$ & {absolute angle between $\sC_i$ \& $\sC_j$, $i\neq j$}\\\hline
\end{tabular}
\end{center}
\caption{Various  invariants of the right surface Cauchy-Green tensor $\bC$, induced by multiple fiber families $\sC_i$ ($i=1,~...,~n_\mrf$). Here, $L_i^{\alpha\beta}:=L_i^\alpha\, L_i^\beta$ (no sum in $i$). }
\label{t:invariantC}
\end{table}

\section{Balance laws} {\label{s:balance}}
In this section, we discuss the balance laws for fibrous thin shells taking into account not only  in-plane stretching and  out-of-plane bending, but also  in-plane bending. 
Like in classical Kirchhoff-Love shell theory, we directly postulate linear and angular momentum balance for our generalized Kirchhoff-Love shell with in-plane bending.  Although the presented theory is not derived from general Cosserat theory,  we show in Appendix \ref{s:app_KLequilibrium} that our set of balance equations is {consistent with} that of Cosserat shell theory.

{We consider that the out-of-plane shear energy and the out-of-plane thickness strain energy are negligible. The first condition amounts to the Kirchhoff-Love kinematical assumption of zero out-of-plane shear strains.  The second condition is satisfied here by the plane stress assumption (i.e.~zero thickness stress), which is commonly used for thin shells. It still allows for thickness changes, e.g.~due to large membrane stretching. Note, however, that these assumptions on the shear strains and thickness stress are mathematically independent of the thickness, so that the governing equations can be written directly in surface form without the need of a thickness variable. Instead, the influence of the thickness appears in  the constitutive equations, see e.g.~\cite{naghdi82,steigmann99b,shelltheo,solidshell}.} 

{In the following, we} first discuss  in detail  the theory with one embedded fiber family $\sC$. The extension to multiple fiber families is straightforward and will be discussed subsequently.

In order to define internal stresses and internal moment tensors, the shell $\sS$ is virtually cut into two parts at position $\bx$ as depicted in Fig.~\ref{f:traction}. On the parametrized cut, denoted by $\sI(s)$, we define the unit tangent vector $\btau:=\partial{\bx}/\partial{s}$ and the unit normal $\bnu:=\btau\times\bn = \nu_\alpha\,\ba^\alpha$ at $\bx$. 

\subsection{Stress tensor}\label{s:stress} 
This section discusses Cauchy stress tensor $\bsig$ for thin shells under 
plane-stress conditions. The influence of angular momentum balance on the stress tensor is then discussed in Sec.~\ref{s:angbalance}. 
\begin{figure}[ht]
\begin{center} \unitlength1cm
\begin{picture}(0,9)
\put(-7.7,3.0){\includegraphics[width=0.6\textwidth]{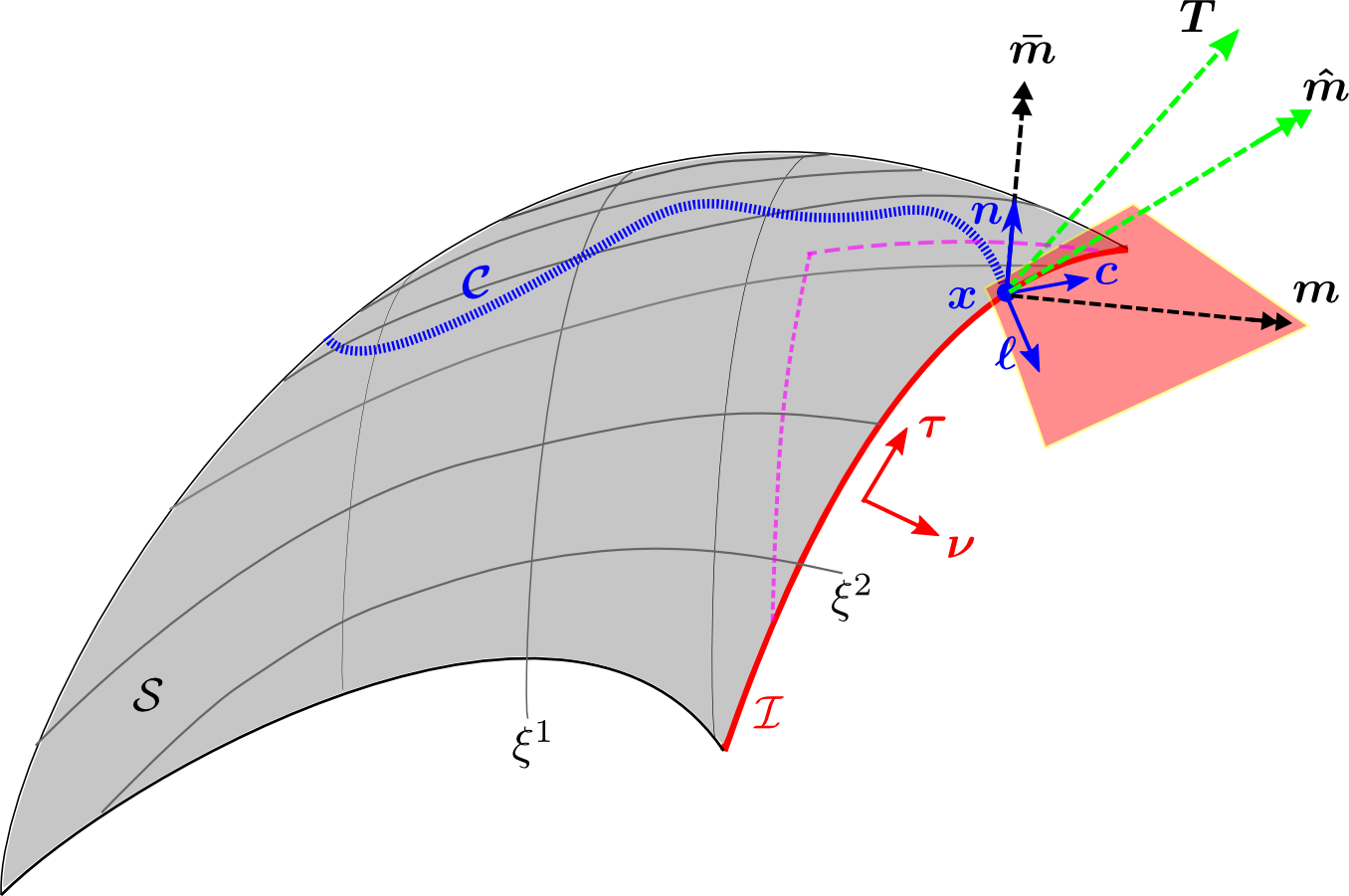}}

\put(-2.0,0.0){\includegraphics[width=0.37\textwidth]{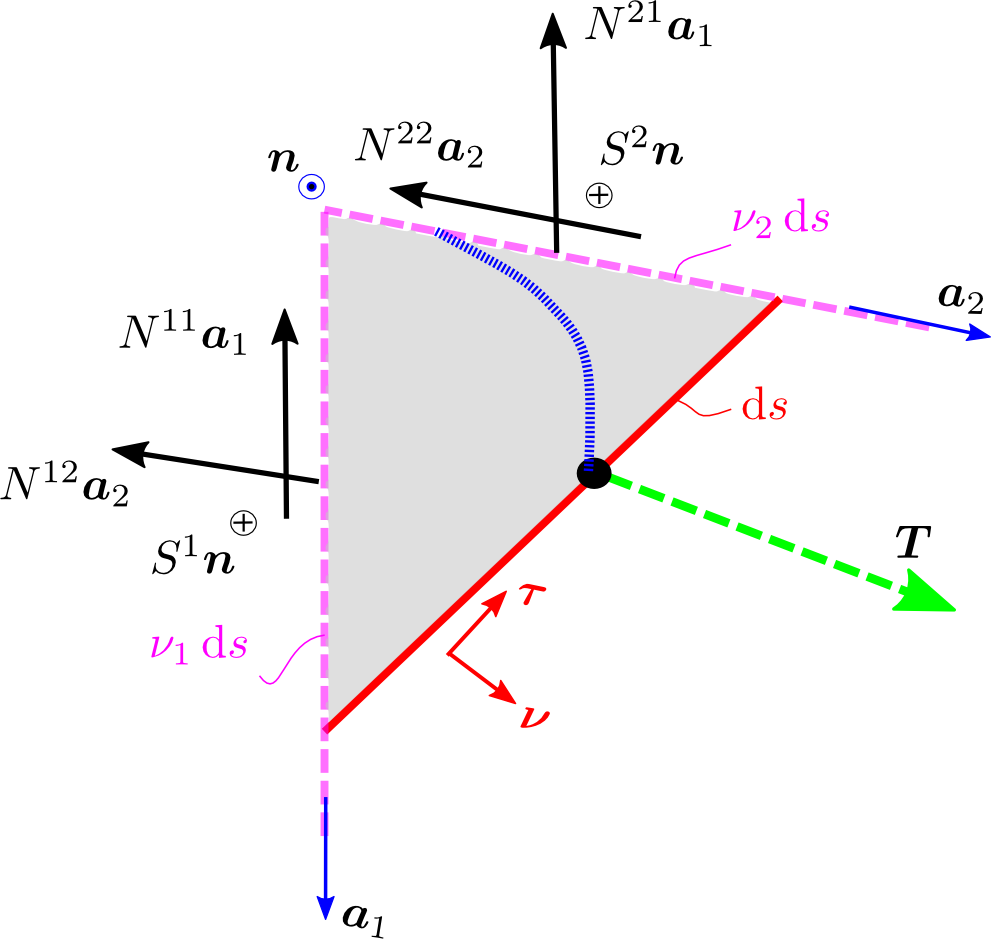}}

\put(2.3,4.0){\includegraphics[width=0.37\textwidth]{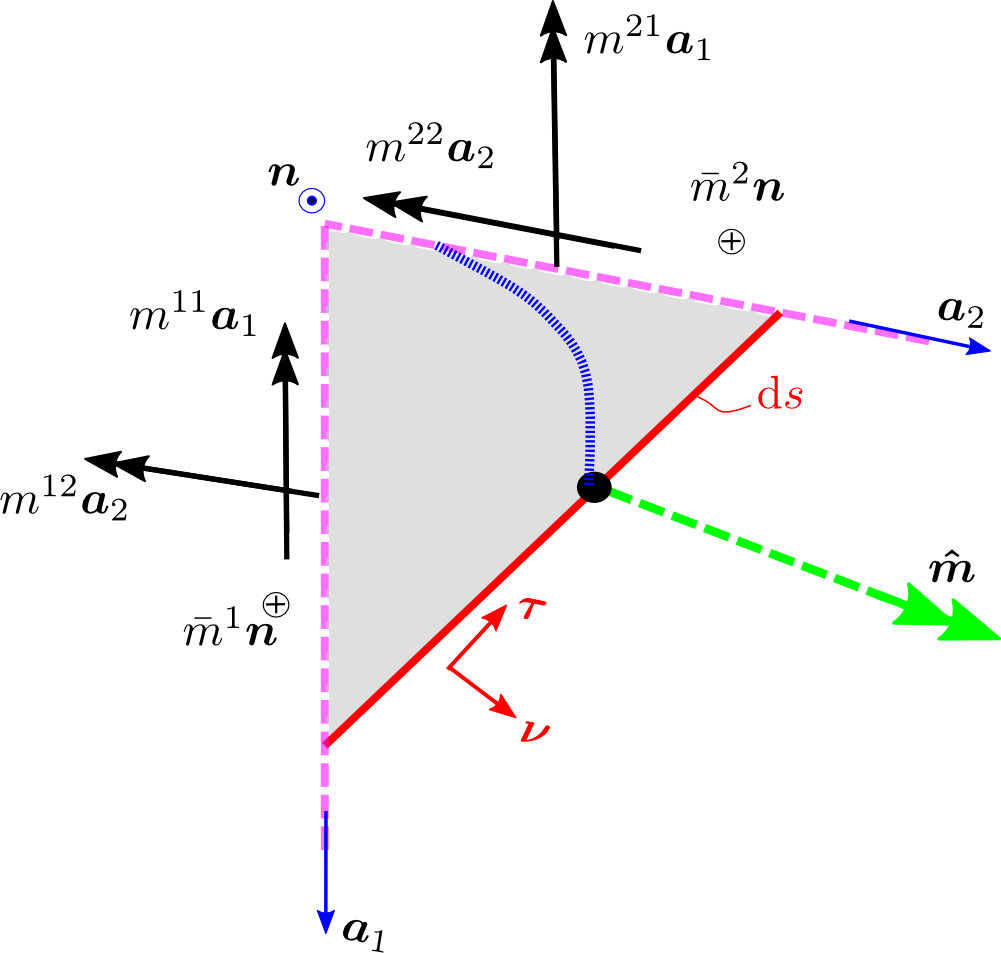}}

\put(-7.0,3){(a)}
\put(-1.0,0.5){(b)}
\put(3.5,4.6){(c)}

\end{picture}
\caption{Internal stresses and moments: (a)~illustration of physical traction vector $\bT$ and moment vector $\bmhat$ acting on the cut  $\sI$ (red curve)  through surface $\sS$ with its embedded fiber $\sC$ (blue curve). Both $\bT$  and  $\bmhat$ are general vectors in 3D space. The moment vector $\bmhat$ can be decomposed into the   component $\bar{\bm}$, causing in-plane bending, and the  component $\bm$, causing out-of-plane bending and twisting.  Vectors $\bell$, $\bc$, $\bnu$, $\btau$, and $\bm$ lie in the tangent plane of $\sS$.
(b) Stress and (c) moment components appearing on a triangular element -- which is in force and moment equilibrium -- under the action of the traction $\bT$ and moment $\bmhat$, respectively. These stress and moment components appear on the cuts that intersect with the tangent plane (pink dash lines). Assuming plane stress conditions, the stress and moment on any cut perpendicular to the surface normal $\bn$ (grey area) are  neglected for the equilibrium of the triangular element.  }
\label{f:traction}
\end{center}
\end{figure}

The traction vector $\bT$ appearing on the cut can have arbitrary direction (see Fig.~\ref{f:traction}b), but we can generally express it with respect to the basis  $\{\ba_1,\ba_2,\bn\}$  as
\eqb{l}
\bT = T^\alpha\,\ba_\alpha + T^3\,\bn~,
\label{e:tractionTdef}
\eqe
where $T^\alpha$ and $T^3$ are the contravariant components of $\bT$. 
 This traction induces the Cauchy stress tensor $\bsig$. Following Cauchy's theorem, the  components of $\bsig$ are balanced by the traction $\bT$ on an infinitesimal triangular element, as shown in Fig.~\ref{f:traction}b. Under  plane-stress conditions, all stresses on any cut perpendicular to surface normal $\bn$ are neglected for the (force) equilibrium of the triangular element. That is, the components of tensor $\bsig$ associated with bases $\bn\otimes\ba_\alpha$ and $\bn\otimes\bn$ are considered to be zero. This results in the asymmetry of tensor $\bsig$ in the form, 
\eqb{l}
\bsig = \Nab\,\ba_\alpha\otimes\ba_\beta +  S^\alpha\, \ba_\alpha\otimes\bn~,
\label{e:stresssig}
\eqe
where $ \Nab$ and $ S^\alpha$ are the membrane stress and out-of-plane shear stress components of   $\bsig$.
Accordingly, the components of $\bsig^\mrT$ 
on the cross-section perpendicular to
$\ba^\alpha$  are defined by 
\eqb{l}
\bT^\alpha :=  \bsig^\mrT\,\ba^\alpha = \Nab\,\ba_\beta + S^\alpha\,\bn~,
\label{e:bTalp}
\eqe
which is illustrated in Fig.~\ref{f:traction}b for an infinitesimal triangular element on $\sS$. With this, we can write
\eqb{l}
\bsig = \ba_\alpha\otimes\bT^\alpha~.
\label{e:bsigT}
\eqe

Further, the force equilibrium of the triangle (see Fig.~\ref{f:traction}b) gives
\eqb{l}
\bT\,\dif s =  (\nu_\alpha\,\Nab)\,\ba_\beta \, \dif s+ (\nu_\alpha \, S^\alpha)\,\bn\,\dif s~ =  \bT^\alpha\,\nu_\alpha\, \dif s~,
\eqe
which implies Cauchy's formula 
\eqb{l}
\bT =
 (\bT^\alpha\otimes\ba_\alpha)\,\bnu =  \bsig^\mrT\,\bnu~,
\label{e:tractionT}
\eqe
since $\nu_\alpha = \bnu\cdot\ba_\alpha$. By comparing Eq.~\eqref{e:tractionT} with Eq.~\eqref{e:tractionTdef}, we can further write

\eqb{lll}
T^\alpha = \nu_\beta\,N^{\beta\alpha}~,\quad \quad
T^3 = \nu_\alpha \, S^\alpha ~.
\label{e:tractionT2}
\eqe

 \begin{remark}
Due to the presence of transverse shear components $S^\alpha$, the traction vector $\bT$ has both in-plane and out-of-plane components as is seen from Eq.~\eqref{e:tractionTdef} and \eqref{e:tractionT2}. As shown in Sec.~\ref{s:angbalance}, in general $\Nab$ in Eq.~\eqref{e:stresssig} is not symmetric according to angular momentum balance. Instead, the so-called effective stress  -- denoted by $\tilde{\sigma}^{\alpha\beta}$ and defined in Eq.~\eqref{e:sigabdef} -- is symmetric according to angular momentum balance. 
  \end{remark}

 \begin{remark}
The format of the Cauchy stress \eqref{e:stresssig} also contains all the stresses within a cut fiber.
Indeed, consider a fiber $\sC$ described by beam theory. According to beam theory, all stresses are neglected on a cut parallel to the fiber. Accordingly, the stress tensor in the fiber $\sC$ --  denoted by $\bsig_{\!\mathrm{fib}}$ --  can be written as
%
\eqb{lll}
\bsig_{\!\mathrm{fib}} = {\sigma}\,\bell\otimes\bell + {s}_c\, \bell \otimes \bc + {s}_\mrn\, \bell\otimes\bn~,
\label{e:bernoullistress}
\eqe
where ${\sigma}$, ${s}_c$, and $ {s}_\mrn$ denote the axial stress and the two shear stresses. By inserting the basis vectors $\bell=\ell^\alpha\,\ba_\alpha$, and $\bc=c^\alpha\,\ba_\alpha$, Eq.~\eqref{e:bernoullistress} becomes
\eqb{lll}
\bsig_{\!\mathrm{fib}} = {N}^{\alpha\beta}\,\ba_\alpha \otimes \ba_\beta +  {S}^\alpha\,\ba_\alpha\otimes\bn~,
\eqe
with
\eqb{lll}
 {N}^{\alpha\beta} \is  {\sigma}\,\ellab + {s}_c\,\ell^\alpha\,c^\beta~,\\[2mm]
 {S}^\alpha  \is {s}_\mrn\,\ell^\alpha~.
\eqe
%
\end{remark}

\subsection{Generalized moment tensor}\label{s:moment1}
This section discusses the moment tensor ${\bmuhat}$ for Kirchhoff-Love shells with in-plane bending. Also, equivalent stress couple tensors $\bmu$ and $\bmubar$ are defined
 by introducing stress couple vectors that are equivalent to the in-plane and out-of-plane bending moments.

At position $\bx$ on cut $\sI(s)$ (see Fig.~\ref{f:traction}c), the bending moment vector, denoted by {$\bmhat$}  (with unit moment per unit length), is allowed to have both in-plane and out-of-plane components.  It can be expressed with respect to  basis $\{\btau,~\bnu,~\bn\}$ as 
\eqb{lll}
{\bmhat} :  = m_\tau\,\btau + m_\nu\,\bnu +  \bar{m}\,\bn ~,
\label{e:bmdef1}
\eqe
where\footnote{Here and henceforth, the new terms  for  in-plane bending that are added to classical  Kirchhoff-Love shell theory, see e.g.~\cite{shelltheo}, are denoted by a bar.}  $\bar{m}\,\bn=:\bmbar$ is the moment causing in-plane bending 
 and
\eqb{lll}
m_\tau\,\btau + m_\nu\,\bnu  =: {\bm} ~,
\label{e:bmoffplanedef1}
\eqe
denotes the combined moment causing out-of-plane bending and twisting.
%
%

Similar to the Cauchy stress tensor \eqref{e:stresssig}, all moments on a cut perpendicular to surface normal $\bn$ are zero under Kirchhoff-Love assumptions. The total moment tensor $\bmuhat$ induced by ${\bmhat}$ can thus be expressed in the form
\eqb{lll}
{\bmuhat} \dis  \mab\ba_\alpha\otimes\ba_\beta + \bar{m}^\alpha\,\ba_\alpha\otimes\bn~.
\label{e:bmustar1}
\eqe
In this form, the first term $ \mab\ba_\alpha\otimes\ba_\beta$ combines both out-of-plane bending and twisting in response to a change in the out-of-plane curvature of $\sS$, while the second term  $\bar{m}^\alpha\,\ba_\alpha\otimes\bn$ is the response to a change in the in-plane curvature of fiber $\sC$.

The components of $\bmuhat^\mrT$, on the cross-section perpendicular to $\ba^\alpha$, thus read as (see Fig.~\ref{f:traction}c)
\eqb{lll}
\bmhat^\alpha \dis \bmuhat^\mrT\,\ba^\alpha =   \mab\ba_\beta + \bar{m}^\alpha\,\bn ~.
\label{e:mhat_map_baalpha}
\eqe
Here, moment vectors $ \mab\ba_\beta$ and $\bar{m}^\alpha\,\bn$ are associated with the angular velocity vector around the in-plane axis
\eqb{lll}
\bn\times\dot{\bn}=(\bn\cdot\dot{\ba}^\alpha)\,\ba_\alpha\times\bn~,
\eqe
and  the out-of-plane axis $\bn = \bell\times\bc = \ell^\alpha\ba_\alpha\times\bc$, respectively. Therefore, it is  mathematically convenient to express $\bmhat^\alpha $ with respect to the basis $\{\ba_1\!\times\!\bn,~\ba_2\!\times\!\bn,~\bn\}$, i.e.
\eqb{lll}
\bmhat^\alpha \is \Mab(\ba_\beta\times\bn)  + \Mbarab\,(\ba_\beta\times\bc) ~\\[3mm]
  \is \bn\times\bM^\alpha \, +\, \bc\times\bMbar^{\alpha}~,
\label{e:defbMa}
\eqe
where $\Mab$ and $\Mbarab$ denote the components of vectors $\bmhat^\alpha$ in directions $\ba_\beta\times\bn$ and $\ba_\beta\times\bc$, respectively. 
Here, we have defined the so-called stress couple vectors for  out-of-plane and in-plane bending\footnote{\label{noteSignM}The sign convention for the moment components follows \cite{steigmann99} and \cite{shelltheo}.}
\eqb{lll}
\bM^\alpha := -\Mab\,\ba_\beta~, \quad $and$ \quad 
\bMbar^\alpha := -\Mbarab\,\ba_\beta~, \quad $with$\quad \Mbarab :=  \bar{m}^\alpha\,\ell^\beta~.
\label{e:defMbaralp}
\eqe
Eq.~\eqref{e:bmustar1} thus becomes
\eqb{lll}
{\bmuhat} \dis  \Mab \ba_\alpha\otimes(\ba_\beta\times\bn) + \bar{M}^{\alpha\beta}\,\ba_\alpha\otimes(\ba_\beta\times\bc)~.
\label{e:bmustar}
\eqe
Similar to Eq.~\eqref{e:tractionT}, moment equilibrium of the triangle (see Fig.~\ref{f:traction}c) results in Cauchy's formula
\eqb{lll}
{\bmhat} = \bmhat^\alpha \,\nu_\alpha = \bn\!\times\!\bM + \bc\!\times\!\bMbar =  \bmuhat^\mrT\,\bnu~,
\label{e:physMmap}
\eqe
where
\eqb{ll}
\bM:= \bM^\alpha\,\nu_\alpha~,\quad $and$ \quad \bMbar:= \bMbar^\alpha\,\nu_\alpha
\label{e:defMbar}
\eqe
%
 are referred to as the stress couple vectors associated with  out-of-plane and in-plane bending, respectively. By comparing Eq.~\eqref{e:physMmap} and \eqref{e:bmdef1}, the stress couple vectors $\bM$ and $\bMbar$ can be related to their moment vector counterparts by
\eqb{ll}
\bm = \bn\times\bM~, \quad $and$ \quad \bmbar = \bc\times\bMbar~, \quad $with$ \quad \bMbar = -\bar{m}\,\bell~.
\label{e:defMbarcross}
\eqe

%
In line with previous works, see e.g.~\cite{shelltheo},  we can also define the  stress couple  tensors,\footnoteref{noteSignM}
%
\eqb{lll}
\bmu:= - \Mab\,\ba_\alpha\otimes\ba_\beta~,\quad $and$\quad \bmubar:= - \Mbarab\,\ba_\alpha\otimes\ba_\beta~,
\label{e:defbmuandbmubar}
\eqe
associated with  out-of-plane and in-plane bending, respectively.  In view of \eqref{e:physMmap}, we obtain the mapping
\eqb{lll}
\bM = \bmu^\mrT\,\bnu~, \quad $and$\quad \bMbar =  \bmubar^\mrT\,\bnu.
\label{e:definedCoupleM_Mbar}
\eqe
%
%
In order to relate the components of the moment vector \eqref{e:bmdef1} to the components of the  stress couple tensors, we can equate equations \eqref{e:bmdef1} and \eqref{e:physMmap}. This results in
\eqb{lll}
m_\nu \dis {\bmhat}\cdot\bnu =  \bm\cdot\bnu = \Mab\,\nu_\alpha\,\tau_\beta~,\\[2mm]
m_\tau \dis {\bmhat}\cdot\btau =   \bm\cdot\btau =-\Mab\,\nu_\alpha\,\nu_\beta~,\\[2mm]
\bar{m} \dis {\bmhat}\cdot\bn =   \bmbar\cdot\bn =  \Mbarab\,\nu_\alpha \,\ell_\beta  =  \bar{m}^\alpha\,\nu_\alpha~,
\label{e:mcomps}
\eqe
where we have used the relations $\ba_\beta\times\bn = \tau_\beta\bnu - \nu_\beta\,\btau$ due to $\ba_\alpha = \tau_\alpha\,\btau + \nu_\alpha\,\bnu$.
%

 \begin{remark}
 Similar to Eq.~\eqref{e:bernoullistress}, the  moment tensor for a fiber described by beam theory -- denoted by $\bmuhat_{\mathrm{fib}}$ --  
 can also be written in the form of Eq.~\eqref{e:bmustar1}, i.e.~%
\eqb{lll}
\bmuhat_{\mathrm{fib}} := {\mu}_\ell\,\bell\otimes\bell + {\mu}_c\,\bell\otimes\bc + \bar{\mu}\, \bell\otimes\bn = {m}^{\alpha\beta}\,\ba_\alpha\otimes\ba_\beta +  \bar{m}^\alpha\,\ba_\alpha\otimes\bn~.
\eqe
Here, $ \mu_\ell$, ${\mu}_c$, and $ \bar{\mu}$ denote twisting, out-of-plane bending, and in-plane bending moments in fiber $\sC$, respectively,  and we have identified 
\eqb{lll}
 {m}^{\alpha\beta} \is {\mu}_\ell\,\ellab + {\mu}_c\,\ell^\alpha\,c^\beta~,\\[2mm]
 \bar{m}^\alpha \is \bar{\mu}\,\ell^\alpha~.
 \label{e:remarkmubar}
\eqe
\end{remark}

 \begin{remark}
Further, inserting $\bar{m}^\alpha$ from Eq.~\eqref{e:remarkmubar} into (\ref{e:defMbaralp}.3) and (\ref{e:mcomps}.3) gives
\eqb{lll}
\Mbarab \is {\bar{\mu}}\,\ellab~\\[2mm]
\bar{m} \is {\bar{\mu}}\,\bell\cdot\bnu~.
\label{e:Mbarabfiber}
\eqe
Therefore, we can conclude that $\Mbarab$ is symmetric. Note that $\bar{\mu}$ does not depend on the cut $\sI$ since it is a component of the internal moment tensor, but $\bar{m}$ does depend on  $\sI$ as seen in Eq.~(\ref{e:Mbarabfiber}.2). For example,~$\bar{m}=0$ when the cut is parallel to the fiber, i.e.~when $\bell\cdot\bnu=0$.
\end{remark}
%



\subsection{Balance of linear momentum}
Consider body forces $\bff$ acting on an arbitrary simply-connected region ${\sR}\subset\sS$. The balance of linear momentum implies that the temporal change of the linear momentum is equal to the resultant of all  acting external forces. That is,
\eqb{l}
\ds \frac{D}{Dt} \int_{\sR}\rho\,\bv\, \dif a  =  \int_{\sR}\bff\, \dif a + \ds \int_{\partial\sR} \bT\, \dif s~, \quad\,\forall\in\sS~,
\label{e:linearmomentumglobal}
\eqe
where $\bv$ denotes the material velocity of the surface. Inserting $\bT$ from Eq.~\eqref{e:tractionT}, and applying the surface divergence theorem and conservation of mass, one obtains the local form of linear momentum balance
\eqb{l}
\mathrm{div}_\mrs\,\bsig^\mrT + \bff = \rho\,\dot{\bv}~,
\label{e:linmomentum}
\eqe
where $\mathrm{div}_\mrs$ denotes the surface divergence operator, defined by $\mathrm{div}_\mrs \bullet := \bullet_{,\alpha}\!\cdot\!\ba^\alpha$. 
Note that Eq.~\eqref{e:linmomentum} can also be written in the form (see \cite{shelltheo}) 
\eqb{l}
\bT^\alpha_{;\alpha} + \bff = \rho\,\dot{\bv}~,
\eqe
%
since
\eqb{lll}
\mathrm{div}_\mrs\bsig^\mrT = \bsig^\mrT_{,\beta}\cdot\ba^\beta = \bT^\alpha_{,\alpha} + \Gamma_{\alpha\beta}^\beta\,\bT^\alpha = \bT^\alpha_{;\alpha}~,
\label{e:div_vs_comma}
\eqe
follows from Eq.~\eqref{e:bsigT}.

\subsection{Balance of angular momentum}\label{s:angbalance}
The balance of angular momentum implies that the temporal change of angular momentum is equal to  the resultant of all external moments. That is,
\eqb{l}
\ds \frac{D}{Dt} \int_{{\sR}}\rho\,\bx\times\bv\, \dif a  =  \int_{{\sR}}\bx\times\bff\, \dif a + \ds \int_{\partial{\sR}} \bx\times\bT\, \dif s +   \ds \int_{\partial{\sR}} {\bmhat}\, \dif s~,
\label{e:angmomentglobal}
\eqe
where ${\bmhat}$ includes both in-plane and  out-of-plane  bending moments acting on $\partial{\sR}$. From Eqs.~\eqref{e:tractionT} and \eqref{e:physMmap}, the surface divergence theorem gives
\eqb{lll}
\ds \int_{\partial{\sR}}\bx\times\bT\, \dif s  \is \ds \int_{{\sR}} \big(\ba_\alpha\!\times\!\bT^\alpha\, +\, \bx\!\times\!\mathrm{div}_\mrs\,\bsig^\mrT\big)\, \dif a~
\eqe
and
\eqb{lll}
 \ds \int_{\partial{\sR}}{\bmhat}\, \dif s  \is \ds \int_{{\sR}} \mathrm{div}_\mrs\, {\bmuhat^\mrT}\, \dif a~.
\eqe
Inserting these equations into~\eqref{e:angmomentglobal} and applying local mass conservation gives
\eqb{l}
\ds\int_{{\sR}}( \ba_\alpha\!\times\!\bT^\alpha\, + \,  \mathrm{div}_\mrs\,{\bmuhat^\mrT}) \, \dif a + \ds  \int_{{\sR}}\bx\times (\mathrm{div}_\mrs\,\bsig^\mrT + \bff  - \rho\,\dot{\bv})\, \dif a = \boldsymbol{0}~.
\label{e:angmomentglobal2}
\eqe
The second integral in Eq.~\eqref{e:angmomentglobal2} vanishes due to \eqref{e:linmomentum}. This leads to the local form of the angular momentum balance,
\eqb{l}
\ba_\alpha\!\times\!\bT^\alpha  \, +\,  \mathrm{div}_\mrs\,{\bmuhat^\mrT} = \boldsymbol{0}~.
\label{e:Mbalance}
\eqe
Keeping Eq.~(\ref{e:Mbarabfiber}.1)  in mind, the surface divergence of tensor $\bmuhat^\mrT$ follows from Eq.~\eqref{e:bmustar} as
\eqb{l}
 %

 \mathrm{div}_\mrs\, {\bmuhat^\mrT} =  \big(-\Mab\,b_\alpha^\gamma   + \bar{m}^\alpha_{;\alpha}\,\ell^\beta\,c^\gamma\big)\,\ba_\beta\!\times\!\ba_\gamma + \big(M^{\beta\alpha}_{;\beta} 
 + {\bar{\mu}}\, \tau_\mrg\,\ell^\alpha -   {\bar{\mu}}\, \kappa_\mrn\,c^\alpha \big)\,\ba_\alpha\!\times\!\bn ~, 

    \label{e:divmu}
\eqe
where we have used $\ba_\alpha = \ell_\alpha\bell + c_\alpha\bc = (\ell_\alpha\,c^\gamma - c_\alpha\ell^\gamma)\,\ba_\gamma\times\bn$.
Inserting Eqs.~\eqref{e:divmu} and \eqref{e:bTalp} into Eq.~\eqref{e:Mbalance} then implies that the so-called effective membrane stress\footnote{Note that we will redefine $\tilde{\sigma}^{\alpha\beta}$ later in Eq.~(\ref{e:Wint01}.2) and \eqref{e:sigdeffinal} for various choices of  the strain measure for in-plane bending.}
\eqb{l}
\tilde{\sigma}^{\alpha\beta}:= \Nab - M^{\gamma\alpha}\,b_\gamma^\beta + \bar{m}^\gamma_{;\gamma}\,\ell^\alpha\,c^\beta 
\label{e:sigabdef}
\eqe
is symmetric, and that the shear stress is given by
\eqb{l}
S^\alpha= - M^{\beta\alpha}_{;\beta} + {\bar{\mu}}\, (\kappa_\mrn \,c^\alpha - \tau_\mrg\,\ell^\alpha)~.
\label{e:Sadef}
\eqe
Thus, angular momentum balance implies the symmetry of the effective stress $\tilde{\sigma}^{\alpha\beta}$ instead of  the Cauchy stress $\Nab$. The latter is only symmetric when there is no  in-plane and out-of-plane bending resistance. 
%
 \begin{remark}
 The last term in stress expression \eqref{e:sigabdef} relates to the in-plane shear force in the fiber, which, in accordance with the assumed Euler-Bernoulli kinematics, follows as the derivative of the in-plane bending moment.
\end{remark}


\subsection{Mechanical power balance}{\label{s:balancepower}}
The mechanical power balance can be obtained from local momentum balance~\eqref{e:linmomentum}. To this end, we can write
\eqb{l}
 \ds\int_{{\sR}} \bv\cdot( \mathrm{div}_\mrs\,\bsig^\mrT+ \bff -  \rho\,\dot{\bv}) \,\dif a~ = 0~.
 \label{e:Pbalance}
\eqe
Here, the divergence term can be transformed by the identity
\eqb{l}
\bv\cdot \mathrm{div}_\mrs\,\bsig^\mrT = \mathrm{div}_\mrs(\bv\,\bsig^\mrT) - \bsig^\mrT:\nabla_{\!\mrs}\,\bv~,
\label{e:divgrad}
\eqe
{by} the divergence theorem 
\eqb{lll}
 \ds\int_{{\sR}}  \mathrm{div}_\mrs(\bv\,\bsig^\mrT) \,\dif a =  \int_{\partial{\sR}}\,\bv\cdot\bsig^\mrT\,\bnu\,\dif s =    \int_{\partial{\sR}}\,\bv\cdot\bT\,\dif s~,
 \label{e:divgradInt}
\eqe
and by using the relation
\eqb{lll}
\bsig^\mrT:\nabla_{\!\mrs}\,\bv = \Nab\,\ba_\beta\!\cdot\!\,\dot{\ba}_\alpha + S^\alpha\,\bn\cdot\dot{\ba}_\alpha~.
\label{e:Sig2DotGradv}
\eqe
Inserting Eq.~\eqref{e:sigabdef} and \eqref{e:Sadef} into Eq.~\eqref{e:Sig2DotGradv} gives
\eqb{lll}
\bsig^\mrT:\nabla_{\!\mrs}\bv =  \ds \frac{1}{2}\,\tilde{\sigma}^{\alpha\beta}\,\dot{a}_{\alpha\beta} -  M^{\alpha\beta}\bn_{;\alpha}\!\cdot\!\dot{\ba}_\beta -  \bar{\mu}  \big(\kappa_\mrn\, \bc\cdot\dot{\bn} - \tau_\mrg\, \bell\cdot \dot{\bn}\big)
-  M^{\beta\alpha}_{;\beta}\bn\!\cdot\!\dot{\ba}_\alpha +   \bar{m}^\alpha_{;\alpha}\,\bell\!\cdot\!\dot{\bc}~.
\label{e:Sig2DotGradv2}
\eqe
Considering Eq.~(\ref{e:curvector}.2), \eqref{e:coderivbc}, and (\ref{e:remarkmubar}.2), the last two terms can be written as 
\eqb{lll}
  M^{\beta\alpha}_{;\beta}\,\bn\cdot\dot{\ba}_\alpha \is (\dot{\bn}\cdot\bM^\alpha)_{;\alpha} + \Mab\,\dot{\bn}_{;\alpha}\cdot\,\ba_\beta~,\\[2mm]
  \bar{m}^\alpha_{;\alpha}\,\bell\cdot\dot{\bc} \is ( \bar{m}^\alpha\bell\cdot\dot{\bc})_{;\alpha} + \bar{m}^\alpha\, \dot{(\overline{c^\beta\,\ell_{\beta;\alpha}})} + \bar{\mu}\,(\kappa_\mrn\,\bc\cdot\dot{\bn} - \tau_\mrg\,\bell\cdot\dot{\bn})~.
  \label{e:Sig2DotGradv3}
\eqe

By taking  Eqs.~\eqref{e:divgrad}, \eqref{e:divgradInt},\eqref{e:Sig2DotGradv2},  \eqref{e:Sig2DotGradv3},     and local mass conservation into account, Eq.~\eqref{e:Pbalance} becomes
\eqb{l}
\dot{K} + P_{\mathrm{int}} = P_{\mathrm{ext}}~,
\label{e:Pbalance2}
\eqe
where 
\eqb{l}
\dot{K}  =  \ds \int_{{\sR}}\,\rho\,\bv\cdot\dot{\bv}\,\dif a
\label{e:Kdot}
\eqe
is the rate of kinetic energy,
\eqb{lll}
 P_{\mathrm{int}} \is\ds\frac{1}{2}\int_{{\sR}}\, \tilde{\sigma}^{\alpha\beta} \,\dot{a}_{\alpha\beta}\,\dif a + \int_{{\sR}}\Mab\,\dot{b}_{\alpha\beta}\,\dif a + \,\int_{{\sR}} \bar{m}^\alpha\, \dot{(\overline{c^\beta\,\ell_{\beta;\alpha}})} \,\dif a ~,
  \label{e:Pint}
 \eqe
 is the internal power, and
 \eqb{lll}
  P_{\mathrm{ext}} \is \ds \int_{{\sR}}\,\bv\cdot\bff\,\dif a + \int_{\partial{\sR}}\,\bv\cdot\bT\,\dif s + \int_{\partial{\sR}}\,\dot{\bn}\cdot\bM\,\dif s + \int_{\partial\sR}\, \dot{\bc}\cdot\bMbar\,\dif s~
    \label{e:Pextok}
\eqe
denotes the external power. The last term in $ P_{\mathrm{int}}$ can be written in alternative forms, as is discussed in the following section. The mechanical power balance \eqref{e:Pbalance2} simplifies to the expression in \citet{shelltheo}, for $\bar m^\alpha \equiv 0$ and $\bMbar = \boldsymbol{0}$.
%

\section{Work-conjugate variables and constitutive equations} {\label{s:workconjugation}}

As seen from the expression~\eqref{e:Pint}, the internal stress power per current area reads
 \eqb{lll}
 \dot{w}_{\mathrm{int}} := \frac{1}{2} \tilde{\sigma}^{\alpha\beta} \,\dot{a}_{\alpha\beta}  + \Mab\,\dot{b}_{\alpha\beta} + \bar{m}^\alpha\, \dot{(\overline{c^\beta\,\ell_{\beta;\alpha}})}~,
  \label{e:Wint0}
 \eqe
 which directly shows work-conjugate pairs.
Accordingly, {$c^\beta\,\ell_{\beta;\alpha}\,\bA^\alpha = \bc\,\bar\nabla_{\!\mrs}\bell\,\bF^{-1}$} is a strain measure for {the change in the in-plane} curvatures.  As shown in Appendix~\ref{s:frameinvariance}, the strain measure $c^\beta\,\ell_{\beta;\alpha}$ is frame invariant under superimposed rigid body motion of the shell. 
However, it is somewhat unintuitive and thus inconvenient for the construction of material models. In the following, we will present two approaches with alternative definitions of this strain measure.

\subsection{Geodesic curvature-based approach}{\label{s:geodesicapproach}}

This approach uses the geodesic curvature $\kappa_\mrg$ instead of $c^\beta\,\ell_{\beta;\alpha}$ within the in-plane bending power. Namely, inserting $\bar{m}^\alpha$ from (\ref{e:remarkmubar}.2) into Eq.~\eqref{e:Wint0} and rearranging terms gives
  \eqb{rll}
 \dot{w}_{\mathrm{int}} \dis \frac{1}{2} {\sigma}^{\alpha\beta} \,\dot{a}_{\alpha\beta}  + \Mab\,\dot{b}_{\alpha\beta} +{ \bar{\mu}}\,\dot{\kappa}_\mrg~,  \\[3mm]
$with $ \quad  \sigma^{\alpha\beta} \dis  \tilde{\sigma}^{\alpha\beta} +{ \bar{\mu}}\,\kappa_\mrg\,\ellab~. 
  \label{e:Wint01}
 \eqe
Here, $\sigab$ represents the components of an effective membrane stress tensor\footnote{\label{note1} Note that, generally $\sigma^{\alpha\beta} \neq  N^{\alpha\beta}=\ba^\alpha\,\bsig\,\ba^\beta$. $\sigab = N^{\alpha\beta}$ only {in special cases, i.e.~}when both in-plane and out-of-plane bending are negligible.}. It is symmetric as both $\tilde{\sigma}^{\alpha\beta}$ and $\ellab$ are symmetric. The internal power \eqref{e:Pint} then becomes
 \eqb{lll}
 P_{\mathrm{int}} \is \ds\frac{1}{2}\int_{\sR_0}\, \tau^{\alpha\beta} \,\dot{a}_{\alpha\beta}\,\dif A + \int_{\sR_0}\Mab_0\,\dot{b}_{\alpha\beta}\,\dif A +  \int_{\sR_0} \bar{\mu}_0\,\dot{\kappa}_\mrg\,\dif A~,
 \label{e:workpairKappa}
\eqe
where {$\sR_0\in\sS_0$} and 
\eqb{lll}
\tauab := J\,\sigab~,\quad\quad
\Mab_0 := J\,\Mab~,\quad\quad
\bar{\mu}_0 := J\,{ \bar{\mu}}~.
\label{e:stress_geodesic_based}
\eqe
These are the nominal quantities corresponding to the physical quantities $\sigab$, $\Mab$, and $ \bar{\mu}$, respectively.
Accordingly, we can assume a stored energy function of an hyperelastic shell in the form
\eqb{l}
W = {\check{W}} (a _{\alpha\beta},~b_{\alpha\beta},~\kappa_\mrg;~h^{\alpha\beta} )~,
\label{e:WEKo}
\eqe 
where $h^{\alpha\beta}$ collectively represents the components of the structural tensor(s), e.g.~$\ell^{\alpha\beta}$, $c^{\alpha\beta}$,  or, $c^\alpha\ell^\beta$, that characterize material anisotropy due to embedded fibers. Eq.~\eqref{e:WEKo} can equivalently be expressed in terms of invariants. All the strain measures $a _{\alpha\beta}$, $b_{\alpha\beta}$, $\kappa_\mrg$, and $h^{\alpha\beta}$ used in Eq.~\eqref{e:WEKo} are frame invariant under superimposed rigid body motions as shown in Appendix~\ref{s:frameinvariance}.  Using the usual arguments of \citet{coleman64}, the constitutive equations can be written as
\eqb{llrlrlr}
\tau^{\alpha\beta} = \ds2\,\pa{ {\check{W}}}{a _{\alpha\beta}}~,\quad\quad
M_0^{\alpha\beta} = \ds\pa{ {\check{W}}}{b _{\alpha\beta}}~,\quad\quad
\bar{\mu}_{0} \is \ds\pa{ {\check{W}}}{\kappa _\mrg}~.
\label{e:consticomKappa}
\eqe

%
 \begin{remark}
Compared to classical Kirchhoff-Love shell theory (see e.g.~\cite{naghdi82,shelltheo}, the effective membrane stress $\sigma^{\alpha\beta}$ in (\ref{e:Wint01}.2) additionally contains the  high order bending term $\bar\mu$ and the in-plane fiber shear term $m^\gamma_{;\gamma}$ (see Eq.~\eqref{e:sigabdef}). For slender fibers, they are negligible since the in-plane bending stiffness is usually much smaller than the membrane stiffness. However, these terms may become significant when there is a large (usually local) change in curvature (e.g.~at shear bands).
\end{remark}


 \begin{remark}
 Although a constitutive formulation following from \eqref{e:workpairKappa} appears elegant,
 expression (\ref{e:consticomKappa}.3) is restricted to a material response expressible in terms of the geodesic curvature $\kappa_\mrg$. Therefore, this setup might be unsuited for complex material behavior, e.g.~due to fiber dispersion \citep{Gasser2006}. In such cases, a more sophisticated structural tensor is usually desired for the in-plane bending response, and it thus may not always be possible to express $W$ in terms of $\kappa_\mrg$.  This  motivates the following director gradient-based approach.
 \end{remark}
 
%
 \begin{remark}
In Eq.~\eqref{e:consticomKappa}, the stress components $\tau^{\alpha\beta}$ and  $M_0^{\alpha\beta}$,  together with the strain components $\auab$ and $\buab$  in Eq.~\eqref{e:WEKo}, are defined in the parameter space $\sP$, where they can be treated as independent variables without forming them into tensors.  It is possible, though, to construct different stress and strain tensors from these components, such that their scalar product results in the same power as in Eq.~\eqref{e:workpairKappa}. For example, $\tau^{\alpha\beta}$ can be the components of the Kirchhoff surface stress tensor  $\hat\btau := \tauab\,\ba_\alpha\otimes\ba_\beta$, 
the second Piola-Kirchhoff  stress tensor $\bS$, or the first Piola-Kirchhoff stress tensor $\bP$ but with different bases. Indeed,  $\bS = \bF^\mrT\,\hat\btau\,\bF = \tauab\,\bA_\alpha\otimes\bA_\beta$, and $\bP = \bF\,\bS= \tauab\,\ba_\alpha \otimes\bA_\beta$.  The strain variables work-conjugate to $\hat\btau$, $\bS$, and $\bP$ are the rate of surface deformation tensor $\bD$, 
the Green-Lagrange surface strain tensor $\bE$ (or $\frac{1}{2}\,\bC$), and  the surface deformation gradient tensor $\bF$, respectively, since $\hat\btau:\bD = \bS:\dot{\bE} = \bS:\frac{1}{2}\dot{\bC} = \bP:\dot\bF = \frac{1}{2}\tauab\,\dot{a}_{\alpha\beta}$. See also Remark 4 in \cite{sauer2019shell}.
\end{remark}

\subsection{Director gradient-based approach}{\label{s:directorbased}}

 
In this approach, the power expression \eqref{e:Wint0} is rewritten by employing relation \eqref{e:c_ab_vs_l_ab} and definitions (\ref{e:defMbaralp}.3) and  \eqref{e:binplane}  as
 \eqb{lll}
 \dot{w}_{\mathrm{int}} := \frac{1}{2} {\sigma}^{\alpha\beta} \,\dot{a}_{\alpha\beta}  + \Mab\,\dot{b}_{\alpha\beta} + \bar{M}^{\alpha\beta}\,\dot{\bar{b}}_{\beta\alpha}~,
 \label{e:workpair0}
 \eqe
where   $\sigma^{\alpha\beta}$ is now defined by\footnoteref{note1}
\eqb{lll}
\sigma^{\alpha\beta} \dis  \tilde{\sigma}^{\alpha\beta}  + (\bar{M}^{\gamma\delta}\,c_{\delta;\gamma})\,\ellab~. 
\label{e:sigdeffinal}
\eqe
Thus, the internal power can be written as
\eqb{lll}
 P_{\mathrm{int}} \is \ds\frac{1}{2}\int_{\sR_0}\, \tau^{\alpha\beta} \,\dot{a}_{\alpha\beta}\,\dif A + \int_{\sR_0}\Mab_0\,\dot{b}_{\alpha\beta}\,\dif A +  \int_{\sR_0}\Mbarab_0\,\dot{\bar{b}}_{\alpha\beta}\,\dif A~,
 \label{e:workpair3}
\eqe
where we have again used \eqref{e:stress_geodesic_based} and defined
\eqb{lll}
\Mbarab_0 := J\,\Mbarab~.
\label{e:stress_director_based}
\eqe
Accordingly, the  stored energy function  can now be given in the form
\eqb{l}
W = {\hat{W}}(a _{\alpha\beta},~b_{\alpha\beta},~\bar{b}_{\alpha\beta};~h^{\alpha\beta} )~.
\label{e:WEKoo}
\eqe 
This function can also be expressed equivalently  in terms of invariants, and $\bar{b}_{\alpha\beta}$ can  be shown to be frame invariant, see Appendix~\ref{s:frameinvariance}. The corresponding constitutive equations now read
\eqb{llrlrlr}
\tau^{\alpha\beta} = \ds2\,\pa{{\hat{W}}}{a _{\alpha\beta}}~,\quad\quad
M_0^{\alpha\beta} = \ds\pa{{\hat{W}}}{b _{\alpha\beta}}~,\quad\quad
\bar{M}_{0}^{\alpha\beta} \is \ds\pa{{\hat{W}}}{\bar{b} _{\alpha\beta}}~.
\label{e:consticomoo}
\eqe

 \begin{remark}
 Compared to \eqref{e:WEKo}, expression  \eqref{e:WEKoo} allows to model more complex in-plane bending behavior using a generalized structural tensor applied to $\bar{b}_{\alpha\beta}$. We therefore consider this setup in the  following sections.
\end{remark}

 \begin{remark}
 \label{rm:work}
 Eq.~\eqref{e:workpair3} can also be written in tensor notation  as
\eqb{lll}
 P_{\mathrm{int}} \is \ds\int_{\sR_0}\, \bS:\dot{\bE}~\dif A - \int_{\sR_0}\bmu_0:\dot{\bK}\,\dif A -  \int_{\sR_0}\bmubar_0: {{{\boldsymbol{\dot{\bar K}}}}}\,\,\dif A,
 \label{e:workpair4}
\eqe
where  $\bE$, $\bK$, and $\bKbar$ are the strain tensors defined by \eqref{e:bEtensor}, \eqref{e:Kten}, and \eqref{e:Ktenb}, respectively, and
\eqb{lll}
\bS := \tauab\,\bA_\alpha\otimes\bA_\beta~,\quad\quad
\bmu_0 := -\Mab_0\,\bA_\alpha\otimes\bA_\beta~,\quad\quad
\bmubar_0 :=  -\Mbarab_0\,\bA_\alpha\otimes\bA_\beta~,
\label{e:stressmoment_abs}
\eqe
are  the effective second Piola-Kirchhoff surface stress tensor, and the nominal  stress couple tensors associated with  out-of-plane and  in-plane bending, respectively. {They are symmetric and follow from the pull-back of tensors $J\,\sigma^{\alpha\beta}\ba_\alpha\otimes\ba_\beta$, $J\,\bmu$ and $J\,\bmubar$ in Eqs.~\eqref{e:sigdeffinal} and \eqref{e:defbmuandbmubar}.}
\end{remark}

\begin{remark}
{
In view of internal power expression \eqref{e:workpair4}, an alternative form of the stored energy function, apart from Eq.~\eqref{e:WEKoo}, is
\eqb{l}
W =\tilde{W}\big(\bE, \bK,  \bKbar; ~\bH\big)~,
\label{e:WEKabs}
\eqe 
where $\bH$ denotes the structural tensor(s). However, since the temporal change of $\bE$, $\bK$, and  $\bKbar$ only depends on $a_{\alpha\beta}$, $b_{\alpha\beta}$, and $\bar b _{\alpha\beta}$, respectively, the energy form \eqref{e:WEKabs} is equivalent to \eqref{e:WEKoo}, i.e.~$\tilde{W}\big(\bE, \bK,  \bKbar\big) = \hat{W}(a _{\alpha\beta},~b_{\alpha\beta},~\bar{b}_{\alpha\beta})$. Therefore, the stress and moment tensors \eqref{e:stressmoment_abs} can be determined either from
 \eqb{llrlrlr}
\bS = \ds\,\pa{\tilde W}{\bE}~,\quad\quad
-\bmu_0  =  \ds\pa{\tilde W}{\bK}~,\quad\quad
-\bmubar_0 = \ds\pa{\tilde W}{\bKbar}~,
\label{e:consticomabs}
\eqe
or from their components $\tauab$, $\Mab_0$, and $\Mbarab_0$ using Eq.~\eqref{e:consticomoo}. 
}
\end{remark}

\begin{remark}
{
Balance equation \eqref{e:Pbalance2} and various constitutive equations, such as  \eqref{e:consticomabs}, \eqref{e:consticomoo}, and \eqref{e:consticomKappa} have been expressed directly in surface form without introducing a thickness variable. The unit of the strain energy $W$ is thus energy per reference area. However, the influence of the thickness is still present -- either in the material parameters of the surface form, or via a through-the-thickness integration of 3D constitutive laws (see e.g.~\cite{kiendl15,solidshell}).  In a surface formulation,  the thickness strain can be still determined from various approaches: One can enforce incompressibility or the plane stress condition \citep{solidshell}, or one can introduce an additional degree-of-freedom \citep{simo90b}. 
}
\end{remark}

\subsection{Comparison with existing second-gradient theory of Kirchhoff-Love shells}

To show the consistency of our proposed theory with the existing second-gradient theory of  \cite{Steigmann2018}, 
we  insert $\dot{\kappa}_\mrg$ obtained from \eqref{e:kgok} into   the power expression \eqref{e:workpairKappa}. This results in the expression (see  Appendix~\ref{s:app_effectivestress})
\eqb{lll}
 P_{\mathrm{int}} \is \ds \int_{\sR_0}\, \btau^{\alpha}\cdot \,\dot{\ba}_{\alpha}\,\dif A + \int_{\sR_0}\Mab_0\,\dot{b}_{\alpha\beta}\,\dif A +  \int_{\sR_0}\Mbarab_{0\gamma}\, \dot{S}_{\alpha\beta}^\gamma\,\dif A~,
 \label{e:workpairGradientTheory0}
\eqe
{where ${S}_{\alpha\beta}^\gamma := {\Gamma}_{\alpha\beta}^\gamma -  \bar{\Gamma}_{\alpha\beta}^\gamma$.}
The last term in  \eqref{e:workpairGradientTheory0} is the power due to in-plane fiber bending. Here, the relative Christoffel symbol ${S}_{\alpha\beta}^\gamma$ is the chosen strain measure for the in-plane curvature, and $\Mbarab_{0\gamma}:= J\,\bar{\mu}\, \ellab\,c_\gamma~$ is the in-plane bending moment corresponding to a change in ${S}_{\alpha\beta}^\gamma$. 

In the first term of \eqref{e:workpairGradientTheory0},  $\btau^{\alpha}:= J\,\sigab\,\ba_\beta$ denotes the effective stress vectors that are  work-conjugate to $\dot{\ba}_\alpha$, with $\sigab$ now being defined by
\eqb{lll}
\sigab := \tilde\sigma^{\alpha\beta} - \bar{\mu}\,\kappa_\mrg\,\ellab +  \bar{\mu}\,(\lambda^{-1}\, L^\alpha_{;\gamma}\,\ell^{\gamma} + S_{\gamma\delta}^\alpha\,\ell^{\gamma\delta})\,c^\beta - \bar{\mu}\,(\lambda^{-1}\,L^\gamma_{;\delta}\,\ell_{\gamma}^{\delta} + S_{\gamma\delta}^{\theta}\,\ell^{\gamma\delta}\,\ell_{\theta})\,\ell^{\alpha}\,c^\beta~,
\label{e:eff_sigab}
\eqe
which is generally unsymmetric.
Therefore, in contrast to  expressions \eqref{e:workpairKappa} and \eqref{e:workpair3},  the effective stress $\tauab:=\btau^\alpha\cdot\ba^\beta$ is generally unsymmetric here. 
With a possible loss of generality,  its symmetrization is adopted in \cite{Steigmann2018}, i.e.~$\tau^{\alpha\beta}:=\frac{1}{2}( \btau^\alpha\cdot\ba^\beta + \btau^\beta\cdot\ba^\alpha)$, so that the internal power becomes
\eqb{lll}
 P_{\mathrm{int}} \is \ds\frac{1}{2}\int_{\sR_0}\, \tau^{\alpha\beta} \,\dot{a}_{\alpha\beta}\,\dif A + \int_{\sR_0}\Mab_0\,\dot{b}_{\alpha\beta}\,\dif A +  \int_{\sR_0}\Mbarab_{0\gamma}\,\dot{S}_{\alpha\beta}^\gamma\,\dif A~,
 \label{e:workpairGradientTheory}
\eqe
which is equivalent to the expression (63) in \cite{Steigmann2018}.\footnote{adapted to the notations used in this paper.}

 \begin{remark}
As shown in Appendix~\ref{s:app_effectivestress}, the  asymmetry of the effective stress \eqref{e:eff_sigab} is due to the fact that it still contains in-plane bending apart from surface stretching. It is unsymmetric even for initially straight fibers in a general setting. Therefore the symmetrization employed in \cite{Steigmann2018} is valid only for special cases.
\end{remark}

%
 \begin{remark}
The power term $\Mbarab_{0\gamma}\,\dot{S}_{\alpha\beta}^\gamma$  in the gradient theory of \cite{Steigmann2018} can become ill-defined for initially curved fibers when $\dot{\Gamma}_{\alpha\beta}^\gamma$ approaches zero even though there is a change in geodesic curvature. This can solely result from the choice of parametrization, as seen e.g.~in Sec.~\ref{s:criticalexample}.
In contrast, the internal power expressions \eqref{e:Pint}, \eqref{e:workpairKappa} and \eqref{e:workpair3} presented above overcome this limitation.
\end{remark}

\subsection{Extension to multiple fiber families}
In case of $n_\mrf$ fiber families $\sC_i$, $i=1,...,~n_{\mrf}$, we define the tangent $\bell_i$ and  director $\bc_i$ for each $\sC_i$. The in-plane curvatures \eqref{e:Ktenb} and the moment  $\bar{m}$ in \eqref{e:bmdef1} are then defined for each fiber family $\sC_i$. Then $ P_{\mathrm{int}} $ in Eq.~\eqref{e:workpair3} simply becomes
\eqb{lll}
 P_{\mathrm{int}} \is \ds\frac{1}{2}\int_{\sR_0}\, \tau^{\alpha\beta} \,\dot{a}_{\alpha\beta}\,\dif A + \int_{\sR_0}\Mab_0\,\dot{b}_{\alpha\beta}\,\dif A + \sum_{i=1}^{n_\mrf} \int_{\sR_0}\Mbarab_{0i}\,\dot{\bar{b}}^i_{\alpha\beta}\,\dif A ~.
 \label{e:workpair5}
\eqe
 In accordance with Eq.~\eqref{e:workpair5}, the form of the stored energy function is extended from \eqref{e:WEKoo} to
\eqb{l}
W = {\hat{W}}\big(a _{\alpha\beta}, b_{\alpha\beta},  \bar{b}^i _{\alpha\beta}; ~h_i^{\alpha\beta}\big)~.
\label{e:WEK}
\eqe 
{This function} can equivalently be expressed in terms of the invariants, e.g.~as 
\eqb{l}
W = {\breve{W}}\big(I_1,~J,~{\Lambda}_i,~{\gamma}_{ij},~H,~\kappa,~\kappa_\mrn^i,~\tau_\mrg^i,~\kappa_\mrg^i\big)~,
\label{e:WEKI}
\eqe 
{since all these invariants are functions of $a _{\alpha\beta}$, $b_{\alpha\beta}$, and  $ \bar{b}^i _{\alpha\beta}$.}
The constitutive equations thus read
\eqb{llrlrlr}
\tau^{\alpha\beta} = \ds2\,\pa{{\hat{W}}}{a _{\alpha\beta}}~,\quad\quad
M_0^{\alpha\beta}  =  \ds\pa{{\hat{W}}}{b _{\alpha\beta}}~,\quad\quad
\bar{M}_{0i}^{\alpha\beta} = \ds\pa{{\hat{W}}}{\bar{b}^i _{\alpha\beta}}~,
\label{e:consticom}
\eqe
where $\bar{M}_{0i}^{\alpha\beta}$ are the components of the {nominal  stress couple tensor associated with in-plane bending of fiber $i$. They correspond} to the change in the in-plane curvature $\bar{b}^i _{\alpha\beta}$ of fiber $i$.

\section{Constitutive examples} {\label{s:examples}}
This section presents constitutive examples for the presented theory considering unconstrained and constrained fibers. We restrict ourselves here to two families of fibers. Note however that our approach allows for any number of fiber families.


\subsection{A simple generalized fabric model} 

A simple generalized shell model for two-fiber-family fabrics that are initially curved and bonded to a matrix is given by  
\eqb{llllll}
W =     W_{\mathrm{matrix}} + W_\mathrm{fib\mbox{-}stretch} + W_\mathrm{fib\mbox{-}bending} +  W_\mathrm{fib\mbox{-}torsion} + W_\mathrm{fib\mbox{-}angle} ~,
   \label{e:eg_Wsimple}
\eqe
where
\eqb{llllll}
W_{\mathrm{matrix}} \is \ds U(J) +  \frac{1}{2}\mu\,(I_1 - 2 - 2\,\ln J) ~,\\[5mm]
W_{\mathrm{fib\mbox{-}stretch}} \is \ds  {  \frac{1}{8}\, \epsilon_{\mrL}}\, \sum\limits_{i=1}^2  \big( {\Lambda}_i - 1 \big)^{2} ~,\\[5mm]
%
 %
 W_{\mathrm{fib\mbox{-}bending}} \is \ds  \frac{1}{2} \,  \sum\limits_{i=1}^2 \Big[ \beta_\mrn\, (K^i_\mrn)^2  +   \beta_\mrg\, (K^i_\mrg)^2  \Big]~,\\[5mm]
 W_{\mathrm{fib\mbox{-}torsion}} \is \ds  \frac{1}{2}\, \beta_{\tau}\,   \sum\limits_{i=1}^2 {(T^i_\mrg)^2}, \\[5mm]
   W_{\mathrm{fib\mbox{-}angle}} \is \ds   { \frac{1}{4}}\,  \epsilon_{\mra}\, \big( {\gamma}_{12} - \gamma^0_{12}\big)^2 
   \label{e:eg_W1}
\eqe
are  the strain energies for matrix deformation, fiber stretching,  out-of-plane and in-plane fiber bending, fiber torsion, and the linkage between the two fiber families, respectively. $U(J)$ is the surface dilatation energy. In the above expression,  ${\sqrt{\Lambda}}$,  $T_\mrg$, $K_\mrn$,   $K_\mrg$, and ${\gamma}_{12}$ denote the fiber stretch, {the norminal change in geodesic fiber torsion, normal fiber curvature and geodesic fiber curvature},  and the relative angle between fiber families, respectively (see  Tabs.~\ref{t:invariantKbar} and \ref{t:invariantC}).  Symbols $\mu$, $\epsilon_{\bullet}$, and $\beta_{\bullet}$ are material parameters. $\epsilon_{\mrL}$ can be taken as zero during fiber compression ($\Lambda_i < 1$) to mimic buckling phenomenologically.

From Eq.~\eqref{e:consticomoo} and \eqref{e:eg_W1}, we then find the effective stress and moment components for the director gradient-based formulation (Sec.~\ref{s:directorbased}), (see Appendix~\ref{s:variation} and \cite{shelltheo} for the required derivatives of kinematical quantities)  as
\eqb{lll}
\tau^{\alpha\beta} \is \ds J \,{\pa{U}{J}}\,a^{\alpha\beta} +  \mu\,(A^{\alpha\beta} - a^{\alpha\beta}) + { \frac{1}{2}\,\epsilon_{\mrL}}\, \sum\limits_{i=1}^2(\Lambda_i-1)\,L_i^{\alpha\beta}  + \epsilon_{\mra}\, \big( {\gamma}_{12} - \gamma^0_{12}\big)\, (L_1^\alpha\,L_2^\beta)^{\mathrm{sym}}~,\\[5mm]
M_0^{\alpha\beta} \is  \ds \beta_\mrn\sum\limits_{i=1}^2\,K^i_\mrn\, L_i^{\alpha\beta} +   \beta_\tau \sum\limits_{i=1}^2\,\, T^i_\mrg\, (c_{0i}^\alpha\,L_i^\beta)^{\mathrm{sym}}~,\\[5mm]
\bar{M}_0^{\alpha\beta} \is  \ds \beta_\mrg\sum\limits_{i=1}^2\,K^i_\mrg\, L_i^{\alpha\beta}~,
\label{e:eg_tauab_W1}
\eqe
where $(\bullet^{\alpha\beta})^{\mathrm{sym}} = \frac{1}{2}(\bullet^{\alpha\beta} + \bullet^{\beta\alpha})$ denotes symmetrization.

\subsection{Fiber inextensibility constraints}

For most textile materials, the deformation is usually characterized by very high tensile stiffness in fiber direction and low in-plane shear and bending stiffness. In this case, one may model the very high tensile stiffness along the fiber direction  $i$ by the inextensibility constraint
\eqb{l}
g_i:= {\Lambda}_{i} - 1 = 0~, \quad\quad \Lambda_i>1~, 
\eqe
where ${\Lambda}_i$ is defined in Tab.~\ref{t:invariantC}. This constraint then replaces the $W_\mathrm{fib\mbox{-}stretch}$ term in \eqref{e:eg_Wsimple}.  To enforce this constraint, we can employ the Lagrange multiplier method in the strain energy function
\eqb{l}
\tilde{W} := W + \ds\sum_{i=1}^{n_\mrf}  q_i\,g_i~,
\label{e:constrainedW}
\eqe
where $q_i$ ($i=1,~...,~n_\mrf$) denote the corresponding Lagrange multipliers. The stress components in this case become 
\eqb{llrlrlr}
\tau^{\alpha\beta} \is \ds2\,\pa{\tilde W}{a _{\alpha\beta}} = \ds2\,\pa{W}{a _{\alpha\beta}} +  \ds\sum_{i=1}^{n_\mrf} 2\, q_i\,L_i^{\alpha\beta}~. 
\label{e:constrainedW_stress}
\eqe
This leads to the same stress and moment components as in (\ref{e:eg_tauab_W1}.1) with the exception that $\epsilon_{\mrL}\,(\Lambda_i-1)/2$ is now replaced by $2\,q_i$.

\section{Weak form}{\label{s:weakform}}
This section presents the weak form for the generalized Kirchhoff-Love shell. 
The weak form is obtained by the same steps as the mechanical power balance in Sec.~\ref{s:balancepower}, simply by replacing velocity $\bv$ by variation $\delta\bx$. This gives
\eqb{l}
G_\mathrm{in} + G_\mathrm{int} - G_\mathrm{ext} = 0 \quad\forall\,\delta\bx\in\sV~,
\label{e:wfu}\eqe
where according to Eq.~\eqref{e:Pbalance2}, \eqref{e:Kdot}, \eqref{e:Pextok}, and \eqref{e:workpair5}
\eqb{lll}
G_\mathrm{in} 
\is \ds\int_{\sS_0} \delta\bx\cdot\rho_0\,\dot\bv\,\dif A~, \\[4mm]
G_{\mathrm{int}} \is \ds\frac{1}{2}\int_{\sS_0}\, \tau^{\alpha\beta} \,\delta{a}_{\alpha\beta}\,\dif A + \int_{\sS_0}\Mab_0\,\delta{b}_{\alpha\beta}\,\dif A + \sum_{i=1}^{n_\mrf} \int_{\sS_0}\Mbarab_{0i}\,\delta{\bar{b}}^i_{\alpha\beta}\,\dif A~,\\[4mm]
G_\mathrm{ext} \is \ds\int_{\sS}\delta\bx\cdot\bff\,\dif a 
+ \ds\int_{{\partial\sS}} \delta\bx\cdot\bT\,\dif s + \ds \int_{\partial\sS} \delta\bn\cdot\bM\,\dif s + \ds \sum_{i=1}^{n_\mrf} \int_{\partial\sS} \delta\bc_i\cdot\bMbar_{\!i}\,\dif s~.
\label{e:Giie}
\eqe
 Using \eqref{e:consticom}, $G_{\mathrm{int}}$ can also be expressed as the variation of potential \eqref{e:WEK} w.r.t. its arguments,
\eqb{lll}
G_{\mathrm{int}} \is \ds\int_{\sS_0} \delta W\, \dif A =   \ds\int_{\sS_0}\, \pa{W}{a_{\alpha\beta}} \,\delta{a}_{\alpha\beta}\,\dif A + \int_{\sS_0} \pa{W}{b_{\alpha\beta}} \,\delta{b}_{\alpha\beta}\,\dif A +  \sum_{i=1}^{n_\mrf}\int_{\sS_0}\pa{W}{\bar{b}^i_{\alpha\beta}}\,\delta{\bar{b}}^i_{\alpha\beta}\,\dif A~.
\label{e:Giie2}
\eqe
For constrained materials, e.g.~\eqref{e:constrainedW}, $G_{\mathrm{int}}$ becomes
\eqb{lll}
G_{\mathrm{int}} =\ds\int_{\sS_0} \delta \tilde{W}\, \dif A \!=\!   \ds\int_{\sS_0}\, \left( \frac{1}{2} \tau^{\alpha\beta} \,\delta{a}_{\alpha\beta}+ \Mab_0\,\delta{b}_{\alpha\beta} +  \sum_{i=1}^{n_\mrf} \Mbarab_{0i}\,\delta{\bar{b}}^i_{\alpha\beta}\right)\dif A + \sum_{i=1}^{n_\mrf}  \int_{\sS_0}\pa{\tilde{W}}{q_i}\,\delta q_i\,\dif A~.
\label{e:Giie3}
\eqe

For the constitutive example in Sec.~\ref{s:examples},  $\tauab$, $\Mab_0$ and $\Mbarab_0$  are given by  \eqref{e:constrainedW_stress}, (\ref{e:eg_tauab_W1}.2) and (\ref{e:eg_tauab_W1}.3), respectively, while ${\partial\tilde W}/{\partial q_i} = g_i$.

The linearization of weak form \eqref{e:Giie}  and its discretization can be found in \citet{shelltextileIGA}. 


\section {Analytical solutions}{\label{s:analyticalexamples}}
This section illustrates the preceding theory by several analytical examples considering simple homogeneous deformation states. They are useful elementary test cases for the verification of computational formulations.

\subsection{Geodesic curvature of a circle embedded in an expanding flat surface}{\label{s:criticalexample}}
The first example presents the computation of geodesic curvature $\kappa_\mrg$ from Eqs.~\eqref{e:kgok} and \eqref{e:kg2} and  confirms that only Eq.~\eqref{e:kgok} 
%
 gives the correct value for initially curved fibers. This illustrates the limitation of  existing second-gradient Kirchhoff-Love shell theory for cases where $\dot{\Gamma}_{\alpha\beta}^\gamma=0$ due to the choice of the surface parametrization.

\begin{figure}[ht]
\begin{center} \unitlength1cm
\begin{picture}(0,6)

\put(-6.6,-0.3){\includegraphics[width=0.85\textwidth]{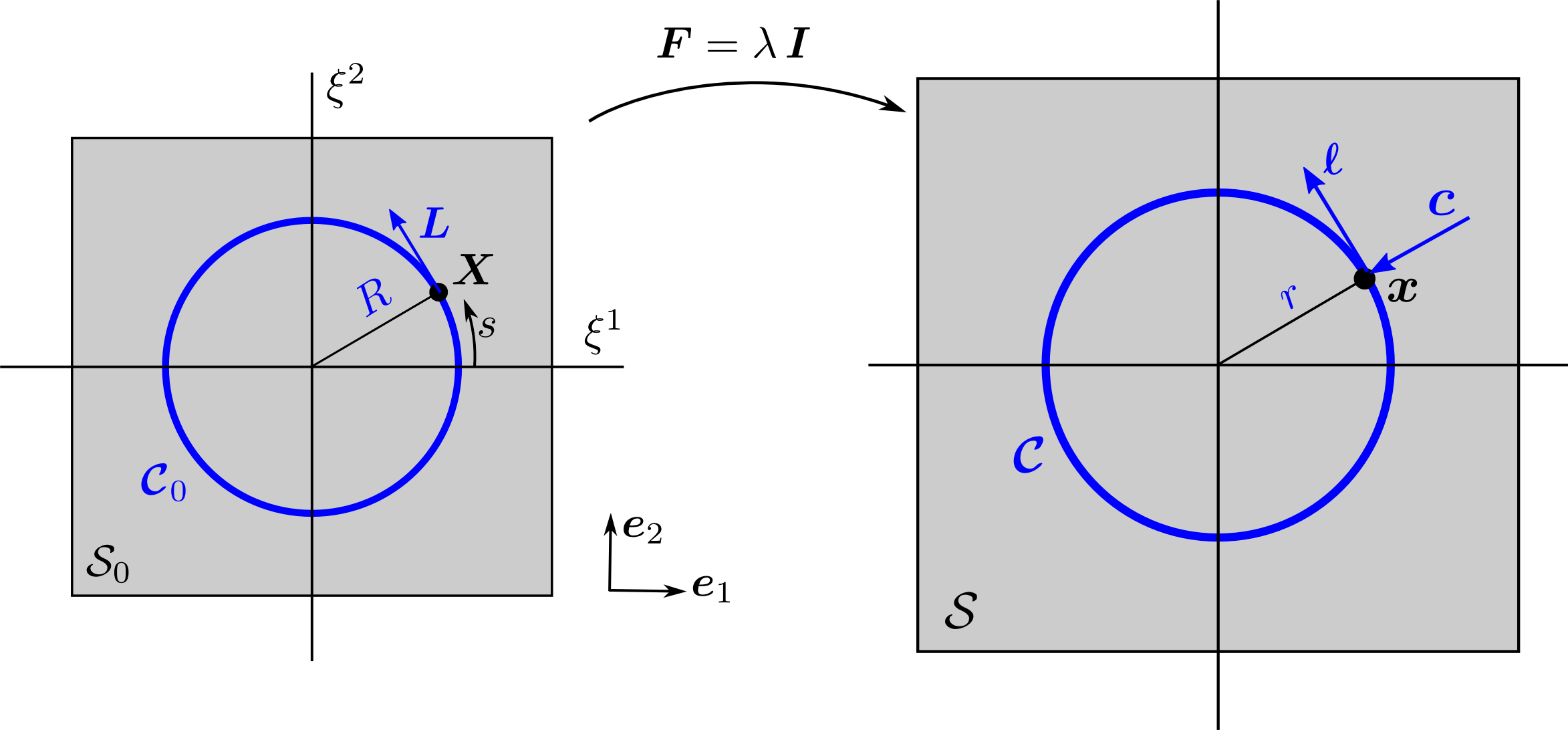}}

\end{picture}
\caption{Computation of geodesic curvature $\kappa_\mrg$: A circular fiber $\sC$ is embedded in the planar surface $\sS$ that is expanded by the homogeneous deformation $\bF=\lambda\,\bI$.}
\label{f:eg1}
\end{center}
\end{figure} 
To this end, we consider a circular fiber $\sC_0\in\sS_0$ with initial radius $R$ expanding according to the deformation gradient $\bF=\lambda\,\bI$ to $\sC\in\sS$ with the current radius $r$ as shown in Fig.~\ref{f:eg1}. Therefore, the geodesic curvatures of $\sC_0$ and $\sC$ are  expected to simply be $\kappa^0_\mrg=1/R$ and $\kappa_\mrg=1/r$, respectively. 

%


With respect to the convective coordinates $(\xi^1=X$, $\xi^2=Y)$  shown in Fig.~(\ref{f:eg1}a), the position vector on any point of $\sS_0$ and $\sS$ can be represented by 
\eqb{lll}
\bX \is \bX_{\!\mrs}(\xi^1,\xi^2) := \xi^1\,\be_1 + \xi^2\,\be_2~,\\[2mm]
{\bx} \is  {\bx_{\mrs}}(\xi^1,\xi^2) := \lambda\,\xi^1\,\be_1 + \lambda\,\xi^2\,\be_2~.
\label{e:eg1surfaceS}
\eqe
The surface tangent  and normal vectors follow from Eq.~\eqref{e:eg1surfaceS} as
\eqb{lll}
\bA_1 \is \be_1~, \quad \quad ~\,\bA_2 = \be_2~, \quad \quad~ $and$ \quad \bN = \be_3~, \\[2mm]
\ba_1 \is \lambda\,\be_1~, \quad \quad \ba_2 = \lambda\,\be_2~, \quad \quad $and$ \quad \bn = \be_3~,
\eqe
so that
\eqb{lll}
\bA^1 \is \be_1~, \quad \quad ~\,\bA^2 = \be_2~, \\[2mm]
\ba^1 \is \ds\frac{1}{\lambda}\,\be_1~, \quad  \quad \ba^2 =\ds \frac{1}{\lambda}\,\be_2~.
\eqe
The surface deformation gradient then reads
\eqb{lll}
\bF = \ba_\alpha\otimes\bA^\alpha = \lambda\,\big(\be_1\otimes\be_1 + \be_2\otimes\be_2\big)~,
\label{e:eg1F}
\eqe
and the surface Christoffel symbols take the form
\eqb{lll}
\bar{\Gamma}_{\alpha\beta}^\gamma := \bA_{\alpha,\beta}\cdot\bA^\gamma = 0~, \quad \quad\Gamma_{\alpha\beta}^\gamma:=\ba_{\alpha,\beta}\cdot\ba^\gamma = 0~.
\eqe

In the reference configuration, the fiber is parametrized by the arc-length coordinate $s$ as
 \eqb{lll}
{\bX}\is {\bX_{\!\mrc}}(s) := \ds R\,\cos {\frac{s}{R}}\,\be_1 + R\,\sin\frac{s}{R}\,\be_2~.
\label{e:eg1fiberC}
\eqe
From Eq.~\eqref{e:eg1fiberC} and \eqref{e:eg1F} follows 
\eqb{lll}
\bL \dis \bX_{\!{\mrc},s} = - \ds \sin\frac{s}{R}\,\be_1 + \cos\frac{s}{R}\,\be_2 ~ = - \frac{Y}{R}\,\bA_1 + \frac{X}{R}\,\bA_2~,\\[4mm]
\lambda\,{\bell} \dis \bF\,\bL = - \ds \frac{Y}{R}\,\ba_1 + \frac{X}{R}\,\ba_2~,\\[4mm]
\bell \dis \ds\frac{ \bF\,\bL }{\lambda} =   - \ds\frac{Y}{R}\,\be_1 + \frac{X}{R}\,\be_2~,\\[4mm]
\bc \dis \ds \bn\times\bell = -\frac{X}{R}\,\be_1 - \frac{Y}{R}\,\be_2~.

\eqe
In this equation $(\xi^1,\xi^2) = (X,Y) \in\sC_0$. Thus,
\eqb{lll}
[\hat{L}^\alpha_{,\beta}] := \ds \frac{1}{\lambda} \, [{L}^\alpha_{,\beta}] = \frac{1}{\lambda\,R}\,  \left[\begin{array}{cc}
0 & -1 \\
1  & 0
\end{array} \right]~, 
\eqe
and
\eqb{lll}
 \quad [\ell^{\alpha}] := [\bell\cdot\ba^\alpha] =  \ds \frac{1}{\lambda\,R}\, \left[\begin{array}{r}
 - Y  \\ 
X  
\end{array} \right]~, \quad  \quad [c_{\alpha}] := [\bc\cdot\ba_\alpha] =   \ds - \frac{\lambda}{R}\, \left[\begin{array}{r}
  X  \\ 
 Y
\end{array} \right]~.
\eqe
Inserting these expressions into Eq.~\eqref{e:kgok}, and using the identity $X^2 + Y^2 = R^2$ gives  the geodesic curvature
\eqb{lll}
\kappa_\mrg^{\Gamma} = 0~,\quad \quad
\kappa_\mrg^{\mathrm{L}} = \ds\frac{1}{\lambda\,R} = \frac{1}{r}~,\quad\quad
\kappa_\mrg = \kappa_\mrg^{\Gamma} + \kappa_\mrg^{\mathrm{L}}= \ds \frac{1}{r}~.
\eqe
For $\sC_0$, setting $\lambda=1$ directly yields $\kappa^0_\mrg = 1/R$.

In contrast, Eq.~\eqref{e:kg2} obviously fails to reproduce the correct geodesic curvatures since the Christoffel symbols are zero everywhere solely due to the choice of the  surface parametrization. Furthermore, since $\dot{\Gamma}_{\alpha\beta}^\gamma = 0$, the in-plane bending term of the internal power  is ill-defined in second-gradient theory \eqref{e:workpairGradientTheory}, whereas we get a well-defined power from \eqref{e:kgok} with \eqref{e:Pint}, \eqref{e:workpairKappa} and \eqref{e:workpair3}. It correctly captures the change in geodesic curvature.

\subsection{Biaxial stretching of a sheet containing diagonal fibers}
\begin{figure}[htp]
\begin{center} \unitlength1cm
\begin{picture}(0,4.0)
\put(-7.0,0){\includegraphics[width=0.3\textwidth]{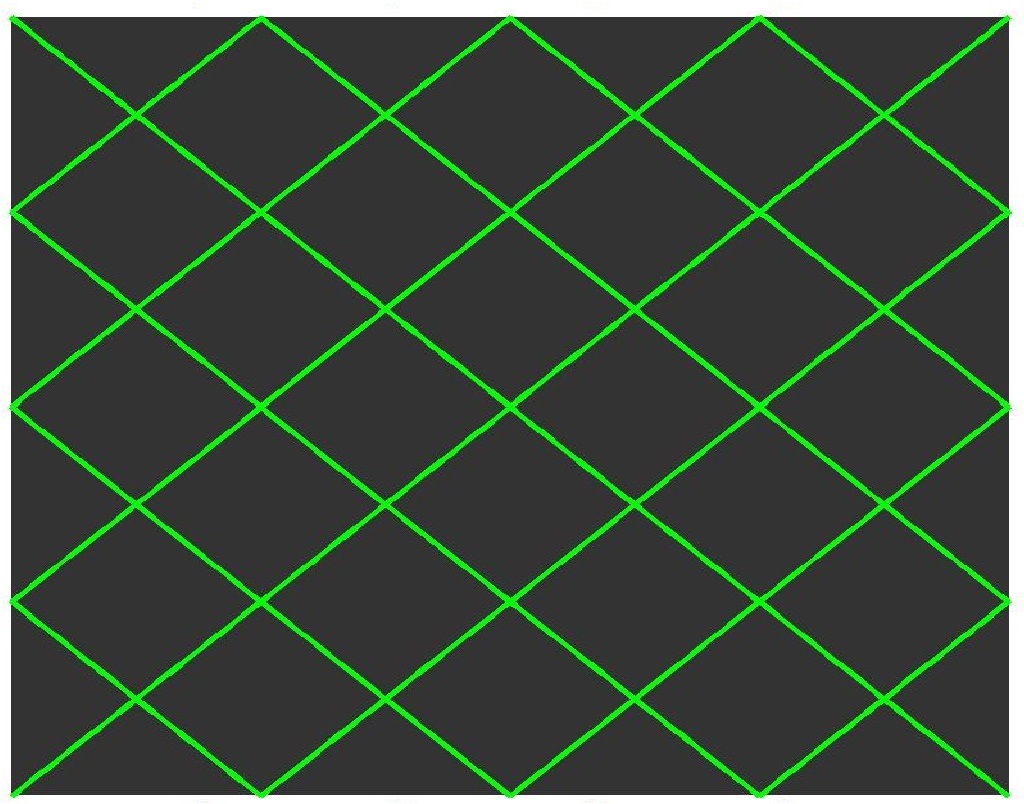}}
\put(-2.0,0){\includegraphics[width=0.08\textwidth]{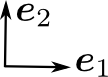}}
\put(1,0){\includegraphics[width=0.40\textwidth]{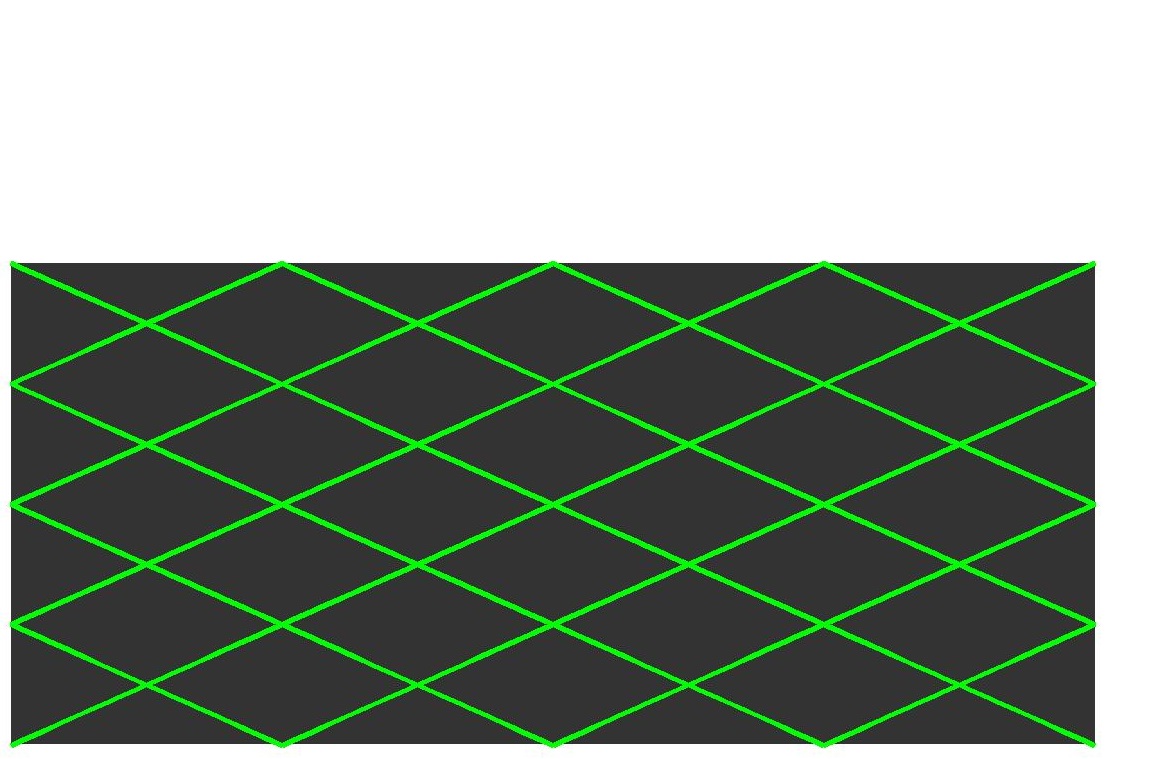}}
\end{picture}
\caption[caption]{Biaxial stretching of a rectangular sheet from dimension $L\times H$ (left) to $\ell\times h$ (right) .
}
\label{f:pureshear}
\end{center}
\end{figure}

The second example presents an analytical solution for homogeneous biaxial stretching of a rectangular sheet from dimension $L\times H$ to $\ell\times h$, such that $\ell = \lambda_{\ell}\,L$ and $h = \lambda_{h}\,H$. The sheet contains matrix material and two fiber families distributed diagonally as shown in Fig.~\ref{f:pureshear}.
 The strain energy function in this example is taken from \eqref{e:eg_W1} as
\eqb{lll}
W = \ds\frac{\mu}{2}\,(I_1 - 2 - \ln J) +  { \frac{1}{8}\, \epsilon_{\mrL}}\, \sum\limits_{i=1}^2  \big( {\Lambda}_i - 1 \big)^{2} ~ +   \frac{1}{4}\,  \epsilon_{\mra}\, \big( {\gamma}_{12} - \gamma^0_{12}\big)^2 ~.
\label{e:ex4_pbW}
\eqe
Therefore the stress components follow as
\eqb{lll}
\tau^{\alpha\beta} =   \ds \mu\,(A^{\alpha\beta} - a^{\alpha\beta}) +  \sum\limits_{i=1}^2 {\tau_i}\,L_i^{\alpha\beta}  + \epsilon_{\mra}\, \big( {\gamma}_{12} - \gamma^0_{12}\big)\, (L_1^\alpha\,L_2^\beta)^{\mathrm{sym}}~,
\label{e:ex4_stress}
\eqe
where $\tau_i:= {\frac{1}{2}\, \epsilon_{\mrL}}\,(\Lambda_i-1)$ denotes the nominal fiber tension.
The parameterization can be chosen such that the surface tangent vectors are $\bA_1=L\,\be_1$, $\bA_2 = H\,\be_2$, $\ba_1=\ell\,\be_1$, and $\ba_2=h\,\be_2$, where $\be_\alpha$ are the basis vectors shown in Fig.~\ref{f:pureshear}. 
{The fiber directions} are $\bL_1 = (L\,\be_1 + H\,\be_2)/D$, $\bL_2 = (L\,\be_1 - H\, \be_2)/D$, 
 where $D^2:= L^2 + H^2$.
We thus  find
\eqb{lll}
[A^{\alpha\beta}] \is \ds  \left[\begin{array}{cc}
1/L^2 &0 \\
0 & 1/H^2
\end{array} \right]~, \quad \quad
[a^{\alpha\beta}] = \ds  \left[\begin{array}{cc}
1/\ell^2 &0 \\
0 & 1/h^2
\end{array} \right]~,\quad\quad
[L_1^{\alpha\beta}] = \ds \frac{1}{D^2} \,\left[\begin{array}{cc}
1 & 1 \\
1 & 1
\end{array} \right]~, \\[6mm]
[L_2^{\alpha\beta}] \is \ds  \frac{1}{D^2} \left[\begin{array}{rr}
1 & -1 \\
-1 & 1
\end{array} \right]~, \quad $and$ \quad 
[L_1^{\alpha}\,L_2^\beta]^\mathrm{sym} = \ds  \frac{1}{D^2} \left[\begin{array}{rr}
1 & 0 \\
0 & -1
\end{array} \right]~.
\label{e:eg4_metric}
\eqe
 Further, we find the fiber stretch and angles
 \eqb{lll}
 \Lambda_\mrf \dis \Lambda_1=\Lambda_2 =\ds (\ell^2 + h^2)/D^2~, \\[3mm]
 \gamma^0_{12} \is \ds A_{\alpha\beta}\, L_1^{\alpha}\,L_2^\beta= (L^2-H^2)/D^2~,\\[3mm] 
\gamma_{12} \is \ds a_{\alpha\beta}\, L_1^{\alpha}\,L_2^\beta = (\ell^2-h^2)/D^2~. 
\label{e:eg4_LamGamm}
 \eqe
Inserting Eq.~\eqref{e:eg4_LamGamm} into Eq.~\eqref{e:ex4_stress} gives the stress tensor $\bsig= J^{-1} \tau^{\alpha\beta}\,\ba_\alpha\otimes\ba_\beta$,  where $J:=\lambda_{\ell}\,\lambda_{h}$~. The resultant reaction forces at the boundaries then follow as
 \eqb{lll}
F_1 =  h\,\be_1\,\bsig\,\be_1 \is \ds \frac{h}{J}\,\left[ \mu\,(\lambda_{\ell}^2-1) +  \frac{2\,\ell^2}{D^2}\, \tau +  \epsilon_{\mra}\,\frac{\ell^2}{D^2}\,(\gamma_{12} -\gamma^0_{12})\right]~,\\[5mm]
F_2 =  \ell\,\be_2\,\bsig\,\be_2 \is \ds \frac{\ell}{J}\,\left[ \mu\,(\lambda_{h}^2-1) +  \frac{2\,h^2}{D^2}\, \tau -  \epsilon_{\mra}\,\frac{h^2}{D^2}\,(\gamma_{12} -\gamma^0_{12})\right]~,
\label{e:eg4_sol}
\eqe
where  $\tau:=\tau_1 = \tau_2= {\frac{1}{2}\,\epsilon_{\mrL}}\,(\Lambda_\mrf-1)$.

 \begin{remark}
 The presented solution \eqref{e:eg4_sol}  includes pure shear by simply setting $\lambda_{h}= 1/\lambda_{\ell}$~. 
\end{remark}

 \begin{remark}
For uniaxial tension e.g.~in the $\be_1$ direction and with free  horizontal boundaries,  condition $F_2=0$ in Eq.~(\ref{e:eg4_sol}.2) gives the solution of $\lambda_h$, which in turn can be inserted to Eq.~(\ref{e:eg4_sol}.1) for the resultant reaction force $F_1 (\lambda_{\ell})$ as
 \eqb{lll}
F_1 = \ds \frac{H}{\lambda_{\ell}\,D^4}\,\big[ D^4\, \mu\,(\lambda_{\ell}^2-1) + ( \epsilon_{\mrL} \!+\! \epsilon_{\mra} )\, \lambda^4_{\ell}\,{L^4} + ( \epsilon_{\mrL}\! -\! \epsilon_{\mra} )\, \lambda^2_{\ell}\,\lambda^2_{h}\,{H^2\,L^2}  +   ( \epsilon_{\mrL}\! -\! \gamma^0_{12}\, \epsilon_{\mra} )\, \lambda^2_{\ell}\, L^2\,D^2  \big]\,,
\eqe
where 
 \eqb{lll}
\lambda_h^2 =  \ds \frac{1}{2\,a}\,\left(-b + \sqrt{b^2 + 4\,a\,\mu}\right)~,
\eqe
while
$a:= \ds\frac{H^2}{D^4}\,\left(\epsilon_\mrL\!+\! \epsilon_\mra \right)$, and $b:= \ds \mu + (\epsilon_\mrL\! -\! \epsilon_\mra)\,\frac{\lambda^2_\ell\, L^2\,H^2}{D^4} - \epsilon_\mrL\,\frac{H^2}{D^2} + \gamma_{12}^0\,\epsilon_\mra\,\frac{H^2}{D^2} $.
\end{remark}

 \begin{remark}
Solution \eqref{e:eg4_sol}  also captures the inextensibility of fibers. In this case either the vertical, or the horizontal boundaries have to be stress free. In the latter case, $F_2=0$ and
 \eqb{lll}
 \lambda_{h} = \ds \sqrt{\frac{D^2}{H^2} - \lambda_{\ell}^2\,\frac{L^2}{H^2}},
 \eqe
 due to $D^2 = \ell^2 + h^2$. In this case the deformation is limited by $\lambda_{\ell}^{\mathrm{max}} = D/L$. From $F_2=0$,   the nominal fiber tension now follows as
 \eqb{lll}
  \tau = \ds \frac{1}{2}\,  \epsilon_{\mra}\,(\gamma_{12} -\gamma^0_{12}) - \frac{1}{2}\,\frac{D^2}{h^2}\, \mu\,(\lambda_{h}^2-1)~.
  \label{e:eg4_q}
\eqe
Inserting \eqref{e:eg4_q} into (\ref{e:eg4_sol}.1) then gives
 \eqb{lll}
F_1  =\ds \frac{h}{J}\,\left[ \mu\,(\lambda_{\ell}^2-1) - \frac{\ell^2}{h^2}\,\mu(\lambda_{h}^2-1) +  2\, \epsilon_{\mra}\,\frac{\ell^2}{D^2}\,(\gamma_{12} -\gamma^0_{12})\right]~.
\eqe
\end{remark}
\subsection{{Picture frame test}}
\begin{figure}[H]
\begin{center} \unitlength1cm
\begin{picture}(0,5.5)
\put(-9.5,0){\includegraphics[width=0.58\textwidth]{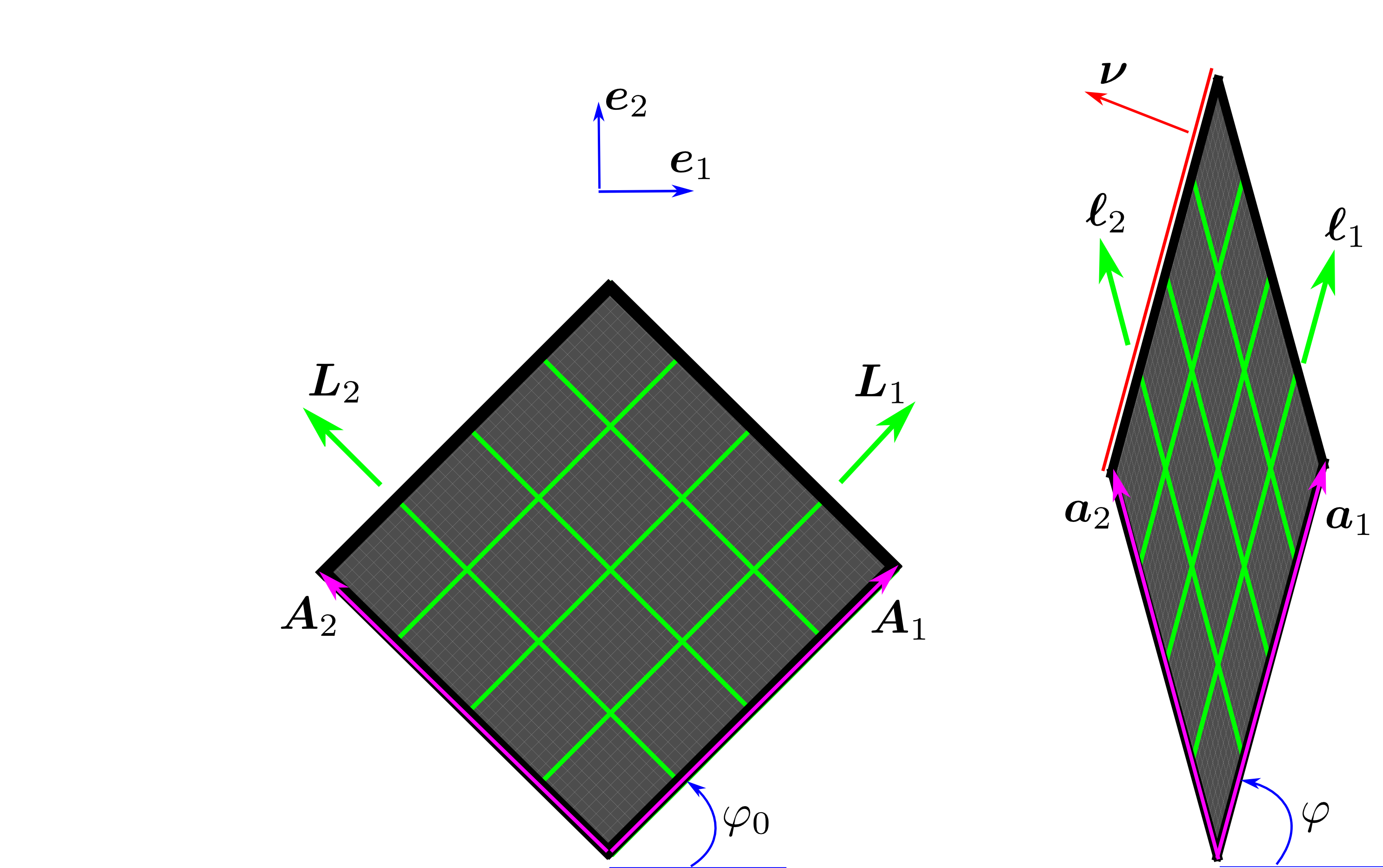}}
\put(0.5,-0.15){\includegraphics[width=0.50\textwidth]{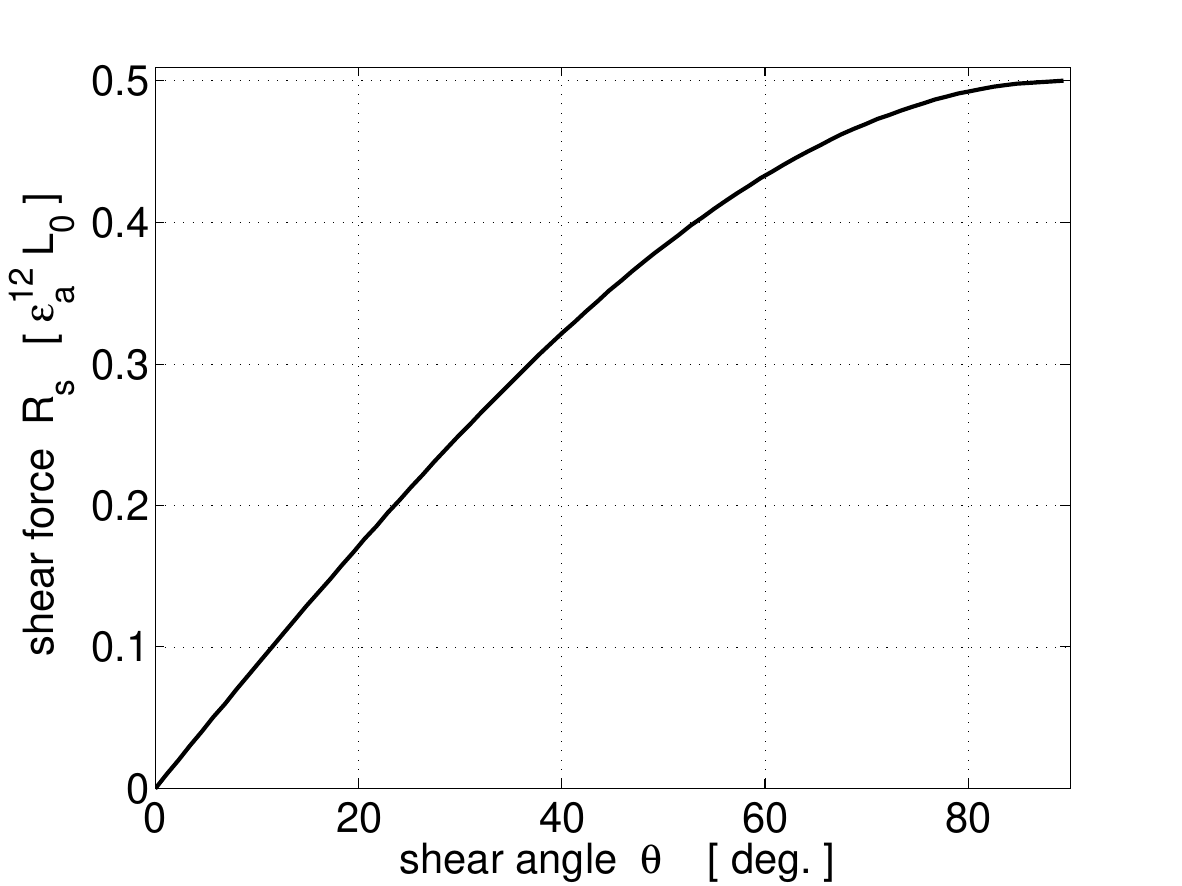}}
\put(-6.0,-0.1){a.}
\put(-1.8,-0.1){b.}
\put(0.65,-0.1){c.}
\end{picture}
\caption[caption]{Picture frame test: (a.) Initial and (b.) deformed configurations containing two fiber families. (c.)~Exact solution of the shear force vs.~shear angle $\theta:= 2\,\varphi- 90^\circ$.  
}
\label{f:pic1}
\end{center}
\end{figure}
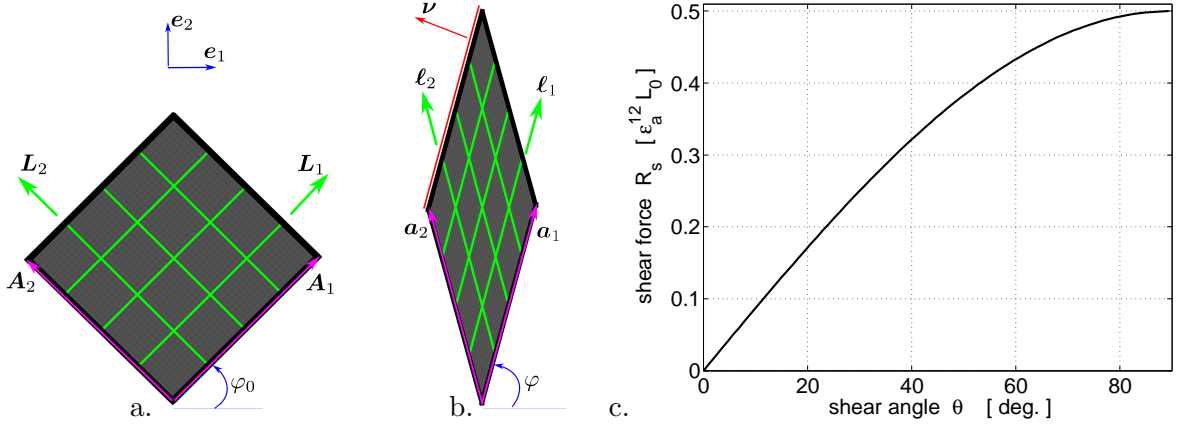

The third example derives an analytical solution for the shear force in the picture frame test of an $L_0\times L_0$ square sheet with  two fiber families as shown in Fig.~\ref{f:pic1}a-b. From the figure, we find the surface tangent vectors
\eqb{lll}
\begin{aligned}
\bA_1~~ \is L_0\,\cos \varphi_0\,\be_1 + L_0\,\sin \varphi_0\,\be_2~,\\[2mm]
\bA_2~~ \is -L_0\,\cos \varphi_0\,\be_1 + L_0\,\sin \varphi_0\,\be_2~,
\end{aligned}
\quad\quad
\begin{aligned}
\ba_1~~ \is L_0\,\cos \varphi\,\be_1 + L_0\,\sin \varphi\,\be_2~,\\[2mm]
\ba_2~~ \is -L_0\,\cos \varphi\,\be_1 + L_0\,\sin \varphi\,\be_2~,
\end{aligned}
\eqe
and the fiber directions
\eqb{lll}
\begin{aligned}
\bL_1~~ \is  \cos \varphi_0\,\be_1 + \sin \varphi_0\,\be_2~,\\[2mm]
\bL_2 ~~\is -\cos \varphi_0\,\be_1 + \sin \varphi_0\,\be_2~,
\end{aligned}
\quad\quad
\begin{aligned}
\bell_1~~ \is  \cos \varphi\,\be_1 + \sin \varphi\,\be_2~,\\[2mm]
\bell_2~~ \is -\cos \varphi\,\be_1 + \sin \varphi\,\be_2~,
\end{aligned}
\eqe
where $\varphi_0 = \pi/4$. Accordingly, the components of tensors $\bC$ and  $(\bL_1\otimes\bL_2)^\mathrm{sym}$ read
\eqb{lll}
[C_{\alpha}^{\beta}] \is  a_{\alpha\gamma}\,A^{\gamma\beta} = \ds  \left[\begin{array}{cc}
1 & -\cos(2\,\varphi) \\
-\cos(2\,\varphi) & 1
\end{array} \right]~,
\quad $and$ \quad
[L_1^{\alpha}\,L_2^\beta]^\mathrm{sym} = \ds \frac{1}{2\,L_0^2}\,  \left[\begin{array}{rr}
0 & 1 \\
1 & 0
\end{array} \right]~,
\label{e:Lab_eg4_solpic}
\eqe
respectively.
From Eq.~(\ref{e:Lab_eg4_solpic}.1),  the surface stretch is found as 
\eqb{ll}
J = \sqrt{\det[ C_{\alpha}^{\beta}]} = \sin(2\,\varphi)~.
\label{e:Lab_eg4_solpic_J}
\eqe
Further, the strain energy function in this example is taken from \eqref{e:eg_W1} as
$W =   \frac{1}{4}\,  \epsilon_{\mra}\, \big( {\gamma}_{12} - \gamma^0_{12}\big)^2$, so that the Cauchy stress components are
\eqb{lll}
[\sigma^{\alpha\beta}] =   \ds \frac{1}{J}\,\epsilon_{\mra}\, \big( {\gamma}_{12} - \gamma^0_{12}\big)\, [L_1^\alpha\,L_2^\beta]^{\mathrm{sym}} = -\ds \frac{1}{2\,L_0^2 }\,\epsilon_{\mra} \cot(2\,\varphi)\,  \left[\begin{array}{rr}
0 & 1 \\
1 & 0
\end{array} \right]~.
\label{e:ex4_stress_picf}
\eqe
Here, we have used Eq.~(\ref{e:Lab_eg4_solpic}.2), Eq.~\eqref{e:Lab_eg4_solpic_J},  $\gamma^0_{12} = \bL_1\cdot\bL_2 = 0$, and $\gamma_{12} = \bell_1\cdot\bell_2 = -\cos(2\,\varphi)$.

Consider the upper left edge with normal vector $\bnu =  -\sin\varphi \,\be_1 + \cos\varphi\,\be_2 = \nu_\alpha\,\ba^\alpha$, where $\nu_1 = \bnu\cdot\ba_1 = 0$, and $\nu_2 = \bnu\cdot\ba_2 = L_0\,\sin(2\,\varphi)$. The traction components on this edge can be computed from
\eqb{lll}
[t^\alpha] = [\sigma^{\alpha\beta}\,\nu_\beta] =   -\ds \frac{1}{2\,L_0}\,\epsilon^{12}_{\mra} \cos(2\,\varphi)\,   \left[\begin{array}{r}
1  \\
0 
\end{array} \right]~.
\eqe
Therefore, the traction vector solely contains the shear contribution 
\eqb{lll}
\bt = t^\alpha\,\ba_\alpha =  t^1\,\ba_1  =  -\ds \frac{1}{2}\,\epsilon^{12}_{\mra} \cos(2\,\varphi)\,  (\cos\varphi\,\be_1 + \sin\varphi\,\be_2)~,
\eqe
so that the shear force (i.e.~the tangent reaction) at the edge of the sheet is
\eqb{lll}
R_\mrs = \ds \bt\cdot \frac{\ba_1}{\norm{\ba_1}}\,L_0 =  -\ds \frac{1}{2}\,\epsilon^{12}_{\mra} \cos(2\,\varphi)\,L_0 = \ds \frac{1}{2}\,\epsilon^{12}_{\mra} \sin(\theta)\,L_0 ~,
\eqe
where {$\theta:= 2\,\varphi- 90^\circ$ denotes the shear angle. This solution is plotted in Fig.~\ref{f:pic1}c.}

\subsection{{Annulus expansion}} 
\begin{figure}[ht]
\begin{center} \unitlength1cm
\begin{picture}(0,5.5)

\put(-7.9,-0.3){\includegraphics[width=1\textwidth]{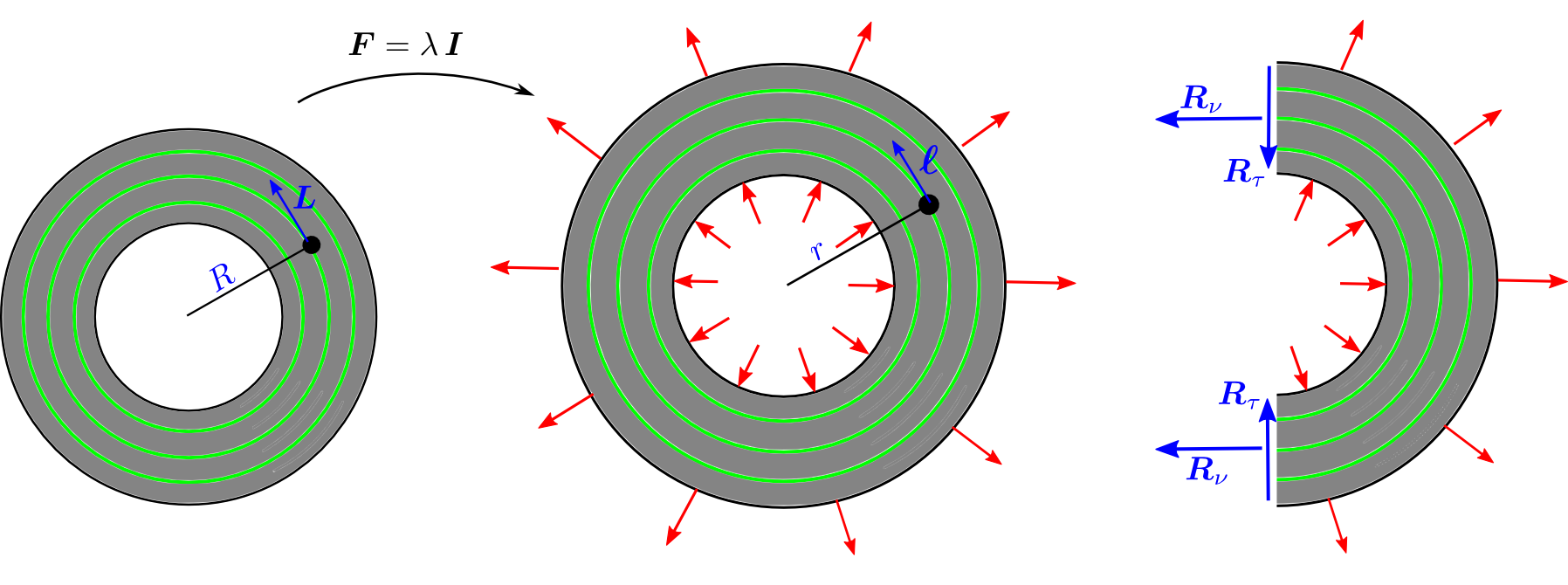}}

\end{picture}
\caption{Annulus expansion: An annulus containing matrix (grey color) and distributed circular fibers (green color) is expanded homogeneously from initial configuration (left) by applying Dirichlet boundary condition on both inner and outer surfaces (middle). The expansion causes the resultant interface forces $\bR_\nu$ and $\bR_\tau$ on the cut through a symmetry plane (right).  }
\label{f:eg2}
\end{center}
\end{figure}

{\label{s:example_diskexpan}}
The fourth example presents an analytical solution for the expansion of an annulus containing distributed circular fibers and matrix material as depicted in Fig.~\ref{f:eg2}.
 The inner and the outer rings with radius $R_\mri$ and $R_\mro$, respectively, are expanded to $r_\mri$ and $r_\mro$  by the constant stretch ${\bar\lambda} = {r_\mri}/{R_\mri} = {r_\mro}/{R_\mro}$.
The strain energy density (per reference area) is taken as
\eqb{llllll}
W =     W_{\mathrm{matrix}} + W_\mathrm{fib\mbox{-}bend} +  W_\mathrm{fib\mbox{-}stretch}~,
\label{e:Weg1}
\eqe
where
\eqb{lll}
W_{\mathrm{matrix}} \is  {\ds \frac{1}{2}\, K \,(J - 1)^2}~, \\[3mm]
W_{\mathrm{fib\mbox{-}bend}} \is \ds {\frac{1}{2}\,\beta\,k_\mrg^2}~.\\[3mm]
W_{\mathrm{fib\mbox{-}stretch}} \is \ds { \frac{1}{8}\,\epsilon_{\mrL}} \,({\Lambda - 1})^2~.
\label{e:eg1_Wmb}
\eqe
Here, { $K(R)$}, $\beta$, and $\epsilon_{\mrL}$ are material parameters for  matrix dilatation, fiber bending, and fiber stretching, respectively, and $J$, $\Lambda$, and $k_\mrg$ are invariants induced by tensors $\bC$ and $\bar\bK$ as listed in Tab.~\ref{t:invariantKbar} and \ref{t:invariantC}.

%
\subsubsection{Kinematical quantities}
According to  Fig.~\ref{f:eg2},  the initial and current configurations as well as the initial fiber direction 
can be described by
 \eqb{lll}
\bX \is \bX(R,\phi) := R\,\cos\phi\,\be_1 + R\,\sin\phi\,\be_2~,\\[2mm]
\bx\is  \bx(R,\phi) := r\,\cos\phi\,\be_1 + r\,\sin\phi\,\be_2~~,\\[2mm]
\bL \is \bL(\phi) := -\sin\phi\,\be_1 + \cos\phi\,\be_2~. 
\label{e:eg2surfaceS}
\eqe
{Here  $r=\lambda\,R$, due to the homogeneous deformation, with $\lambda$ being the fiber stretch}. From this, we find the covariant tangent vectors
 \eqb{lll}
\bA_1 = \ds\pa{\bX}{R} = \cos\phi\,\be_1 +\sin\phi\,\be_2~,\\[4mm]
\bA_2 = \ds\pa{\bX}{\phi} = -R\, \sin\phi\,\be_1 + R\,\cos\phi\,\be_2~,\\[4mm]
\ba_1 = \ds\pa{\bx}{R} = \lambda \cos\phi\,\be_1 + \lambda \sin\phi\,\be_2~,\\[4mm]
\ba_2 = \ds\pa{\bx}{\phi} = -r\, \sin\phi\,\be_1 + r\,\cos\phi\,\be_2~, 
\eqe
and the constant surface normal $\bn = \bN = \be_3$ during deformation.
From the tangent vectors, we get
\eqb{lll}
[A_{\alpha\beta}] \is \ds \left[\begin{array}{cc}
1 &0 \\
0 & R^2
\end{array} \right]~, \quad \quad
[A^{\alpha\beta}] = \ds \left[\begin{array}{cc}
1 &0 \\
0 & 1/R^2
\end{array} \right]~,\\[6mm]
[a_{\alpha\beta}] \is \ds \left[\begin{array}{cc}
\lambda^2 &0 \\
0 & r^2
\end{array} \right]~, \quad \quad
[a^{\alpha\beta}] = \ds \left[\begin{array}{cc}
1/\lambda^2 &0 \\
0 & 1/r^2
\end{array} \right]~,
\label{e:eg1_metric}
\eqe
and thus the contravariant tangent vectors take the form
 \eqb{lll}
\bA^1 = \cos\phi\,\be_1 +\sin\phi\,\be_2~,\\[4mm]
\bA^2 = \ds -\frac{1}{R}\, \sin\phi\,\be_1 + \frac{1}{R}\,\cos\phi\,\be_2~,\\[4mm]
\ba^1 = \ds \frac{1}{\lambda} \,\cos\phi\,\be_1 + \frac{1}{\lambda}\, \sin\phi\,\be_2~,\\[4mm]
\ba^2 = \ds -\frac{1}{r}\, \sin\phi\,\be_1 + \frac{1}{r}\,\cos\phi\,\be_2~.
\eqe

The initial and current fiber direction thus can be expressed as
 \eqb{lll}
 \bL \is L^\alpha\,\bA_\alpha~, \\[3mm] 
\bell \is \ds \frac{1}{\lambda}\,\bF\,\bL = \ell^\alpha\,\ba_\alpha~, 
%
\eqe
with 
\eqb{lll}
 [L^{\alpha}] := [\bL\cdot\bA^\alpha] =  \left[\begin{array}{c}
0  \\ 
1/R
\end{array} \right]~,
 \quad  
 $and$
 \quad [\ell^{\alpha}] := [L^\alpha/\lambda] =  \left[\begin{array}{c}
0 \\ 
1 /r
\end{array} \right]~.
\label{e:eg1_ella}
\eqe

The components of the structural tensors $\bL\otimes\bL$ and $\bell\otimes\bell$ thus read
\eqb{lll}
[L^{\alpha\beta}] = \ds \left[\begin{array}{cc}
0 &0 \\
0 & 1/R^2
\end{array} \right]~, \quad \quad
[\ell^{\alpha\beta}] = \ds \left[\begin{array}{cc}
0 &0 \\
0 & 1/r^2
\end{array} \right]~,
\quad\quad
[\ell^{\alpha}_{\beta}] = \ds \left[\begin{array}{cc}
0 &0 \\
0 & 1
\end{array} \right]~.

\label{e:eg1_ellotimesell}
\eqe

Further, the fiber director is obtained as
\eqb{lll}
\bc \is \bn\times\bell = - \cos\phi\,\be_1 - \sin\phi\,\be_2 = c_\alpha\,\ba^\alpha~,
\eqe
with 
\eqb{lll}
 [c_{\alpha}] := [\bc\cdot\ba_\alpha] =  \left[\begin{array}{c}
-\lambda \\ 
0
\end{array} \right]~.
\label{e:eg1_c_a}
\eqe
Therefore, its derivatives read
\eqb{lll}
\bc_{,1} \dis \ds \pa{\bc}{R} = \boldsymbol{0}~,\\[4mm]
\bc_{,2} \dis \ds\pa{\bc}{\phi} = \sin\phi\,\be_1 - \cos\phi\,\be_2~.
\eqe
From Eq.~\eqref{e:binplane2} then follows the components of the in-plane curvature tensor in the current configuration as
 \eqb{lll}
[\bar{b}_{\alpha\beta}] \is \ds -\frac{1}{2}\,[\bc_{,\alpha}\cdot\ba_\beta + \bc_{,\beta}\cdot\ba_\alpha] = \ds \left[\begin{array}{cc}
0 &0 \\
0 & r
\end{array} \right]~.
\eqe

Similarly, we find the in-plane curvature tensor in the initial configuration as
 \eqb{lll}
[\bar{B}_{\alpha\beta}] = \ds \left[\begin{array}{cc}
0 &0 \\
0 & R
\end{array} \right]~.
\eqe
Accordingly, the current geodesic curvature can be computed from Eq.~\eqref{e:kg1} as
\eqb{lll}
\kappa_\mrg = \bar{b}_{\alpha\beta}\,\ellab = 1/r~.
\label{e:eg1_kg}
\eqe
Similarly, we can verify the initial geodesic curvature $\kappa^0_\mrg = 1/R$~. These results lead to  $k_\mrg = \kappa_\mrg\,\lambda - \kappa_\mrg^0 = 0$, and thus
%
$W_{\mathrm{fib\mbox{-}bend}} = 0$ due to the particular choice of strain energy~(\ref{e:eg1_Wmb}.2). This  means that the change in the geodesic curvature $\kappa_\mrg$ in this example is purely due to fiber stretching and not due to fiber bending.
%
\subsubsection{Analytical expression for the reaction forces}
From Eq.~\eqref{e:Weg1}, we find the effective membrane stress
\eqb{rll}
J\,\sigma^{\alpha\beta} =\ds 2\,\pa{W}{a_{\alpha\beta}} = \ds  {K}\,(J - 1)\,J\,a^{\alpha\beta} + \frac{1}{2}\, \epsilon_{\mrL}\,(\lambda^2-1)\,L^{\alpha\beta}~.
\label{e:eg1_stressmethod1}
\eqe
Here, the component of the Cauchy stress tensor is $N^{\alpha\beta} = \sigab$ since $M^{\alpha\beta}=\bar{M}^{\alpha\beta}=0$  (see Eq.~\eqref{e:sigdeffinal} with \eqref{e:sigabdef}). Its mixed components thus read
\eqb{lll}
[N^\alpha_\beta] =  \ds { {K}\,(J - 1)\,  \left[\begin{array}{cc}
1 &0 \\
0 &  1
\end{array} \right] }+  \frac{\epsilon_{\mrL}}{2\,J} \,(\lambda^2-1)\, \left[\begin{array}{cc}
0 &0 \\
0 &  \lambda^2
\end{array} \right]~,
\eqe
where we have inserted $A^{\alpha\beta}$ and $L^{\alpha\beta}$ from \eqref{e:eg1_metric} and \eqref{e:eg1_ellotimesell}, respectively. It can be verified that with a homogeneous deformation $J=\lambda^2$, the local equilibrium equation $\mathrm{div}_\mrs\,\bsig = \mathrm{div}_\mrs\,(N^{\alpha\beta}\,\ba_\alpha\otimes\ba_\beta) = \boldsymbol{0}$ is satisfied for the functionally graded {surface bulk} modulus
\eqb{lll}
K(R) = \ds \frac{1}{2}\,\epsilon_{\mrL}\,\ln R~.
\label{e:eg2:mulog} 
\eqe
Further, the resultant reaction force at the cut in Fig.~\ref{f:eg2} (right) can be computed by
\eqb{lll}
\bR_\nu = \ds\int_{r_\mri}^{r_\mro}\, (\bnu\otimes\bnu)\,\bT\,\dif r = R_\nu\,\bell~, \\[5mm]
\bR_\tau =  \ds\int_{r_\mri}^{r_\mro}\, (\btau\otimes\btau)\,\bT\,\dif r = R_\tau\,\bc~,
\label{e:ex2Reac}
\eqe
since $\bnu=\bell$ and $\btau=\bc$. Here, $\bT= N^{\alpha\beta}\,\nu_\beta\,\ba_\alpha$ denotes the  traction acting on the cut. Therefore, we get
\eqb{lll}
\bT\cdot\bell = N^\alpha_\beta\,\ell_{\alpha}^{\beta} = \ds ({\frac{1}{2}\,\epsilon_{\mrL}} + K)\,(J-1)~,\\[3mm]
\bT\cdot\bc = \ds N^\alpha_\beta\,c_\alpha\,\ell^{\beta} = 0~.
\eqe
Inserting this into \eqref{e:ex2Reac} and taking \eqref{e:eg2:mulog} into account,  finally results in
\eqb{lll}
R_\nu = \ds {\frac{1}{2}\,\epsilon_{\mrL}}\,(J-1)\,{\bar\lambda}\,\big(R_\mro\,\ln R_\mro - R_\mri\,\ln R_\mri\big)~, \\[5mm]
R_\tau = 0~.
\eqe

\subsection{{Pure bending of a flat rectangular sheet}}
\begin{figure}[htp]
\begin{center} \unitlength1cm
\begin{picture}(0,5)

\put(-7.9,0){\includegraphics[width=1\textwidth]{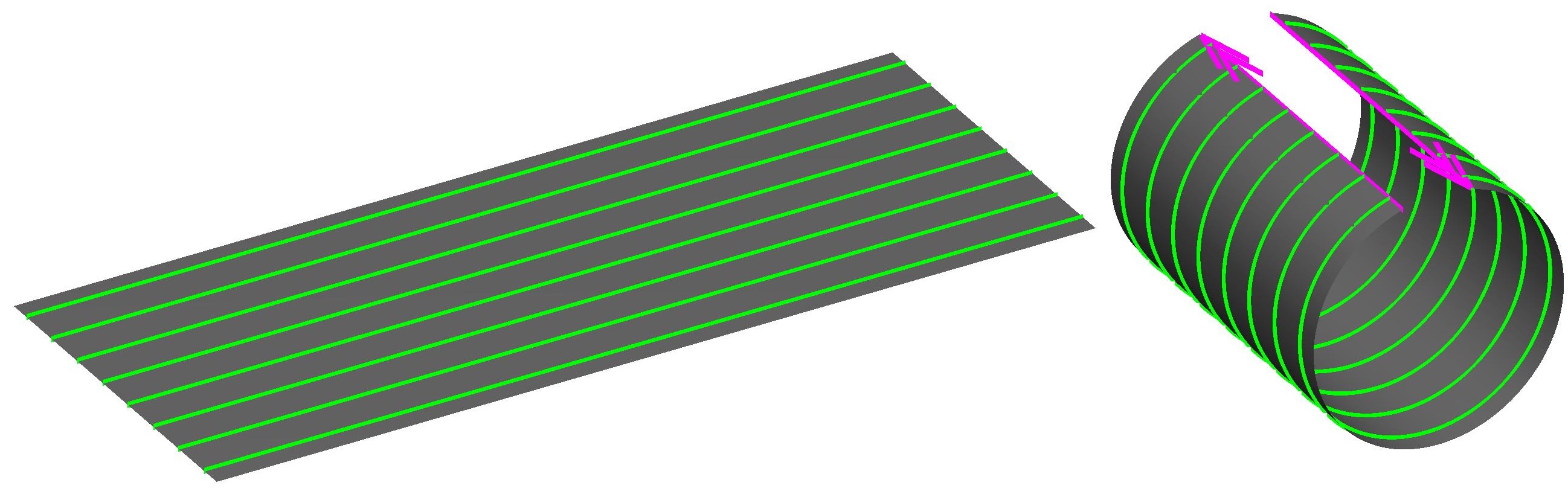}}
\end{picture}
\caption[caption]{Pure bending: Deformation of a rectangular sheet (left) into a circular arc (right).
}
\label{f:purebend1}
\end{center}

%
\end{figure}

The last example presents the analytical solution for the pure bending of a flat rectangular sheet $\sS_0$ with dimension $S\times L$,  subjected to an external bending moment $M_{\mathrm{ext}}$ along the two shorter edges,  as shown in Fig.~\ref{f:purebend1}. This problem was solved by \cite{shelltheo} for an isotropic material, and here we additionally consider the influence of fiber bending.  For simplicity the fibers are aligned along the bending direction. The strain energy function is taken from \eqref{e:eg_W1}, which reduces to
\eqb{lll}
W = \ds\frac{\mu}{2}\,(I_1 - 2 - \ln J) + \frac{\beta}{2}\,K_\mrn^2~%
\label{e:ex3_pbW}
\eqe
in this example. 


\subsubsection{Kinematical quantities}
We extend the kinematical quantities derived in \cite{shelltheo} to account  for embedded fibers. The sheet is parametrized by $\xi^1\in[0,~S]$ and $\xi^2\in[0,~L]$. 
Bending induces (high-order) in-plane deformation, such that the current configuration  $\sS$  has dimension $s\times \ell$. The stretches along the longer and shorter edges are $\lambda_{\ell}=\ell/L$ and $\lambda_{s}=s/S$, respectively, and we have relations $\theta:=\kappa_{\ell}\,\lambda_{\ell}\,\xi$ and $r:=1/\kappa_{\ell}$, where $\kappa_{\ell}$ is the homogenous curvature of the sheet.

With this, the sheet in the initial and current configurations, and the initial fiber direction can be described by 
\eqb{lll}
\bX(\xi,\eta) \is\xi\,\be_1 + \eta\,\be_2~,\\[3mm]
\bx(\xi,\eta) \is r\sin\theta\,\be_1 + \lambda_{s}\,\eta\,\be_2 + r\,(1-\cos\theta)\,\be_3~,\\[3mm]
\bL \is \be_1~.
\label{e:pureMA}
\eqe
Therefore,  the initial and current tangent vectors, and current normal vector are
\eqb{llll}
\bA_1 \is \ds\frac{\partial\bX}{\partial\xi} = \be_1~, \quad \quad   \quad   \quad   \quad  \quad  \quad  \quad  \quad  \quad    \bA_2 = \ds\frac{\partial\bX}{\partial\eta}= \be_2~, \\[4mm]
\ba_1 \is \ds\frac{\partial\bx}{\partial\xi} 
= \lambda_{\ell}\,\big(\cos\theta\,\be_1 + \sin\theta\,\be_3\big)~, \quad\quad
\ba_2 = \ds\frac{\partial\bx}{\partial\eta} = \lambda_{s}\,\be_2~,\\[5mm]
\bn \is - \sin\theta\,\be_1 + \cos\theta\,\be_3~.
\eqe 
From these, we find the components of the structural tensor 
\eqb{l}
[L^{\alpha\beta}] = [ \bA^\alpha\,(\bL\otimes\bL)\,\bA^\beta ] =  \left[\begin{array}{cc}
1 & 0 \\
0 & 0
\end{array} \right], 
\label{e:ex3_Lab}
\eqe
 the surface metrics 
\eqb{l}
[A_{\alpha\beta}] = \left[\begin{array}{cc}
1 & 0 \\
0 & 1
\end{array} \right], \quad [A^{\alpha\beta}] = \left[\begin{array}{cc}
1 & 0 \\ 
0 & 1
\end{array} \right],
\label{e:ex3_Aab}
\eqe
\eqb{l}
[\auab] = 
\left[\begin{array}{cc}
\lambda_{\ell}^2 & 0 \\ 
0 & \lambda^2_s
\end{array} \right], \quad 
[\aab] = \left[\begin{array}{cc}
\lambda_{\ell}^{-2} & 0 \\ 
0 & \lambda^{-2}_s
\end{array} \right],
\label{e:ex3_aab}
\eqe
the surface stretch~$J=\lambda_{\ell}\lambda_{s},$ and the components of the curvature tensor
\eqb{l}
[\buab] = \left[\begin{array}{cc}
\kappa_{\ell}\lambda_{\ell}^2 & 0 \\ 
0 & 0
\end{array} \right],\quad 
[b^{\alpha}_{\beta}] 
= \left[\begin{array}{cc}
\kappa_{\ell} & 0 \\ 
0 & 0
\end{array} \right], \quad [\bab] 
= \left[\begin{array}{cc}
\kappa_{\ell}\lambda_{\ell}^{-2} & 0 \\ 
0 & 0
\end{array} \right],\quad 
H=\ds\frac{\kappa_{\ell}}{2}~.
\label{e:babEx}
\eqe
Therefore, the nominal change in the normal curvature is 
\eqb{lll}
K_\mrn = b_{\alpha\beta}\,L^{\alpha\beta} = \ds\kappa_{\ell}\,\lambda_{\ell}^2
\eqe

\subsubsection{Analytical relation between  external moment and mean curvature}
From Eq.~\eqref{e:ex3_pbW}, we find
\eqb{lll}
J\,\sigab \is \ds 2\, \pa{W}{\auab} = \mu\,(A^{\alpha\beta} - a^{\alpha\beta})~,\\[4mm]
J\,M^{\alpha\beta} \is \ds  \pa{W}{\buab} = \beta\,K_\mrn\,L^{\alpha\beta}~.
\eqe
{According to Eqs.~\eqref{e:sigabdef} and \eqref{e:sigdeffinal} with $\bar{M}^{\alpha\beta} = \bar{m}^\alpha_{;\alpha} = 0$,} the components of the Cauchy stress tensors read
\eqb{lll}
N^\alpha_\beta = \sigma^{\alpha\gamma}\,a_{\gamma\beta} + M^{\alpha\gamma}\,b_{\gamma\beta}~.
\eqe
Inserting here $A^{\alpha\beta}$, $a^{\alpha\beta}$, and $L^{\alpha\beta}$ from Eqs.~\eqref{e:ex3_Aab}, \eqref{e:ex3_aab}, and \eqref{e:ex3_Lab} gives
\eqb{lll}
[N^\alpha_\beta] = \ds \frac{\mu}{J}\,\left[\begin{array}{cc}
\lambda_{\ell}^2-1 & 0 \\ 
0 & \lambda_{s}^2-1
\end{array} \right] +   \frac{\beta}{J}\,\kappa_{\ell}^2\,\lambda_{\ell}^4 \left[\begin{array}{cc}
1 & 0 \\ 
0 & 0
\end{array} \right]~.
\eqe
By assuming free edges, we have $N^1_1 = N^2_2 = 0$.  That is
\eqb{lll}
N_1^1 \is  \ds \frac{\mu}{J}\,(\lambda_{\ell}^2-1) +   \frac{\beta}{J}\,\kappa_{\ell}^2\,\lambda_{\ell}^4 = 0~,\\[4mm]
N_2^2 \is   \ds \frac{\mu}{J}\,(\lambda_{s}^2-1) = 0~.
\label{e:ex3_Nab}
\eqe
From the second equation follows $\lambda_{s} = 1$. The distributed external moment $M_\mathrm{ext}$ is in equilibrium to the  distributed internal moment $m_{\tau}$ -- see Eq.~(\ref{e:mcomps}.2) -- on any cut   parallel to the shorter edge.  We therefore have
\eqb{lll}
M_\mathrm{ext} = - m_{\tau} = M^{\alpha\beta}\,\nu_\alpha\,\nu_\beta = \ds\frac{\beta}{J}\,\kappa_{\ell}\,\lambda_{\ell}^4~,
\label{e:ex3_Mext}
\eqe
where $\bnu=\ba_1/\lambda_{\ell}$, $\nu_1=\lambda_{\ell}$, and $\nu_2=0$. By combining Eqs.~\eqref{e:ex3_Mext} and (\ref{e:ex3_Nab}.1), we get an expression for $\lambda_{\ell}$ in terms of $M_\mathrm{ext}$,
\eqb{lll}
 \lambda_{\ell} = \ds\sqrt{ \frac{1}{2} + \sqrt{ \frac{1}{4} - \frac{1}{\mu\,\beta} \,M^2_{\mathrm{ext}} }  } $~, with the condition $M^2_{\mathrm{ext}}\leq \ds\frac{1}{4}\,\mu\,\beta~.
\eqe
With this result and \eqref{e:ex3_Mext}, we obtain the relation between  external moment and the mean curvature,
\eqb{lll}
 H = \ds\frac{M_{\mathrm{ext}}}{2\,\beta\,\lambda_{\ell}^4}~.
\eqe

\section{Conclusion}{\label{s:conclusion}}

We have presented a generalized Kirchhoff-Love shell theory capable of capturing in-plane bending of fibers embedded in the surface. The formulation is an extension of classical Kirchhoff-Love shell theory that can handle multiple fiber families, possibly with  initial curvature. We use a direct approach to derive our theory assuming Kirchhoff-Love kinematics, plane-stress conditions, and lumped mass at the mid-surface. Like classical Kirchhoff-Love shell theory, no additional degrees-of-freedom are required such as the independent director fields of Cosserat shell theory, and the postulated set of balance laws for the shell  only include  linear and angular momentum balance.


Our thin shell kinematics is fully characterized by the change of three symmetric  tensors: $a_{\alpha\beta}$,  $b_{\alpha\beta}$, and the newly added $\bar{b}_{\alpha\beta}$ -- denoting  the in-plane curvature tensor. These tensors capture the change of the tangent vectors $\ba_1$, $\ba_2$, and the in-plane fiber director $\bc$. Tensor $\bar{b}_{\alpha\beta}$ and director vector $\bc$ are defined for each fiber family separately. The corresponding stress and moments work-conjugate to  $a_{\alpha\beta}$,  $b_{\alpha\beta}$, and $\bar{b}_{\alpha\beta}$,
are the effective membrane stress $\sigma^{\alpha\beta}$, the out-of-plane stress couple $M^{\alpha\beta}$, and the in-plane stress couple $\bar{M}^{\alpha\beta}$, respectively. The symmetry of $\sigma^{\alpha\beta}$ follows from angular momentum balance.  With these work-conjugated pairs, general constitutive equations for
hyperelastic materials are derived. 

The introduction of $\bar{b}_{\alpha\beta}$ for the in-plane curvature measure is an advantage over  existing second-gradient shell theory. Since it is a second order tensor instead of a third order tensor, its induced invariants and their physical meaning can be identified in a straightforward manner from the defined kinematics. Furthermore, it makes in-plane bending fully analogous to out-of-plane bending for both the internal and the external power. These features are advantageous in constructing material models and evaluating Neumann boundary conditions.

Finally  this work also provides the weak form, which is required for a finite element formulation based on $C^1$-continuous surface discretizations that are presented in future work \citep{shelltextileIGA}.
Several  analytical examples are presented to serve as useful elementary test cases for the verification of nonlinear computational formulations.  The presented analytical examples also confirm that the model correctly describes the mechanics of shells with initially curved fibers.

\appendix

\section{Variation of various kinematical quantities} {\label{s:variation}}
This section provides the variation of several kinematical quantities defined in Sec.~\ref{s:kinematics}. They are required for the derivation of the mechanical work balance (Sec.~\ref{s:workconjugation}), the stresses and moments from a stored energy function (Sec.~\ref{s:examples}),  the weak form (Sec.~\ref{s:weakform}), and the linearization of the weak form \citep{shelltextileIGA}.\footnote{
The material time derivative and the linearization of a given quantity $\bullet$ follow directly from its variation $\delta \bullet$, by replacing $\delta\bullet$ by either $\dot{\bullet}$ or $\Delta\bullet$, respectively.
%
} 
We focus here mostly on the new variables introduced above, while the expressions for existing ones can be found elsewhere, e.g.~in \cite{shelltheo}.  

Consider a variation of position $\bx$ by $\delta\bx$. Accordingly, the variation of the tangent vectors reads $\delta\ba_\alpha = \delta\bx_{,\alpha}$. 
From Eq.~\eqref{e:a_ab}, the variation of $\delta{a}_{\alpha\beta}$ is
\eqb{l}
\delta{a}_{\alpha\beta} =\delta\ba_\alpha\cdot\ba_\beta + \ba_\alpha\cdot\delta\ba_\beta~.
\label{e:deltaa_ab}
\eqe
Further, from Eq.~(\ref{e:curvector}.2), we find
\eqb{lll}
\delta\ba_{\beta;\alpha} = \bn\otimes\bn\,(\delta\ba_{\alpha,\beta} - \Gamma_{\alpha\beta}^\gamma \,\delta\ba_\gamma)~,
 \eqe
 where we have used the variation (see e.g.~\cite{wriggers-contact}), 
 \eqb{lll}
 \delta\bn = -\ba^\alpha\,(\bn\cdot\delta\ba_\alpha) ~.
 \label{e:var_bn}
 \eqe
From Eq.~\eqref{e:bellhatF} we have $\lambda^2 = (\bF\,\bL)\cdot(\bF\,\bL) = L^{\alpha\beta}\,a_{\alpha\beta}$. It then follows that,
\eqb{l}
\delta\lambda = \ds\frac{1}{2\, \lambda}\, L^{\alpha\beta}\,\delta \auab~.
\label{e:deltalambdaC}
\eqe
With $\delta\lambda$, the variation of $\bell$ can be derived from Eq.~\eqref{e:bellhatF}  as 
\eqb{l}
\delta\bell = (\bone - \bell\otimes\bell)\,\ella\,\delta\ba_\alpha~.
\label{e:deltaell}
\eqe
The variation of components $\ell_\alpha=\bell\cdot\ba_\alpha$ and $ \ell^\alpha=\bell\cdot\ba^\alpha$ read
\eqb{lll}
\delta\ell_\alpha =    \ell^\beta\,c_\alpha\,\bc\cdot\delta\ba_\beta + \bell\cdot\delta\ba_\alpha~,
\eqe  
and
\eqb{lll}
\delta\ell^\alpha =- \ellab\,\bell\cdot\delta\ba_\beta~,
\label{e:va_ellovalpha}
\eqe  
where we have used  \eqref{e:deltaell} and (see e.g.~\cite{shelltheo})
  \eqb{lll}
  \delta\ba^\alpha = (a^{\alpha\beta}\,\bn\otimes\bn - \ba^\beta\otimes\ba^\alpha)\,\delta\ba_\beta~.
  \label{e:var_baovalpha}
 \eqe
With $\ellab = \ell^\alpha\,\ell^\beta$ and Eq.~\eqref{e:va_ellovalpha}, we further find
 \eqb{lll}
\delta\ell^{\alpha\beta} = - \ellab\,\ell^{\gamma\delta}\,\delta a_{\gamma\delta}~.
\label{e:var_ellab}
\eqe  
%
%
The variation of the in-plane director $\bc$ is
\eqb{l}
\delta\bc = -(\bn\otimes\bc)\,\delta\bn - (\bell\otimes\bc)\,\delta\bell~,
\label{e:deltac1}
\eqe
which follows from identities $\bc\cdot\bn = 0$ and $\bell\cdot\bn=0$. Inserting Eq.~\eqref{e:var_bn} and \eqref{e:deltaell} into Eq.~\eqref{e:deltac1} gives
\eqb{l}
\delta\bc =  \big(c^\alpha\, \bn\otimes\bn  - \ell^\alpha\, \bell\otimes\bc \big)\,\delta\ba_\alpha~.
\label{e:deltac3}
\eqe
%
Using this result and Eq.~\eqref{e:var_baovalpha}, the variation of the components of $\bc$ read
\eqb{ll}
\delta c_\alpha = - \ell_\alpha^\beta\,\bc\cdot\delta\ba_\beta + \bc\cdot \delta\ba_\alpha~,
\label{e:var_c_alpha}
\eqe
and
\eqb{ll}
\delta c^\alpha = - \ellab \,\bc\cdot\delta\ba_\beta - c^\beta\,\ba^\alpha\cdot \delta\ba_\beta~.
\label{e:var_covalpha}
\eqe
From Eq.~\eqref{e:Gama_c_and_ell} follows
\eqb{ll}
\delta\Gamma_{\!\alpha\beta}^{\mrc} = \bc\cdot \delta\ba_{\alpha,\beta} + \ba_{\alpha,\beta}\cdot\delta\bc~.
\label{e:varGc_ab}
\eqe
%
Further, taking the variation of $\hat{L}^{\alpha}_{,\beta}$ (see Eq.~\eqref{e:Lhatdefine}) and using result \eqref{e:deltalambdaC} gives
\eqb{l}
\delta\hat{L}^\alpha_{,\beta} = - \hat{L}^\alpha_{,\beta}\,\ell^\gamma\,\bell\cdot\delta\ba_\gamma = \hat{L}^\alpha_{,\beta}\,\ell_{\gamma}\,\delta\ell^{\gamma}~.
\label{e:deltaLhatab}
\eqe
%
%
From Eq.~(\ref{e:coderivbcDef}.2), the variation of $c^\beta_{;\alpha}$ reads
\eqb{lll}
\delta c^\beta_{;\alpha} = -\ell^{\beta\gamma}\,\delta\Gamma^\mrc_{\alpha\gamma} - \ell^{\beta}\,(\Gamma_{\alpha\gamma}^\mrc + \ell_{\gamma}\,c_\delta\,\hat{L}^\delta_{,\alpha})\,\delta\ell^{\gamma} - (c_\gamma\, \hat{L}^\gamma_{,\alpha} + \ell^{\gamma}\,\Gamma^\mrc_{\gamma\alpha})\,\delta\ell^{\beta} - \ell^{\beta}\,\hat{L}^\gamma_{,\alpha}\,\delta c_{\gamma}~.
\label{e:delta_cbeta_semic_alp}
\eqe
%
%
From Eq.~(\ref{e:coderivbc}.1) follows 
\eqb{l}
\delta\bc_{,\alpha} = \ba_\beta\,\delta c^\beta_{;\alpha} + c^\beta_{;\alpha}\,\delta\ba_\beta + c^\beta\,\delta\ba_{\beta;\alpha} + \ba_{\beta;\alpha}\,\delta c^\beta~.
\label{e:delta_dc_alpha}
\eqe
The variation of the out-of-plane curvature is (see e.g.~\cite{shelltheo})
\eqb{l}
\delta{b}_{\alpha\beta} = \bn\cdot\delta\ba_{\alpha,\beta} - \Gamma_{\alpha\beta}^\gamma\,\bn\cdot\delta\ba_\gamma~.
\label{e:deltab_ab}
\eqe
The variation of the in-plane curvature follows from Eq.~\eqref{e:binplane2} as   
%
%
\eqb{lll}
\delta\bar{b}_{\alpha\beta} = -\frac{1}{2}\,(\delta\ba_\alpha \cdot \bc_{,\beta}  + \ba_\alpha\cdot\delta\bc_{,\beta}+ \delta\ba_\beta\cdot\bc_{,\alpha}  + \ba_\beta\cdot\delta\bc_{,\alpha})~.
\label{e:deltabbarab}
\eqe
%
%
The variation of the normal curvature follows from Eq.~\eqref{e:Kn_bab}  and \eqref{e:var_ellab} as 
\eqb{lll}
\delta\kappa_\mrn = \ds\pa{\kappa_\mrn}{\auab}\,\delta\auab + \pa{\kappa_\mrn}{{b}_{\alpha\beta}}\,\delta{b}_{\alpha\beta} + { \pa{\kappa_\mrn}{\bar{b}_{\alpha\beta}}\,\delta\bar{b}_{\alpha\beta}}  ~,
\label{e:var_kn}
\eqe
with
\eqb{lll}
 \ds\pa{\kappa_\mrn}{\auab} = -\kappa_\mrn\,\ellab~, \quad \quad 
 \ds \pa{\kappa_\mrn}{{b}_{\alpha\beta}} = \,\ellab~,\quad$and$ \quad 
 \ds \pa{\kappa_\mrn}{\bar{b}_{\alpha\beta}} = 0~.
\eqe
Similarly,  from Eqs.~(\ref{e:kgok0}.1) and \eqref{e:var_ellab},  the variation of the geodesic curvature gives  
\eqb{lll}
\delta\kappa_\mrg = \ds\pa{\kappa_\mrg}{\auab}\,\delta\auab + { \pa{\kappa_\mrg}{{b}_{\alpha\beta}}\,\delta{b}_{\alpha\beta} } +  \pa{\kappa_\mrg}{\bar{b}_{\alpha\beta}}\,\delta\bar{b}_{\alpha\beta}~,
\label{e:var_kg}
\eqe
with
\eqb{lll}
 \ds\pa{\kappa_\mrg}{\auab} = -\kappa_\mrg\,\ellab~,\quad \quad
 { \pa{\kappa_\mrg}{{b}_{\alpha\beta}} = 0}~,
 \quad
 $and$ \quad
 \ds \pa{\kappa_\mrg}{\bar{b}_{\alpha\beta}} = \ellab~.
 \label{e:variationk_g}
\eqe
From Eq.~\eqref{e:taug_bab}, \eqref{e:va_ellovalpha}, and \eqref{e:var_covalpha}, the variation of the geodesic torsion reads
\eqb{lll}
\delta\tau_\mrg = \ds\pa{\tau_\mrg}{\auab}\,\delta\auab + \pa{\tau_\mrg}{{b}_{\alpha\beta}}\,\delta{b}_{\alpha\beta} + \pa{\tau_\mrg}{\bar{b}_{\alpha\beta}}\,\delta\bar{b}_{\alpha\beta}~,
\eqe
with $\ds\pa{\tau_\mrg}{\bar{b}_{\alpha\beta}} = 0$, and
\eqb{lll}
 \ds\pa{\tau_\mrg}{\auab} \is \ds -\frac{1}{2}\,\kappa_\mrn\,(\ell^{\alpha} \,c^\beta + \ell^{\beta}\,c^\alpha) - \frac{1}{2}\,\tau_\mrg\,(c^{\alpha\beta} + \ellab)  ,\\[4mm]
 \ds \pa{\tau_\mrg}{{b}_{\alpha\beta}} \is  \ds \frac{1}{2} (\ell^{\alpha} \,c^\beta + \ell^{\beta}\,c^\alpha) ~.
 %
  %
\eqe
Therefore, from Eqs.~\eqref{e:kpvskg}, \eqref{e:var_kn} and \eqref{e:var_kg},  the variation of the principal curvature of the fiber is
\eqb{lll}
\delta\kappa_\mrp = \ds\pa{\kappa_\mrp}{\auab}\,\delta\auab +  \pa{\kappa_\mrp}{{b}_{\alpha\beta}}\,\delta{b}_{\alpha\beta} + \pa{\kappa_\mrp}{\bar{b}_{\alpha\beta}}\,\delta\bar{b}_{\alpha\beta}~,
\eqe
with
\eqb{rll}
  \ds\pa{\kappa_\mrp}{\auab} = -\kappa_\mrp\,\ellab~, \quad\quad
  \kappa_\mrp\, \ds \pa{\kappa_\mrp}{{b}_{\alpha\beta}} = \kappa_\mrn\,\ellab~,\quad\quad
  $and$\quad 
  \kappa_\mrp\, \ds \pa{\kappa_\mrp}{\bar{b}_{\alpha\beta}} =\kappa_\mrg\,\ellab~.
\eqe
%
%
%
%
\section{Frame invariance of various strain measures}{\label{s:frameinvariance}}
In this appendix, we show that the strain measures presented in our theory, such as $\auab$, $\buab$,  $c^\beta\,\ell_{\beta;\alpha}$, $\kappa_\mrg$, $\bar{b}_{\alpha\beta}$, together with the components of the structural tensors $c^{\alpha\beta}$, $\ellab$, and $c^\alpha\,\ell^\beta$, are invariant under superimposed rigid body motions of shell surface $\sS$.
To this end, let the shell surface \eqref{e:x_shell} be translated and rigidly rotated by
\eqb{ll}
\bx^+ = \bx_0 + \bQ\,\bx~, \quad $with$ \quad \bQ^\mrT\,\bQ = \bone~,
\label{e:bx_plus}
\eqe
where $\bx_0 \in\mathbb{E}^3$ and $\bQ\in SO(3)$ are a constant translation vector and a rotation tensor. From (\ref{e:bx_plus}.1), (\ref{e:curvector}.1), \eqref{e:tangentsl} and \eqref{e:belldef},  we have
\eqb{lll}
\ba_\alpha^+ = \bx^+_{,\alpha} = \bQ\,\ba_\alpha~,\\[3mm]
\ba_{\alpha,\beta}^+ = \bx^+_{,\alpha\beta} = \bQ\,\ba_{\alpha,\beta}~,\\[3mm]
\bell^+ = \bx^+_{,s} = \bQ\,\bell~,\\[3mm]
\bell^+_{,\alpha} = \bQ\,\bell_{,\alpha}~.
\label{e:rot_aell}
\eqe
Here, Eq.~(\ref{e:rot_aell}.4) follows from Eq.~(\ref{e:rot_aell}.3). Using the identity $\bQ\,\ba\times\bQ\,\bb = \bQ\,(\ba\times\bb)$, together with \eqref{e:normalmsl} and \eqref{e:vectorcdef}, we further obtain the following relations
\eqb{lll}
\bn^+ = \bQ\,\bn~,\\[3mm]
\bc^+ =  \bQ\,\bc~,\\[3mm]
\bc^+_{,\alpha} = \bQ\,\bc_{,\alpha}~,
\label{e:rotnc}
\eqe
where the last equation follows directly from the second one. The strain measure $a_{\alpha\beta}$ (see Eq.~(\ref{e:a_ab}.1) is frame invariant since  
 \eqb{lll}
  a^{+}_{\alpha\beta} = \ba^{+}_{\alpha}\cdot \ba^{+}_\beta = (\bQ\,\ba_\alpha)\cdot(\bQ\,\ba_\beta)  = \ba_\alpha\cdot(\bQ^\mrT\bQ\,\ba_\beta) = \ba_\alpha\cdot\ba_\beta = \auab~,
  \label{e:invariance_auab}
 \eqe   
 where we have employed Eq.~(\ref{e:rot_aell}.1) and (\ref{e:bx_plus}.2).
Similarly, the frame invariance of $\buab$, the strain measure for out-of-plane curvature  from Eq.~\eqref{e:bab2}, follows as
 \eqb{lll}
  b^{+}_{\alpha\beta} = \bn^{+}\cdot \ba^{+}_{\alpha,\beta} = (\bQ\,\bn)\cdot(\bQ\,\ba_{\alpha,\beta})  = \bn\cdot(\bQ^\mrT\bQ\,\ba_{\alpha,\beta}) = \bn\cdot\ba_{\alpha,\beta} = \buab~.
 \eqe   

The frame invariance is also true for the in-plane curvature measure $\bar{b}_{\alpha\beta}$ (see Eq.~\eqref{e:binplane2}), i.e.
 \eqb{lll}
  \bar{b}^{+}_{\alpha\beta} = -\frac{1}{2} (  c^+_{\beta;\alpha} +  c^+_{\alpha;\beta}) = \bar{b}_{\alpha\beta}~,
    \label{e:invariance_barbuab}
 \eqe   
since 
 \eqb{lll}
 c^+_{\beta;\alpha} = \ba_\beta^+\cdot\bc^+_{,\alpha} = (\bQ\,\ba_\beta)\cdot(\bQ\,\bc_{,\alpha}) = \ba_\beta\cdot(\bQ^\mrT\bQ\,\bc_{,\alpha}) =  c_{\beta;\alpha}~,
 \eqe
which follows from \eqref{e:c_a_comma_b}, (\ref{e:rot_aell}.1), and (\ref{e:rotnc}.3).

Similarly, the covariant derivative $\ell_{\beta;\alpha}$ from Eq.~\eqref{e:covardervEll} is frame invariant, since
\eqb{lll}
 \ell^+_{\beta;\alpha} = \ba_\beta^+\cdot\bell^+_{,\alpha} = (\bQ\,\ba_\beta)\cdot(\bQ\,\bell_{,\alpha}) = \ba_\beta\cdot(\bQ^\mrT\bQ\,\bell_{,\alpha}) =  \ell_{\beta;\alpha}~,
     \label{e:invariance_covarianell}
 \eqe
 due to (\ref{e:rot_aell}.1) and (\ref{e:rot_aell}.4).

Furthermore, following result \eqref{e:invariance_auab}, it can be shown that $\ba^\alpha_+ = \bQ\,\ba^\alpha$. With this, Eq.~(\ref{e:rot_aell}.3) and  (\ref{e:rotnc}.2), we can easily show the frame invariance is also true for components $\ell^\alpha = \bell\cdot\ba^\alpha = \ell^\alpha_+ $ and  $c^\alpha = \bc\cdot\ba^\alpha=c^\alpha_+$. Consequently, we have the frame invariance for all the components of the structural tensors as
\eqb{lll}
c_+^{\alpha\beta} \is c^\alpha_+\,c^\beta_+ = c^{\alpha\beta}~,\\[3mm]
\ell_+^{\alpha\beta} \is \ell^{\alpha}_+\,\ell^{\beta}_+ = \ell^{\alpha\beta}~,\\[3mm]
(c^\alpha\,\ell^{\beta})_+\is c^{\alpha}_+\,\ell^{\beta}_+ = c^\alpha\,\ell^{\beta}~.
  \label{e:invariance_structural}
\eqe
This implies that all the invariants, constructed from the strain measures $\auab$, $\buab$, and $\bar{b}_{\alpha\beta}$ by these structural tensors (see Tab.~\ref{t:invariantKbar}), are frame invariant. Take for example the geodesic curvature \eqref{e:kg1}.  We find
\eqb{lll}
\kappa_\mrg^+ = -\ellab_+\,\bar{b}_{\alpha\beta}^+ =   -\ellab\, \bar{b}_{\alpha\beta} = \kappa_\mrg~,
\eqe
due to Eq.~\eqref{e:invariance_barbuab} and (\ref{e:invariance_structural}.2). Also, the strain measure $c^\beta\,\ell_{\beta;\alpha}$ (used e.g.~in Eq.~\eqref{e:Wint0}) is frame invariant, i.e.~ 
\eqb{lll}
(c^\beta\,\ell_{\beta;\alpha})^+ = c_+^\beta\,\ell^+_{\beta;\alpha} = c^\beta\,\ell_{\beta;\alpha}~,
\eqe
following from $c^\beta_+ = c^\beta$ and Eq.~\eqref{e:invariance_covarianell}.

\section{On deriving Kirchhoff-Love theory from Cosserat theory} \label{s:app_KLequilibrium}
In this appendix, we show that the {balance} equations of our generalized  Kirchhoff-Love  shell theory with in-plane fiber bending are consistent with the more general {balance} equations of Cosserat shell theory  \citep{naghdi82}.

To this end, we consider a Cosserat shell $\sS$ characterized by material point $\bx\in\sS$ and a single director field $\bd$ attached to $\bx$. The  global balance equations of $\sS$ are postulated as (see e.g.~\cite{Green74,naghdi82,steigmann99})
\eqb{lll}
\ds \frac{D}{Dt} \int_{{\sR}}\rho\,\bv_\mre\, \dif a  \is \ds  \int_{{\sR}}\bff\, \dif a + \ds \int_{\partial{\sR}} \bT\, \dif s~, \\[7mm] 
\ds \frac{D}{Dt} \int_{{\sR}}\rho\,\bw_\mre\, \dif a  \is \ds  \int_{{\sR}} ({\ml + \mk})\, \dif a + \ds \int_{\partial{\sR}} \bMtilde\, \dif s~, \\[7mm]
\ds \frac{D}{Dt} \int_{{\sR}}\rho\,\big(\bx\!\times\!\bv_\mre \,+\, \bd\!\times\!\bw_\mre\big) \, \dif a  \is  \ds \int_{{\sR}}\,\big(\bx\!\times\!\bff + \bd\!\times\! {\ml}\big) \, \dif a + \ds \int_{\partial{\sR}} \big(\bx\!\times\!\bT +  \bd\!\times\!{\bMtilde}\big)\, \dif s~,
\label{e:coss_globalequi}
\eqe
where  $\bff$, $\bT$, $\bMtilde$, $\ml$, and $\mk$ denote the body force (per area), the traction vector, the (general) stress couple vector,  the assigned director couple (unit force per area), and the intrinsic director couple (unit force per area), respectively, and where
\eqb{lll}
\bv_\mre \dis \bv + \alpha\,\bw,\\[3mm]
\bw_\mre \dis \alpha\,\bv+\beta\,\bw
\eqe
are the effective velocities contributing to the material and director momentum, respectively. Here, $\alpha$ and $\beta$ denote 
{the inertial coefficient} associated with the director velocity $\bw:=\dot{\bd}$. The three equations in \eqref{e:coss_globalequi} correspond to the linear momentum balance, the director momentum balance, and the angular momentum balance of the Cosserat shell.

Now we are in a position to verify that the equilibrium equations in our presented theory are consistent with the equations  of Cosserat shell theory \eqref{e:coss_globalequi}.

First, we apply the kinematics  and moment definition of classical Kirchhoff-Love thin shells (without in-plane bending) to Eq.~\eqref{e:coss_globalequi}.
%
I.e.~we set $\bd=\bn$ and $\bMtilde=\bM$. 
As seen in Sec.~\ref{s:moment1}, Kirchhoff-Love assumptions together with plane-stress conditions allow us to define the out-of-plane bending vector $\bM$ as in Eq.~(\ref{e:defMbar}.1), which
satisfies the condition
\eqb{llll}
\bM\cdot\bn = 0~,
\label{e:bMcdotbn}
\eqe
due to Eq.~(\ref{e:defMbar}.1) and (\ref{e:defMbaralp}.1). 
Further, since we consider thin shells, the mass along the shell thickness can be lumped at the mid-surface. I.e.~the inertia associated with the director field is neglected, which implies $\alpha = 0$ and $\beta=0$. Furthermore, we assume that no external director loads are applied on the thin shell, i.e.~$\mk=\ml=\boldsymbol{0}$.
%
%
%
%
%
With these assumptions, Eq.~\eqref{e:coss_globalequi} becomes
\eqb{llll}
\ds \frac{D}{Dt} \int_{{\sR}}\rho\,\bv\, \dif a  \is \ds  \int_{{\sR}}\bff\, \dif a + \ds \int_{\partial{\sR}} \bT\, \dif s~, \\[7mm] 
 \boldsymbol{0}   \is \ds  \ds \int_{\partial{\sR}} \bM\, \dif s~, \\[7mm]
\ds \frac{D}{Dt} \int_{{\sR}}\rho\,\bx\!\times\!\bv \, \dif a  \is  \ds \int_{{\sR}}\,\bx\!\times\!\bff  \, \dif a + \ds \int_{\partial{\sR}} \big(\bx\!\times\!\bT +  \bn\!\times\!{\bM} \big)\, \dif s~.
\label{e:KL_globalequi0}
\eqe

Second, we extend the equilibrium equations \eqref{e:KL_globalequi0}  in order to describe in-plane bending of fibers embedded within the shell surface.  The kinematics and the definitions of stress and moment for in-plane bending of our presented theory in Secs~\ref{s:kinematics} \& \ref{s:balance} are directly applied to the extended equations.
%
%
%
%
%

To this end, we model a fiber as a beam, where all material points on its cross section follow Euler-Bernoulli kinematics and the  mass is lumped at the center line of the beam. This allows us to introduce an additional director field $\bc$ associated with in-plane bending, analogous to 
{$\bn=\bd$,} that satisfies $\bc\cdot\bn = 0$ and contributes no inertia.
 We assume that the momentum balance of director $\bc$ is independent (decoupled) from the surface director field $\bn$. As there is no stress on any cut  parallel to the beam center line, we also consider no external director loads -- similar to $\mk$ and $\ml$ --  acting on the fiber. Therefore, the set of balance equations in Eq.~\eqref{e:KL_globalequi0} simply expands to
\eqb{llll}
\ds \frac{D}{Dt} \int_{{\sR}}\rho\,\bv\, \dif a  \is \ds  \int_{{\sR}}\bff\, \dif a + \ds \int_{\partial{\sR}} \bT\, \dif s~, \\[7mm] 
 \boldsymbol{0}   \is \ds  \ds \int_{\partial{\sR}} \bM\, \dif s~, \\[7mm]
  \boldsymbol{0}   \is \ds  \ds \int_{\partial{\sR}} \bMbar\, \dif s~, \\[7mm]
\ds \frac{D}{Dt} \int_{{\sR}}\rho\,\bx\!\times\!\bv \, \dif a  \is  \ds \int_{{\sR}}\bx\!\times\!\bff  \, \dif a + \ds \int_{\partial{\sR}} \big(\bx\!\times\!\bT +  \bn\!\times\!{\bM} + \bc\!\times\!\bMbar\big)\, \dif s~
\label{e:KL_globalequi}
\eqe
for in-plane fiber bending. Here the third equation is  the linear momentum balance of the director field $\bc$, while vector $\bMbar$ denotes the stress couple  associated with in-plane bending, which is defined by Eq.~(\ref{e:defMbarcross}.3) in our theory, and thus it always points in fiber direction $\bell$, i.e. 
\eqb{llll}
\bMbar\cdot\bc = 0~.
\label{e:bMbarcdotbc}
\eqe
The local balance equations can then be obtained by inserting Eqs.~\eqref{e:tractionT} and \eqref{e:definedCoupleM_Mbar} into  \eqref{e:KL_globalequi}, considering \eqref{e:bTalp}, \eqref{e:defMbar}, \eqref{e:defMbaralp} and mass conservation. This gives
\eqb{lll}
\bT^\alpha_{;\alpha} \,  +\,  \bff \is \rho\,\dot{\bv}~,\\[4mm]
 \bM^\alpha_{;\alpha} \is \boldsymbol{0}~,\\[4mm]
 \bMbar^\alpha_{;\alpha} \is \boldsymbol{0}~,\\[4mm]
\ba_\alpha\!\times\!\bT^\alpha  \, +\,  \bn_{,\alpha}\!\times\!\bM^\alpha   \, +\,  \bc_{,\alpha}\!\times\!\bMbar^\alpha \, \is \boldsymbol{0}~.
\label{e:KL_localequi}
\eqe
Since we can write
\eqb{lll}
\bn_{,\alpha}\!\times\!\bM^\alpha \is (\bn\!\times\!\bM^\alpha)_{;\alpha} - \bn\!\times\!\bM^\alpha_{;\alpha}
 ~\\[3mm] 
 \bc_{,\alpha}\!\times\!\bM^\alpha \is (\bc\!\times\!\bMbar^\alpha)_{;\alpha} - \bc\!\times\!\bMbar^\alpha_{;\alpha}~,
\eqe
the last three equations in \eqref{e:KL_localequi} can be rewritten into 
\eqb{lll}
\ba_\alpha\!\times\!\bT^\alpha  \, +\, \bmhat^\alpha_{;\alpha}= \boldsymbol{0}~,
\label{e:localMomentumCosserat}
\eqe
where 
 $\hat\bm^\alpha := \bn\!\times\!\bM^\alpha + \bc\!\times\!\bMbar^\alpha$ is known from Eq.~\eqref{e:defbMa}. 
 Eq.~\eqref{e:localMomentumCosserat} is identical to the  local momentum balance Eq.~\eqref{e:Mbalance}, since   $\bmhat^\alpha_{;\alpha} = \mathrm{div}_\mrs\,\bmuhat^\mrT$, 
 which follows directly from Eq.~\eqref{e:div_vs_comma} when replacing $\bsig$ and $\bT^\alpha$ with $\bmuhat= \ba_\alpha\otimes\hat\bm^\alpha$ and $\bmhat^\alpha$, respectively
 %
 %
 %
  %
  %
  %
  
%
Therefore, we can conclude that the linear and angular momentum balance of our presented theory is equivalent to the set of balance equations of Cosserat theory under Kirchhoff-Love kinematics ($\bd = \bn$), suitable external loads ($\ml = \mk = \boldsymbol{0}$), and mass lumping at the mid-surface ($\alpha=0$ and $\beta=0$). In other words, the linear and angular momentum balance equations \eqref{e:linearmomentumglobal} and \eqref{e:angmomentglobal} of our presented theory fully characterize the equilibrium of the generalized Kirchhoff-Love shell with in-plane bending.
%

\begin{remark}
Note  here, that the total (equivalent) moment $\bmhat$ (see  Eq.~\eqref{e:physMmap}) can be defined from the couple vectors as $\bmhat = \hat\bm^\alpha\,\nu_\alpha = \bn\times\bM^\alpha\,\nu_\alpha  + \bc\times\bMbar^\alpha\,\nu_\alpha  =: \bm + \bmbar $, where $\bm = \bn\times\bM$ and $\bmbar=\bc\times\bMbar$ are the moment vectors causing out-of-plane and in-plane bending, respectively.  In contrast to the shell theory with in-plane bending of \cite{Steigmann2018},  these moment vectors are energetically equivalent to couple vectors $\bM$ and $\bMbar$ due to \eqref{e:bMcdotbn} and \eqref{e:bMbarcdotbc}. This is due to the fact that $\bM$ and $\bMbar$ in our theory only do work on  local rotations of the normalized vectors $\bn$ and $\bc$, respectively. Physically, this implies that $\bM$ and $\bMbar$ are not doing work when stretching the material along $\bn$ and $\bc$, respectively. 
\end{remark}

\section{Effective membrane stress in the existing second-gradient theory of Kirchhoff-Love shells}\label{s:app_effectivestress}

This section presents the intermediate steps to derive the effective membrane stress \eqref{e:eff_sigab} appearing in the existing second-gradient theory of Kirchhoff-Love shells \eqref{e:workpairGradientTheory0}, where the change in the relative Christopher symbol $S_{\alpha\beta}^\gamma$ is used as the strain measure of in-plane bending. It also shows that the resulting effective membrane stress is unsymmetric for general materials, even for initially straight fibers.

To this end, we first take the time derivative of the geodesic curvature \eqref{e:kgok}. This gives
\eqb{lll}
\dot{\kappa}_\mrg = \dot{\kappa}_\mrg^{\Gamma} + \dot{\kappa}_\mrg^{\mathrm{L}}~,
\label{e:timederivativekappa_mrg0}
\eqe
where
\eqb{lll}
\dot{\kappa}_\mrg^{\Gamma} \is  \dot{\ell}^{\alpha\beta}\,c_\gamma\, S_{\alpha\beta}^\gamma + \ellab\,\dot{c}_\gamma\, S_{\alpha\beta}^\gamma + \ellab\,c_\gamma\,\dot{S}_{\alpha\beta}^\gamma~,\\[3mm]
\dot{\kappa}_\mrg^{\mathrm{L}} \is \dot{\lambda}^{-1}\,c_\alpha\,\ell^\beta\,L^\alpha_{;\beta} + {\lambda}^{-1}\,\dot{c}_\alpha\,\ell^\beta\,L^\alpha_{;\beta} + {\lambda}^{-1}\,c_\alpha\,\dot{\ell}^\beta\,L^\alpha_{;\beta} ~.
\label{e:timederivativekappa_mrg}
\eqe
Here, the time derivatives  $\dot{\ell}^{\alpha\beta}$, $\dot{c}_\alpha$, $\dot\lambda$, and $\dot{\ell}^\alpha$ can be obtained from replacing $\delta\bullet$ by $\dot{\bullet}$ in expressions \eqref{e:var_ellab}, \eqref{e:var_c_alpha}, \eqref{e:deltalambdaC}, and \eqref{e:va_ellovalpha}, respectively. By taking these into account, Eq.~\eqref{e:timederivativekappa_mrg} becomes
\eqb{lll}
\dot{\kappa}_\mrg^{\Gamma} \is  - 2\,\kappa_\mrg^{\Gamma}\,\ellab\,\ba_\beta\cdot\dot\ba_\alpha + (\ell^{\gamma\delta}\,S_{\gamma\delta}^\alpha - \ell^{\gamma\delta}\,\ell_\theta\,S_{\gamma\delta}^\theta\,\ell^\alpha)\,c^\beta\,\ba_\beta\cdot\dot\ba_\alpha + \ellab\,c_\gamma\,\dot{S}_{\alpha\beta}^\gamma~,\\[3mm]
\dot{\kappa}_\mrg^{\mathrm{L}} \is - 2\,\kappa_\mrg^{\mathrm{L}}\,\ellab\,\ba_\beta\cdot\dot\ba_\alpha + {\lambda}^{-1}\, (\ell^\gamma\,L^\alpha_{;\gamma} - \ell_\gamma^\delta\,L^\gamma_{;\delta} \, \ell^\alpha )\,c^\beta\,\ba_\beta\cdot\dot\ba_\alpha ~.
\label{e:timederivativekappa_mrg2}
\eqe
With this, inserting \eqref{e:timederivativekappa_mrg0} into \eqref{e:Wint01} gives
  \eqb{rll}
 \dot{w}_{\mathrm{int}} \dis {\sigma}^{\alpha\beta}\,\ba_\beta\cdot\dot{\ba}_{\alpha}  + \Mab\,\dot{b}_{\alpha\beta} +   \bar{\mu}\,\ellab\,c_\gamma\,\dot{S}_{\alpha\beta}^\gamma ~, 
 \eqe
 where $\sigma^{\alpha\beta}$ is defined by Eq.~\eqref{e:eff_sigab}. It is valid for initially curved fibers.  
 For initially straight fibers,  $\dot{\kappa}_\mrg^{\mathrm{L}}$ in \eqref{e:timederivativekappa_mrg0} vanishes, so that $\sigma^{\alpha\beta}$ becomes 
\eqb{lll}
\sigab := \tilde\sigma^{\alpha\beta} + \bar{\mu}\, (\kappa^{\mathrm{L}}_\mrg-\kappa^{\Gamma}_\mrg) \,\ellab + \bar{\mu}\,\Big[(\ell^{\gamma\delta}\, S_{\gamma\delta}^\alpha)\,c^\beta -  (\ell^{\gamma\delta}\,S_{\gamma\delta}^{\theta}\,\,\ell_{\theta})\,\ell^{\alpha}\,c^\beta \Big]~,
\label{e:eff_sigab2}
\eqe
which is unsymetric for general materials. The asymmetry is due to the fact that the derivative  $\dot{c}_\alpha$ of fiber director components $c_\alpha$ appearing in (\ref{e:timederivativekappa_mrg}.1) contains not only surface stretching, but also in-plane bending. In other words, $\dot{c}_\alpha \neq \ds \pa{ c_\alpha}{a_{\beta\gamma}} \dot{a}_{\beta\gamma}$.

\vspace{1cm}
{\Large{\bf Acknowledgements}}

The authors are grateful to the German Research Foundation (DFG)
for supporting this research under grants  IT 67/18-1 and SA1822/11-1. 

\bigskip
\bibliographystyle{apalike}
\bibliography{sauerduong,bibliography}

\end{document}